\documentclass{aa}

\usepackage{graphicx}
\usepackage[varg]{txfonts}
\usepackage{natbib}

\bibpunct{(}{)}{;}{a}{}{,} 

\newcommand{\tx}{T_{\rm X}}
\newcommand{\kms}{km\,s$^{-1}$}
\newcommand{\Msun}{M$_{\odot}$} 
\newcommand{\teff}{T_{\rm eff}}
\newcommand{\ml}{M$_{\odot}$\,yr$^{-1}$}
\newcommand{\Tx}{T_{\rm X}}

\newcommand{\Lx}{L_{\rm X}}
\newcommand{\nh}{N_{\rm H}}
\newcommand{\BD}{BD\,+30}
 
\newcommand{\ovii}{\ion{O}{vii}}
\newcommand{\oviii}{\ion{O}{viii}}
\newcommand{\nex}{\ion{Ne}{x}}
\newcommand{\neix}{\ion{Ne}{ix}}
\newcommand{\mgxii}{\ion{Mg}{xii}}
\newcommand{\mgxi}{\ion{Mg}{xi}}
\newcommand{\cvi}{\ion{C}{vi}}
\newcommand{\cv}{\ion{C}{v}}
\newcommand{\nvii}{\ion{N}{vii}}

\defcitealias{sandin16}{Paper\,I}
\defcitealias{1996A&A...309..648Z}{ZP96}

\def\changed{}
\begin{document}

\title{Hot bubbles of planetary nebulae with hydrogen-deficient winds}

\subtitle{II. Analytical approximations with application to BD\,$+30^\circ3639$}

\author{R. Heller\inst{1,2}  \and R. Jacob\inst{1} \and 
        D. Sch\"onberner\inst{1} \and M. Steffen\inst{1}       
        }

\institute{
Leibniz-Institut f\"ur Astrophysik Potsdam, An der Sternwarte 16, 14482 Potsdam, Germany\\
 \email{msteffen@aip.de, deschoenberner@aip.de}\label{inst1}
\and Max-Planck-Institut f\"ur Sonnensystemforschung, Justus-von-Liebig-Weg 3, 
     37077 G\"ottingen, Germany\\
     \email{heller@mps.mpg.de}
          }

\titlerunning{Hot bubbles of planetary nebulae with hydrogen-deficient winds II.}

\authorrunning{Heller et al.}

\date{Received 22 January 2018 / Accepted 13 September 2018}

\abstract
{  The first high-resolution X-ray spectroscopy of a planetary nebula, BD\,+30$^\circ$\,3639,
  opened the possibility to study plasma conditions and chemical compositions of X-ray emitting 
   ``hot'' bubbles of planetary nebulae in much greater detail than before. 
}   
{  We investigate (i) how diagnostic line ratios 
   are influenced by the bubble's thermal structure and chemical profile, (ii) whether the
   chemical composition inside the bubble of BD\,+30$^\circ$\,3639 is consistent with the
   hydrogen-poor composition of the stellar photosphere/wind, and (iii) whether 
   hydrogen-rich nebular matter has already been added to the bubble of BD\,+30$^\circ$\,3639
   by evaporation.
   }
{  We applied an analytical, 1-D model for wind-blown bubbles with temperature
   and density profiles based on self-similar solutions including thermal conduction.
    We constructed also heat-conduction bubbles with a chemical stratification.  
   The X-ray emission is computed using the well-documented CHIANTI code.   
   These bubble models are used to re-analyse the high-resolution X-ray spectrum from the
   hot bubble of BD\,+30$^\circ$\,3639.
   }
{ We found that our 1-D \changed{heat-conducting} bubble models reproduce the observed line
  ratios much better than \changed{plasmas with single} electron temperatures. 
  In particular, all the temperature- and abundance-sensitive line ratios are consistent with
  BD\,+30\degr\,3639 X-ray observations for (i) an intervening column density of neutral
  hydrogen, $ \nh = 0.20_{-0.10}^{+0.05}\times\!10^{22}\rm\ cm^{-2} $, (ii) a 
  characteristic bubble X-ray temperature of ${ \tx = 1.8\pm 0.1~ }$MK together with (iii) a 
  very high neon \changed{mass fraction of about 0.05, virtually} as high as \changed{that 
  of oxygen.} 
  For  lower values of $ \nh, $ it cannot be excluded that the hot bubble of \BD\degr\,3639 
  contains a small amount of ``evaporated'' (or mixed) hydrogen-rich nebular matter.  
  Given the possible range of $ \nh, $ the fraction of \changed{evaporated} hydrogen-rich 
  matter cannot exceed 3\,\% of the bubble mass. 
  }
{ The diffuse X-ray emission from BD\,+30$^\circ$\,3639 can well be explained by 
  models of wind-blown bubbles with thermal conduction and \changed{a chemical composition
  equal to that of the hydrogen-poor and carbon-, oxygen-, and neon-rich stellar surface.}
}

\keywords{conduction -- planetary nebulae: general --  
          planetary nebulae: individual: \object{BD\,$+30^\circ3639$} --
          stars: abundances -- X-rays: stars    }

\maketitle

\section{Introduction}
\label{sec:introduction}

In the late 1990s, space-based observations of the X-ray spectra \changed{from} planetary
 nebulae  have become possible.  Interpretation of nebular X-ray data from the 
 XMM-{\it Newton} and \textit{Chandra} telescopes then allowed essential new insights 
 into the evolution, chemistry, and structure of planetary nebulae
    \citep{2001ApJ...553L..69C, 2001ApJ...553L..55G, 2002A&A...387L...1G, guerrero.12,
2008ApJ...672..957K,2009ApJ...690..440Y,2009ApJ...695..834N,2010ApJ...721.1820M,
2011AJ....142...91R, ruiz.13, 2013ApJ...766...26M}.
\changed{Ongoing efforts have nowadays culminated in the targeted \textit{Chandra} Planetary
 Nebula Survey \citep[ChanPlaNS,][]{2012AJ....144...58K, 2014ApJ...794...99F,
  2015ApJ...800....8M}.}
  
\changed{ChanPlaNS is a volume-limited survey in which only planetary nebulae within
     a distance of $ \simeq\! 1.5 $~kpc were considered.  Altogether, 59 objects could be 
     compiled, either newly observed or retrieved (14) from the \textit{Chandra} archival
     data.  In general, X-rays have been detected as coming either from a point source at the
     position of the central star or as spatial extended (diffuse) emission emerging from the
     central ``cavity'' of the nebula, with detection
     rates of about 36\,\% for point sources and about 27\,\% for diffuse emissions
     \citep{2014ApJ...794...99F}.   
     {A small number of objects have both kinds of emission.} 
     Typically, the diffuse X-ray emission occurs in nebulae
     with a nested shell morphology where the X-rays are confined by the inner rim.   
     This diffuse emission is associated with rather compact nebulae with radii 
     $ \la\!0.15 $~pc only and occurs for about 60\,\% of them 
     \citep[][figure~3 therein]{2014ApJ...794...99F}.  It is still not clear
     whether this restriction of the diffuse X-ray emission to more compact objects
     is due to intrinsic wind properties as claimed by \citeauthor{2014ApJ...794...99F}
     or to observational selection because 
     the X-ray intensity may fall below the \textit{Chandra} detection limit during the
     continued expansion of the bubble (see figure~6 in \citealt{ruiz.13} or figure~12
     in \citealt{TA.16}). }

\changed{Interesting is the question about possible differences between the detection rates 
    of diffuse X-rays from nebulae with normal O-type or hydrogen-poor Wolf-Rayet 
    ([WR] spectral type) central stars.  
    According to \citet{2014ApJ...794...99F}, all four [WR]-type objects with rather
    compact nebulae observed so far with \textit{Chandra} show diffuse X-ray emission, 
    i.e. a detection rate of 100\,\%, which strongly contrasts the lower detection rate of
    corresponding nebulae with O-type central stars.}

An object of particular interest is BD\,$+30^\circ3639$ \citep[BD$+30$ for short, also known 
as ``Campbell's Star'';][]{1893PASP....5..204C}.  It was detected in X-rays by ROSAT 
\citep{krey.92} and turned out to be the brightest X-ray source \changed{of all 
 planetary nebulae.}
 \citet{arnaudetal.96} analysed an X-ray spectrum taken with the {ASCA} satellite and found a
 prominent emission of \ion{Ne}{ix} at 0.9~keV, suggesting a \changed{typical 
  temperature of about ${\approx\!3}$~million K (MK)} for the X-ray emitting plasma.
 Observations with the Advanced CCD Imaging Spectrometer (ACIS) aboard the \textit{Chandra}
 space telescope \citep{2000ApJ...545L..57K} showed a well-resolved X-ray emitting region and
 indicated that its ``hot bubble'' (HB)
 is asymmetric. The same observations led \citet{2000ApJ...545L..57K} to confirm 
 the HB's characteristic X-ray temperature of ${T_\mathrm{X}\approx3}$~MK. 
Using observations with the \textit{Suzaku} satellite, \citet{muaretal.06} were able to 
estimate the ratios carbon-to-oyxgen (C/O) and neon-to-oxygen (Ne/O) which exceeded the solar 
\changed{ratios} by factors of at least 30 and 5, respectively.  
   In the context of a multi-wavelength study of BD\,+30, \citet{FK.16} also re-analysed the
   existing \textit{Chandra} spectrum by means of a \changed{one}-temperature thermal plasma
   model and came up with $ \tx \simeq 2.6 $~MK for the X-ray emitting plasma. 

 \citet{2009ApJ...690..440Y} were able to re-observe BD$+30$ with \textit{Chandra} to take the
  first X-ray gratings spectrum of a planetary nebula, using the Low Energy Transmission
  Gratings in combination with the Advanced CCD Imaging Spectrometer (LETG/ACIS-S). 
  These high-resolution data allowed a detailed chemical and thermal characterisation of
  BD$+30$'s HB.   Based on a one-temperature plasma model,
   \citet{2009ApJ...690..440Y} derived a characteristic X-ray temperature of 2.3 MK.  
   An even better fit to the observation was achieved using a two-component   
   plasma model in which \changed{the two components were found} to have a 
   temperature of ${T_\mathrm{X} = 1.7}$ and 2.9~MK, respectively.  
 \changed{This finding implies that there exists a distinct temperature gradient in \BD`s HB.}     
   For the element ratios C/O and Ne/O, \citet{2009ApJ...690..440Y} found similar excesses
   as \citet{muaretal.06}, viz. factors relative to solar of about 15--45 and 3.3--5.0,
   respectively.

In a similar fashion, \citet{2009ApJ...695..834N} fitted a two-component plasma model to the
 \textit{Chandra} spectrum of BD$+30$ and found best-fit values of the X-ray temperatures of
  1.9 and 3.0 MK, respectively. 
  Beyond that, \citet{2009ApJ...695..834N} used observations of BD$+30$ with \textit{Chandra}'s
  LETG Spectrometer to constrain the temperature jump at the contact discontinuity between the
  HB and the nebular rim.  They concluded that the jump should be $\ga\!930\,000$\,K, thereby
  also constraining the efficiency of heat conduction \changed{and/or mixing of matter across
  the bubble-nebula interface (see below).} 

   The investigations of \citet{2009ApJ...690..440Y} and \citet{2009ApJ...695..834N}
   support the expectation that the chemical composition of the bubble gas reflects
   the photospheric composition of BD\,$+30$ which is extremely hydrogen-poor 
   and rich in helium, carbon, and oxygen:   
\changed{The authors found very high C/O and Ne/O abundance ratios which are not observed
   in the nebula but which are consistent with the stellar surface abundances found by
   \citet{leuetal.96} and \citet{2007ApJ...654.1068M}.}

   \changed{The works of \citet{2009ApJ...690..440Y} and \citet{2009ApJ...695..834N}} suggest
   also the presence of a significant radial temperature gradient across the bubble which,
   if confirmed, makes \changed{single}-temperature approaches questionable.  As shown by
   \citet{2008A&A...489..173S}, heat conduction is \changed{an important physical process
   for the determinion of  the bubble structure:} 
   Conduction naturally leads to a \changed{typical}      
   temperature distribution inside a HB with a very steep temperature
   gradient at the conduction front (i.e. the bubble-nebula interface), as energy is
   transported outwards from the hot inner stellar wind to the much cooler nebular region. 
   
   \changed{There exist observational indications that stellar winds are inhomogeneous, i.e.
   ``clumpy'' (cf. \citealt{2007ApJ...654.1068M}).
   Instead (or in addition) of heat conduction, ``mass-loading'' by either
   hydrodynamic ablation or conductive evaporation of ``clumps'' can thus play a role.  
   The analytical studies of 
   \citet{pit.01a, pit.01b} show that the bubble density (temperature) will be increased 
   (decreased) with respect to the adiabatic case.  The detailed bubble structure depends,
   however, critically on the assumed boundary conditions (wind  parameters and density 
   profile of the environment) and the clump distribution.   These studies did not include
   heat conduction, but we expect that it would dominate anyway
   because heat-conduction changes the bubble structure ``instantaneously'' 
   \citep{1996A&A...309..648Z}. }

  Another possibility to reduce the temperature and to increase the density of  
  X-ray emitting plasma is mixing of the bubble matter with cool nebular gas across the 
  bubble-nebula interface (``contact discontinuity'') by Rayleigh-Taylor instabilities. 
  \changed{The first ``pilot'' 2-D simulations of \citet[][]{SS.06} cover, however, only a 
  very limited time span ($\la$\,300~yr) and simple boundary conditons, and are thus not really
  suited for drawing conclusions concerning the temporal evolution of the mixing efficiency.}
 
\changed{Much more realistic 2-D simulations have been presented by 
   \citet{TA.14, TA.16, TA.16b}.  They are based on post-AGB evolutionary tracks and realistic 
   wind models in a similar manner as the simulations by \citet{villa.02} or
   \citet{peretal.04} and show clearly that mixing of bubble and nebular matter across the
   bubble-nebula interface generates a region of gas with intermediate temperatures 
   ($ \sim$\,MK) and densities, well suited to emit X-rays of the observed properties. 
   The inclusion of heat conduction increases the amount of  gas with these properties,
   and thus also the X-ray emission measure of the bubble, considerably.}
           
\changed{A closer inspection of the \citeauthor{TA.16}'s models shows, however, that they
   seem to overestimate the mixing process for the following reason:  the formation of a
   bright, sharply bounded nebular rim is obviously not possible 
   \citep[see figure~11 in][]{TA.16b}.  This is cleary in contrast to the observations where 
   nebular rims, once formed, persist over the whole nebular lifetime despite the obvious
   occurrence of instabilities.   Moreover, 1-D simulations are
   quite successful in explaining the rim-shell morphology of planetary nebulae and their
   evolution with time \citep[cf.][]{schoenetal.14}.  We conclude that mixing, although it
   certainly does exist, cannot possibly generate sufficient amounts of gas responsible
   for the X-ray emission from planetary nebulae.  Instead, it is mandatory to invoke heat
   conduction for explaining the observations.}         
           
   In general, thermal conduction changes the {\em global} structure of the bubble. 
   Furthermore, the ``evaporation'' of nebular gas increases steadily 
   the mass of the bubble during the evolution and dominates soon the bubble's mass budget.     
   If the bubble gas has a different chemical composition than the enclosing nebula, as
   is the case considered here, a chemical discontunity will move from the conduction front
   inwards.   If it is possible to detect, or to constrain, the position of a chemical
   discontinuity or the amount of evaporated matter, important insights as to the  
   formation and evolution of \changed{[WR]} central stars could be derived, 
   apart from the proof that heat conduction is effective.
      
   In \citet[][hereafter Paper I]{sandin16}, detailed \mbox{1-D} 
   rad\-iation-hydrodynamics
   simulations of \changed{planetary-nebula models in which a stellar wind with a typical
   hydrogen-poor [WR]-composition} collides with a circumstellar envelope of hydrogen-rich
   composition were performed.  These models describe the formation and evolution of 
   hydrogen-poor bubbles inside normal nebulae including heat conduction and the 
   associated evaporation of hydrogen-rich gas into the bubble.   
\changed{It turned out that heat conduction (i) delays the formation of a bubble consisting 
   of hydrogen-poor but carbon- and oxygen-rich matter considerably because of the high
   efficiency of radiation cooling 
   around $ 0.1 $~MK \citep[see figure~1 in][]{ML.02}, and (ii) does not lead necessarily to
   immediate evaporation \citepalias[compare figures~5 and 6 with figure 7 in][]{sandin16}. }
   
\changed{With all these still unsettled problems in mind, we considered it worthwhile to
   re-examine existing studies concerning the brightest X-ray source amongst planetary nebulae,
   \BD.}  
   As the hydrodynamical model simulations are very time consuming, an extensive
   grid of model sequences with various choices of the relevant parameters is prohibitive.  
   In the present paper, therefore, we present analytical, self-similar spherical  models of 
   hot bubbles with thermal conduction based on the formulations of
   \citet[][hereafter ZP96]{1996A&A...309..648Z, ZP.98}, but with various chemical 
   compositions (hydrogen-rich, hydrogen-poor, and stratified), and examine their evolution.
   Questions to be addressed are the following:  
\begin{itemize} 
\item
   Are one- (or two-)component temperature models sufficient for analysing the X-ray
   spectrum from a hot bubble whose structure is dominated by thermal conduction?
\item
   How is the X-ray spectrum and how are diagnostic line ratios, such as 
   \ion{O}{viii}/\ion{O}{vii} and \ion{Ne}{x}/\ion{Ne}{ix}, influenced by the bubble's 
   temperature profile and chemical composition? 
\item   
   Is the bubble abundance distribution consistent with the extremely 
   hydrogen-poor and carbon-rich chemistry at the surface of the star? 
\item  Are there any indications that nebular hydrogen-rich matter evaporated into the 
       hydrogen-poor bubble of BD\,+30 by heat conduction? 
\end{itemize}  
 
  The paper is organised as follows: Section \ref{sec:methods} explains the basic
  ingredients of our ZP96 bubbles controlled by thermal conduction, i.e. how they 
  are constructed, how we compute their X-ray spectra, other relevant facts for 
  interpreting the X-ray lines, and also the simplifications of our modelling.
  In Sect.~\ref{sec:test} we discuss the differences of the analytical ZP96 bubbles to
  those computed by means of radiation-hydrodynamics simulations and show that  
  analytical bubbles are able to reproduce the X-ray observations for BD\,+30.
  Section~\ref{sec:results} is devoted to the analysis 
  of the high-resolution X-ray line spectrum of BD+30 published in 
  \citet{2009ApJ...690..440Y} based on ZP96 bubbles with homogeneous hydrogen-deficient WR
  chemical composition.  
  In Sect.~\ref{sec:inhomogeneous.bubble}, we introduce bubbles with inhomogeneous chemical
  composition and address the question whether the hot bubble
  of \BD\ may already contain a small amount of evaporated hydrogen-rich nebular matter.
  After the discussion (Sect.~\ref{sec:discussion}), the paper closes in
  Sect.~\ref{sec:conclusions} with a summary and the conclusions.
  
   Preliminary results of this work have been presented at IAU Symposium No. 323
   \citep{schoenetal.17}.


\begin{table*}[!th]

\centering
  \caption{Parameter grid of our bubble simulations.}
  \label{tab:grid}
  \renewcommand\arraystretch{1.4}
  \begin{tabular}{c|c|c|c|c|c}
    \hline \hline

    & Age [yr] & $\dot{M}_\mathrm{sw}\ [M_\odot\,\mathrm{yr^{-1}}]$ & $\varv_\mathrm{sw}$
  [\kms] & $C$  [erg\,cm$^{-1}$\,s$^{-1}$\,K$^{-7/2}$]& \\

    \hline

& 200, \ldots, 1000 & $10^{-7}$, $5{\times}10^{-7}$, \ldots, $10^{-4}$ & 10, 20, 25, 30, 40  & 
   $(3.0, \, 4.5, \, 6.0){\times}10^{-7}$ \\
 
     \hline
 
Total values & 9 & 7 & 5 & 3 & = 945 combinations \\
 
 \hline
    
  \end{tabular}
\end{table*}

\section{Methods}
\label{sec:methods}

\subsection{Temperature and density structure of a bubble}
\label{subsec:structure}
 
  The equations of ZP96 are based on the work of \citet{1977ApJ...218..377W} on bubbles 
  formed by spherically interacting winds with inclusion of heat conduction 
  but take into account also the wind evolution of the 
  central star.  They provide us with the temperature and ion density structures as a function 
  \changed{of time and bubble radius.} Most importantly, this analytical, self-similar 
  model is much faster to implement numerically than a full hydrodynamical code.   
  It is assumed (1) that the HB is hot enough (${\gtrsim\!10^6}$\,K) to be isobaric, (2) that 
  the optical nebula (more precisely the nebular rim) is  very thin compared 
  to the extension of the HB (ZP96 call this the ``thin-shell approximation''\,\footnote
  {This condition means that the radius of the conduction front we are 
  interested in is about equal to the swept-up shell (i.e. of the rim). 
  The slow wind is characterising the flow of matter ahead of the rim, which is here
  the ionised shell of the planetary nebula where usually most of the nebular matter resides
  and which is driven by thermal pressure caused by ionisation heating.}),
    and (3) that the rim expands much faster than the outer slow wind.
\changed{We add that the ZP96 models explicitly account for the mass accretion through
    the outer boundary (conduction front) due to evaporation of ambient (rim) matter.}         

  The efficiency of heat conduction is characterised by the conduction constant $C$, 
  which ranges between ${3\times}10^{-7}$ erg\,cm$^{-1}$\,s$^{-1}$\,K$^{-7/2}$ for a 
  hydrogen-free WR composition (\changed{see next section}) and $6{\times}10^{-7}$
  erg\,cm$^{-1}$\,s$^{-1}$\,K$^{-7/2}$ for a bubble consisting of hydrogen-rich matter.   
  The equations of ZP96 can only be solved for constant $C$,  
  whereas in realistic bubbles with a chemical gradient, $C$ depends on the radial distance 
  to the central object, $r$.  Hence we used  also an intermediate
  value of $4.5{\times}10^{-7}$ erg\,cm$^{-1}$\,s$^{-1}$\,K$^{-7/2}$ applicable for 
  bubbles with a stratified composition,
  i.e. for those where the outer bubble layers consist of H-rich instead of WR matter. 
  For the dependence of the heat transfer efficiency $ C $ on
  effective ion charge $Z$, or plasma abundance, see \citet{2012IAUS..283..494S} and
  \citetalias{sandin16}.  \changed{In this context we want to remark that Christer Sandin 
  (priv. commun.) pointed out that very often a conduction constant $ C $ is used in the
  literature which is a factor of two too high.}\footnote
  {The conduction constant $C$ is defined via the coefficient $D$ of the heat
   equation ${\vec{q} = D\,\nabla T_{\rm e}}$, ${D= C\,T_{\rm e}^{5/2}}$, where  
   $\vec{q}$ is the heat flux and $T_{\rm e}$ the electron temperature. Following 
   \citetalias[][\changed{figures~2 and 8} therein]{sandin16}, it turns out that, for a 
   (pure) hydrogen plasma, $C = 6.0\times\!10^{-7}$ erg\,cm$^{-1}$\,s$^{-1}$\,K$^{-7/2}$, for 
   ${\ln\Lambda = 30}$.  This is in contrast to the value $1.2\times\!10^{-6}$
   erg\,cm$^{-1}$\,s$^{-1}$\,K$^{-7/2}$ given by \citet{castor.75} and which is 
   predominantly but erraneously used 
   in the literature (but see also \citealt{borketal.90} who used the value given here).
   The difference is obvious due to the neglect of the factor $\epsilon$ (= 0.42) which 
   has been recommended by \citet{spitzer}; see also \citet{cokee.77}.  The effective 
   charge $Z$ of the WR composition used here (cf. Table~\ref{tab:abundances}) is about 4
   for the case of full ionisation,
   hence the conduction efficiency is reduced by a factor of about 2 (see table~1 in
   \citetalias{sandin16}), and  ${C = 3\!\times\!10^{-7}}$
   erg\,cm$^{-1}$\,s$^{-1}$\,K$^{-7/2}$ follows for the WR plasma.
   }

A range of boundary conditions and input parameters is required to solve for the HB temperature
  and density structures. These constraints are given by the velocity of the fast wind from the
  central star, $\varv_\mathrm{fw}(t)$, the mass-loss rate mediated by this wind,
 $\dot{M}_\mathrm{fw}(t)$, the constant velocity of the slow outer wind, $\varv_\mathrm{sw}$,
  and the constant mass-loss rate of the outer wind, $\dot{M}_\mathrm{sw}$ (see ZP96 for
  details). We numerically solve their differential equation (\changed{equation~9} in ZP96) for
  the dimensionless temperature $\xi$, starting from ${\xi\approx1}$ (at the outer boundary of
  the HB).  Together with \changed{equation~(5)} in ZP96 for the position of the reverse wind
  shock, the full structure of the bubble follows after conversion of the dimensionless
  similarity solution into real physical units.

   Constancy of $\dot{M}_\mathrm{sw}$ and $v_\mathrm{sw}$ means that the ambient density 
   immediately ahead of the rim falls off with distance as $r^{-2}$.  
   Hydrodynamical models show that the shell is a rarefaction wave such that the 
   flow velocity at its inner edge (i.e. ahead of the rim) does only change by tens of \kms\
   during the whole evolution while the stellar wind luminosity varies by orders of
   magnitude at the same time. 
   Thus, we feel that the assumption of constant slow-wind parameters is an acceptible
   compromise. 
   
   The range of the mass-loss rate and flow velocity of the slow wind that we used as 
   outer boundary conditions for the bubble evolution is given in Table~\ref{tab:grid}.  
   Together with three choices of the conduction parameter and the chosen ages, a total
   of 945 parameter combinations follows.  The actual number of models is, however, lower
   because of parameter degeneracy due to the fact that the bubbles' inner and outer radii
   (and their temperature and density distributions as well) are functions of 
   $\dot{M}_\mathrm{sw}/v_\mathrm{sw}$ 
   \citepalias[\changed{equations}~10--12 in][]{1996A&A...309..648Z}.  
   
   {\citet{HS.87} found also self-similar solutions for wind-blown bubbles with
   heat conduction. The difference to the ZP96 models used here are the boundary conditions:
   \citeauthor{HS.87} applied variable outer boundary conditions but kept the inner, i.e.
   the stellar wind paramters, fixed.  This approach is applicable to the case
   of massive \changed{O-type stars} whose winds remain virtually constant over the expansion
   time of the bubble \changed{and} cannot be used here where the stellar parameters, i.e. 
   also the stellar wind, change on timescales that are \changed{comparable or even shorter}
   than the expansion time of the wind-blown bubble. }

\subsection{Parametrisation of the \changed{system} }                                                                      
\label{subsec:parameters}

  ZP96 used the 0.605 \Msun\ central-star model sequence of \citet{B.95} to parametrise 
  the time dependence of the stellar wind parameters.   Here we want to model bubbles
  with hydrogen-deficient, helium- and carbon-enriched compositions as they are expected
  for planetaries around nuclei with \changed{[WR] spectral-type} characteristics.  
  As evolutionary 
  calculations are not available so far, we must resort to empirical relations for
  central-star and wind evolution.  We used therefore recent results of spectral analyses 
  of a large sample of [WR] central stars \citep[see, e.g.,][]{TH.15}, which are compiled in
  \changed{table~4 and visualised in figure~4} of Paper~I.    This figure shows that (i) the
  wind luminosity of [WR] central stars increases much faster with effective temperature
  than is predicted by \citet{pauldrach.88} for the case of normal, hydrogen-rich 
  central stars while (ii) the wind speed is considerably lower.  For instance, at the 
  stellar temperature of BD\,+30, ${\simeq\! 50\,000}$ K, [WR] central stars have 
  mass-loss rates higher by about 2 orders-of-magnitude, i.e. of about $10^{-6}$ \ml\ 
  (instead of $10^{-8}$ \ml) while the wind speed is of only about 800 \kms\ (instead of 
  1700 \kms).   Altogether, the important wind luminosity is about 10--20 times the 
  typical value expected for hydrogen-rich central stars 
  (cf. \changed{figure~4} in \citetalias{sandin16}). 
    
  Instead of the 0.605 \Msun\ post-AGB evolutionary track we applied here the 0.595~\Msun\ 
  track introduced by \citet[][see \changed{figure~1} therein]{schoenetal.05}. 
  Of course, the evolutionary speed of this model across 
  the HR diagram is based on hydrogen-burning together with the \citet{pauldrach.88} mass-loss 
  prescription, both of which are surely inadequate for a description of AGB remnants with a
  hydrogen-poor/free stellar surface.   In order to comply with the expected faster evolution
  of an AGB remnant which burns helium and which emits a comparatively more powerful wind,
  we accelerated the evolution of our 0.595~\Msun\ model by a factor of 5.5.  
  This acceleration ensures that the post-AGB age of the model is 730 years at 
  $\teff = 50\,000$~K, close enough to the estimated kinematical age of BD\,+30 of 800 years 
  \citep{2002AJ....123.2676L}.   This new evolutionary track is then considered to be an 
  approximate representative for the evolution of the central star of BD\,+30.\footnote
{We note that the ''kinematical age'' as used in the literature is actually the present
 time scale of evolution and may not represent the ``true'' post-AGB age.
 A thorough discussion of various definitions of kinematical time scales is given in
 \citet{schoenetal.14}. }     

\begin{figure}
\vskip -2mm
\hskip-8mm
\includegraphics[width=1.1\linewidth]{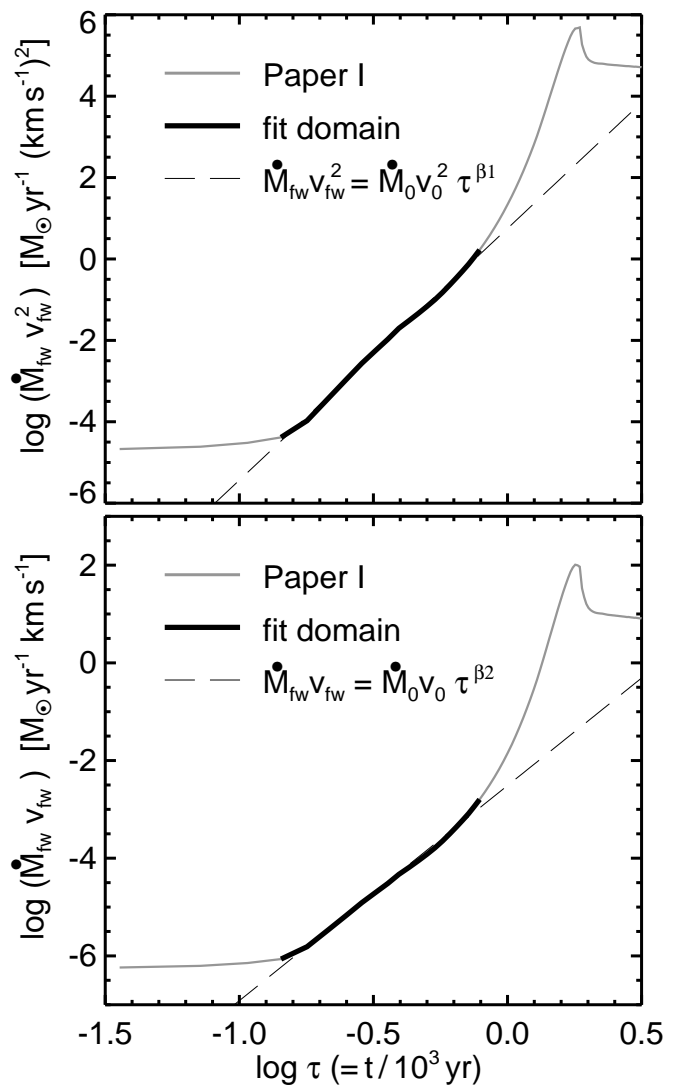}
\vskip-3mm
\caption{\label{fig:fit}     
         Variations of the quantities $\dot{M}_{\rm fw}{v}_{\rm fw}^2$ (\emph{top}) and 
         $\dot{M}_{\rm fw}{v}_{\rm fw}$ (\emph{bottom}) as a function of the
         dimensionless time parameter $\tau = t/10^3$\,yr.
         These quantities are adapted from \citetalias{sandin16} and linked to the
         parameters of a post-AGB model of 0.595~M$_\sun$ whose evolution speed
         across the HR diagram is increased by a factor of 5.5 (see text for explanation).
         The thick lines indicate the domain where the evolution of both quantities is
         well approximated by a linear fit (dashed) in the log-log plane.        
         }
         \vspace{-2mm}
\end{figure} 
  
  We proceed then by linking the mass-loss model as presented by the thin lines in figure~4 
  of \citetalias{sandin16} to the stellar parameters of our new 0.595~\Msun\ post-AGB model
  sequence.   For simplicity, this mass-loss model is an appropriately scaled 
  \citeauthor{pauldrach.88} model that reproduces the observed mass-loss parameters
  of BD\,+30 quite well and is characterised by $\dot{M}_{\rm fw}$ and  ${v}_{\rm fw}$.
  \citetalias{1996A&A...309..648Z} introduced  the quantities
  $\dot{M}_{\rm fw}{v}_{\rm fw}^2$ and $\dot{M}_{\rm fw}{v}_{\rm fw}$ whose time evolution   
  is presented in Fig.~\ref{fig:fit} in terms of the dimensionsless parameter 
   $\tau = t/10^3$\,yr for the case studied in this paper.   
  
  Figure~\ref{fig:fit} demonstrates clearly that during the most important part of 
  the modelled evolution, i.e. between about 160 and 800 years, both quantities can well be
  approximated by the power laws indicated in the figure.   
  These fits  fix the parameters $\beta_1=6.19$, $\beta_2=4.4$
  ($\beta=\beta_1/3+1=3.06$ for $\beta_\mathrm{p}=-2$), which are necessary for solving 
  \changed{equation}~(9) of ZP96.  The position of the reverse wind shock, the inner boundary
  of the bubble, is then fixed, too, by \changed{equation}~(5) of ZP96.
  Moreover, $\dot{M}_0{v}_0^2=5.62$ M$_\odot$\,yr$^{-1}$\,(\kms)$^2$
  and $\dot{M}_0{v}_0=3.07{\times} 10^{-3}$ M$_\odot$\,yr$^{-1}$\,\kms.       
  Having the wind evolution fixed, the only free parameters left are then (1) the age of
  the bubble/nebula, (2) density and velocity of the ambient matter (= nebular shell), given by 
  the constant values of $\dot{M}_\mathrm{sw}$ and $\varv_\mathrm{sw}$,  
  and (3) the coefficient of thermal conduction, $C$, 
  \changed{necessary for solving the dimensionless temperature equation of ZP96.}

\begin{table}
  \caption{\label{tab:abundances}  
           Chemical compositions of the stellar photosphere/wind (WR) and the nebular 
           gas (PN) used in this work, arranged by order of atomic number $Z$ as
           mass fractions (Cols. 3 and 5) and (logarithmic) number fractions ($\epsilon$)
           relative to hydrogen (Cols. 4 and 6).
           For comparison, the nebular abundances of BD\,+30 (number fractions, Col.~7)
           are given as well   \citep{pott.06}. }

\vskip -2mm
  \centering
  \tabcolsep=4.0pt
  \begin{tabular}{r l c r @{\hspace{5mm}} c r c}

    \hline \hline\noalign{\smallskip}

$Z$ & El. & \multicolumn{2}{c}{WR} & \multicolumn{2}{c}{PN} & PN$_{\rm BD\,+30}$
 \\[1.5pt]
  \cline{3-4}  \cline{5-6} \noalign{\smallskip}              
          &     & Mass & Num.    & Mass & Num.           & Num.  \\ 
          
    \hline\noalign{\smallskip}

 1  &  H   & 1.990(--02) & 12.00  &  6.841(--01) & 12.00  & 12.00\enspace \\
 2  &  He  & 4.080(--01) & 12.71  &  2.979(--01) & 11.04  & 11.11\enspace \\
 3  &  Li  & 5.931(--11) &  2.64  &  5.931(--11) &  1.10  & \\
 4  &  Be  & 1.537(--10) &  2.94  &  1.537(--10) &  1.40  & \\
 5  &  B   & 2.604(--09) &  4.09  &  2.604(--09) &  2.55  & \\
 6  &  C   & 4.950(--01) & 12.30  &  6.328(--03) &  8.89  &  8.86\\
 7  &  N   & 1.000(--05) &  7.56  &  2.334(--03) &  8.39  &  8.04\\
 8  &  O   & 5.200(--02) & 11.22  &  4.851(--03) &  8.65  &  8.66\\
 9  &  F   & 4.682(--07) &  6.10  &  4.682(--07) &  4.56  & \\
10  &  Ne  & 2.200(--02) & 10.70  &  1.402(--03) &  8.01  &  8.28\\
11  &  Na  & 3.336(--05) &  7.87  &  3.336(--05) &  6.33  & \\
12  &  Mg  & 6.272(--04) &  9.12  &  6.272(--04) &  7.58  & \\
13  &  Al  & 5.405(--05) &  8.01  &  5.405(--05) &  6.47  & \\
14  &  Si  & 6.764(--04) &  9.01  &  6.764(--04) &  7.55  & \\
15  &  P   & 5.925(--06) &  6.99  &  5.925(--06) &  5.45  & \\
16  &  S   & 2.386(--04) &  8.58  &  2.386(--04) &  7.04  &  6.81\\
17  &  Cl  & 5.028(--06) &  6.86  &  5.028(--06) &  5.32  & \\
18  &  Ar  & 7.820(--05) &  8.00  &  7.280(--05) &  6.46  &  6.72\\
19  &  K   & 3.498(--06) &  6.66  &  3.498(--06) &  5.12  & \\
20  &  Ca  & 6.232(--05) &  7.90  &  6.232(--05) &  6.36  & \\
21  &  Sc  & 4.513(--08) &  4.71  &  4.513(--08) &  3.17  & \\
22  &  Ti  & 3.403(--06) &  6.56  &  3.403(--06) &  5.02  & \\
23  &  V   & 3.458(--07) &  5.54  &  3.458(--07) &  4.00  & \\
24  &  Cr  & 1.651(--05) &  7.21  &  1.651(--05) &  5.67  & \\
25  &  Mn  & 9.153(--06) &  6.93  &  9.153(--06) &  5.39  & \\
26  &  Fe  & 1.200(--03) &  9.04  &  1.200(--03) &  7.50  & \\
27  &  Co  & 3.327(--06) &  6.46  &  3.327(--06) &  4.92  & \\
28  &  Ni  & 7.084(--05) &  7.79  &  7.084(--05) &  6.25  & \\
29  &  Cu  & 6.995(--07) &  5.75  &  6.995(--07) &  4.21  & \\
30  &  Zn  & 1.767(--06) &  6.14  &  1.767(--06) &  4.60  & \\
\hline
  \end{tabular}
  \vskip-2mm
  \tablefoot{The mass fractions are normalised such that their sum equals unity,            
             the logarithmic number fractions fulfill 
             ${\epsilon_{\rm EL} = 12\,+\,\log (N_{\rm El}/N_{\rm H})}$, where $N_{\rm El}$
             is the number density of the element in question. 
             The helium, carbon, and oxygen values of the WR composition in Cols.~3
             and 4 correspond to those measured in the wind of BD\,+30 by
             \citet{2007ApJ...654.1068M}.  \changed{Though the number abundances in 
             Col.~4 are normalised to hydrogen in the usual way, their values still reflect
             the hydrogen-poor composition from Col.~3.}  The PN composition is typical for
             \changed{planetary nebulae} in the disk of the Milky Way 
             \changed{\citep[e.g.,][]{KB.94}}
             \changed{and essentially solar with the exception of carbon and nitrogen
             because many nebulae (including BD\,+30) are somewhat self-enriched by 
             processed matter.}
             All elements not accessible to observations are assumed to have solar 
             mass fractions.            
          } 
          \vskip-2mm   
\end{table}

\begin{figure*}
\vskip -1mm
\includegraphics[width=0.98\textwidth]{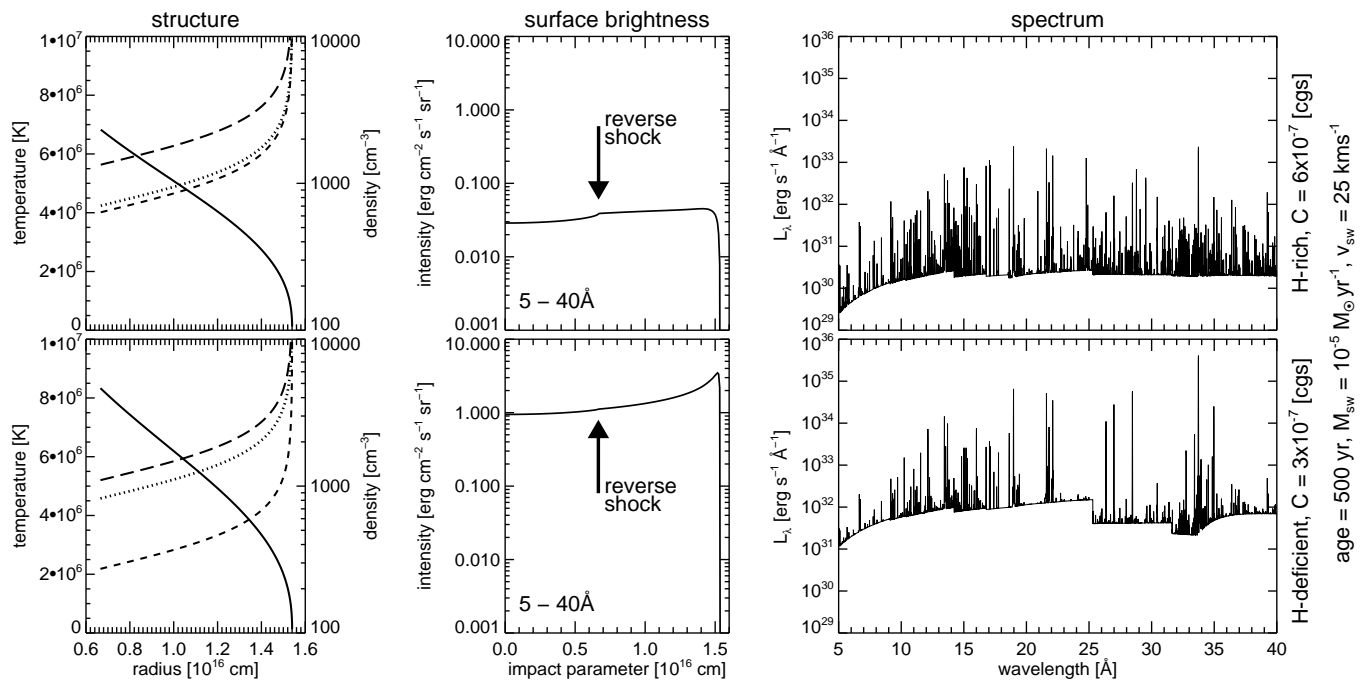}
\caption{\label{fig:bubble.example}   
         Examples of bubble structures (\emph{left}) and X-ray emissions (\emph{middle
         and right}) for hydrogen-rich (PN, \emph{top panels}) and 
         hydrogen-poor (WR, \emph{bottom panels}) composition.
         The inner boundaries are at the position of the reverse wind shock 
         ($0.67{\times}10^{16}$ cm), the outer boundaries at the (heat) conduction front at 
         $1.54{\times}10^{16}$ cm.  The planetary nebula proper is right 
         adjacent to this front.     Both HBs have a similar parameterisation: age = 500 yr,
         ${\dot{M}_{\rm sw} = 10^{-5}}$ \Msun\,yr$^{-1}$, ${\varv_{\rm sw}= 25}$ \kms, 
         but ${C = 6{\times}10^{-7}}$ (\emph{top})
         and ${C = 3{\times}10^{-7}}$ erg\,cm$^{-1}$\,s$^{-1}$\,K$^{-7/2}$ (\emph{bottom}).
         The bubble structures are characterised by the radial runs of electron
         temperature (solid), total particle number density (long dashed), ion number
         density (dashed), and electron density (dotted).  We note the linear scale for
         the temperature, but the logarithmic scale for the densities.  
         The characteristic X-ray temperatures of the bubbles are 2.9 MK (PN) and 2.4 MK (WR), 
         respectively.  The \changed{intrinsic} X-ray surface brightnesses of these two bubble
         models are displayed in the \emph{middle panels} and  correspond to the emission
         (spectral luminosity density) in wavelength bands of 5--40 \AA\
         (0.3--2.5 keV) shown in the \emph{right panels}.  The positions of the respective
         wind reverse shocks (= inner bubble boundaries) are indicated by the arrows.
         The X-ray luminosities in the
         given wavelength band are 3.7$\times 10^{32}$ erg\,s$^{-1}$ (PN) and 
         1.5$\times 10^{34}$ erg\,s$^{-1}$ (WR).
         }
\vspace{-1.5mm}
\end{figure*}

  Once the radial runs of temperature and density are solved for the selected $ C $, the 
  density is split into ion and electron densities according to the two abundance 
  distributions (either ``WR'' or ``PN'', depending on $ C $) listed in
  Table~\ref{tab:abundances}.
  Complete ionisation is assumed, which is not quite correct (see, e.g.,
  Fig.~\ref{fig:ion_fracfig2}) but a reasonable approximation for our purpose.
  
   The WR element distribution corresponds \changed{for the main elements He, C, and O} 
   closely to the photospheric composition of BD\,+30 after \citet{2007ApJ...654.1068M}, 
   \changed{while the PN element mixture is typical for Galactic-disk planetary nebulae.}  
   \citet{leuetal.96} arrived at nearly the same abundances for \BD\ as
   \citet{2007ApJ...654.1068M}.
   We note, however, that \citet{CMS.06} report a smaller C/O ratio of about 5 instead of 12 
   as used here,  and a correspondingly higher helium fraction. The photospheric hydrogen
   content of BD\,+30, \changed{if there is any, is uncertain,}    
   and we assume here a mass fraction of 2\,\%, i.e. a reduction by a factor of 34.    
   Except for the assumed small hydrogen content, the WR elemental mixture used here is 
   typical for the intershell region of AGB stars, exposed by some still unknown
   evolutionary process.\footnote
   {We kept a finite hydrogen content because the evolutionary history of \BD\ is not known
     a priori.  Anyway, the small amount used by us is also compatible with the assumption of 
     no hydrogen at all. 
   }
    
   Since we had to compute complete X-ray spectra with all existing lines included,
   all those elements that are not easily accessible to observations were supplemented,
   assuming solar mass fractions (see Table~\ref{tab:abundances}).  
   Due to the high oxygen content of the WR mixture, all these elements
   have abundances relative to oxygen of about one tenth solar (by number)!   
   Exceptions exist for nitrogen and neon.  
   Complete CNO hydrogen burning converts virtually all C and O nuclei
   into nitrogen, which  is later ``burned'' into neon ($^{22}$Ne) within the 
   pulse-driven, convective helium-burning shell, i.e. there is a one-to-one correspondence
   between the CNO ashes ($^{14}$N) and the neon produced during a thermal pulse.  
   Depending on the efficiency of the 3rd
   dredge-up, very high neon abundances may be produced within the intershell region.
   {The neon of the WR mixture is thus essentially $^{22}$Ne, in contrast to the  PN mixture
    where $^{20}$Ne is the dominant isotope. }      
 
   From the \changed{CNO abundances of the PN mixture (cf. Table~\ref{tab:abundances}, Col. 6), 
   we derived} an intershell neon abundance of 0.022 (by mass) for the stellar
   photosphere/wind, which we then adopted for our WR composition. 
   The nitrogen abundance was set virtually to zero (mass fraction of $10^{-5}$).
   We note in this context that \citet{2007ApJ...654.1068M} estimated a neon mass
   fraction of ${\sim\!2}$\,\%, consistent with the value used here
   (Table~\ref{tab:abundances}).

   The photospheric nitrogen content of BD\,+30 is not known. For a number of
   late-type [WR] central stars, however, nitrogen abundances of up to a few percent 
   (by mass) have been found \citep[cf. review of][]{TH.15}.  Such high amounts of nitrogen
   are a signature of simultaneous non-equilibrium burning and mixing of hydrogen at the
   interface between the envelope and the stellar core as it occurs if a thermal
   pulse happens when the central star evolves across the Hertzsprung-Russell diagram
   towards the white-dwarf stage (``late'' or ``very-late thermal pulse'').   
   Our choice of a virtually zero nitrogen abundance implies complete hydrogen burning 
   without mixing and complete loss of all unburned hydrogen such that the nitrogen-free,
   but neon-rich, intershell layers become exposed. 

   Figure\,\ref{fig:bubble.example} (left) illustrates the physical structure of a
   typical middle-aged HB, computed with the same parameters but different 
   heat conduction constants $C$ according to the two abundance sets used.  
   We see temperature distributions typical for heat conduction, with a very steep
   gradient at the conduction front at $1.54{\times}10^{16}$ cm distance from the
   star (located at the origin). 
   The bubble with the hydrogen-deficient composition is hotter with a somewhat steeper
   temperature gradient because of the smaller heat conduction efficiency.  The
   pressure in both bubbles is the same because of the identical boundary conditions, hence 
   the particle densities (ions plus electrons) must be different.   

  We repeat that the parameters for the slow wind, $\dot{M}_\mathrm{sw}$ and 
  $\varv_\mathrm{sw}$, into which the bubble and 
  its swept-up shell (i.e. the nebular rim) with the forward shock expands are
  assumed to remain constant throughout the whole bubble evolution.  In reality, the
  bubble and its swept-up shell (rim) is expanding into an expanding shell
  that has earlier been set up by photoionisation driven by the hot central star.
  Density and velocity immediately ahead of the forward shock are expected to
  change with time, i.e. $\dot{M}_\mathrm{sw}$ and $\varv_\mathrm{sw}$
  do not remain constant.  Also, their present values are not known, unless  
  a fully radiation-hydrodynamics model fitted to BD\,+30
  becomes available. We therefore adopted quite large ranges for $\dot{M}_\mathrm{sw}$
  and $\varv_\mathrm{sw}$ (cf. Cols.~3 and 4 in Table~\ref{tab:grid}) in order to cover
  all possible values of these parameters.

  From the physical point of view, the bubble's  mass increases by stellar wind matter
  passing through the reverse shock and by ``evaporation'' of nebular matter through the 
  heat conduction front, where the latter contribution to the bubble's mass budget even
  dominates in the later stages of evolution 
  \citep[][\changed{figures~6} and 8 therein]{2008A&A...489..173S}.  
  However, the chemical composition within the bubble is implicitly assumed to remain the 
  same during progress of time.  This is of no concern for models containing normal matter
   since wind and  evaporated gas from the nebula have the same composition.  
  Our bubble models with WR  composition are therefore physically inconsistent: 
  The evaporated matter is normal, 
  hydrogen-rich PN matter, and  since the latter is more important for the bubble's mass
  budget, a composition ``discontinuity'' will develop inside the bubble and will move 
  slowly inwards with time.   
  The actual position of this chemical discontinuity depends on the relative sizes 
  of the wind's mass input and the evaporated mass driven by thermal conduction 
  and how both develop with time.  
  
  In the context of these considerations,  
  the construction and time evolution of ZP96 bubbles with a homogeneous 
  hydrogen-poor chemical composition implicitly means that the evaporated matter has the 
  same hydrogen-poor composition as well, which, of course, is unrealistic. 
  Nevertheless, we study the properties of
  such models with homogeneous WR composition for two reasons: (i) The qualitative dependence
  of the bubble properties on the boundary conditions is independent of the 
  assumed chemical composition; (ii)  they are the basis for bubbles with
  inhomogeneous composition, i.e. bubble models with additional amounts of hydrogen-rich
  matter.    The construction and use of chemically inhomogeneous bubbles is, however,
  postponed to Sect.~\ref{sec:inhomogeneous.bubble}. 
   
  Despite the limitations necessary to derive analytical similarity solutions
  for heat-conducting wind-blown bubbles, 
  they turned out to be a very useful tool for investigating the physical properties of 
  these bubbles and analysing their X-ray emission in terms of temperature and 
  chemical composition, as we demonstrate in Sect.~\ref{sec:results}.

\subsection{\changed{Computation} of X-ray spectra with CHIANTI}
\label{subsec:chianti}

 We use CHIANTI \citep[v6.0.1,][]{1997A&AS..125..149D,2009A&A...498..915D} to 
\changed{compute the X-ray emission spectra of our ZP96 bubbles.}  
 One of the key distinctions between our analysis and earlier assumptions of isothermal plasma
 components lies in the capability of the ZP96 theory to model the temperature variation over  
 the HB radius. Different gas temperatures translate into different HB regions, where a given
  chemical element will show different stages of ionisation 
\changed{(see Fig.~\ref{fig:ioneq}).}   
These ionisation fractions are independent of the electron density.\footnote
{\changed{Since the beginning of our work, new versions of the CHIANTI software package have
 appeared, the latest beeing v8. We compared the relevant ionisation fractions generated by 
 the actual CHIANTI version with the corresponding fractions used throughout the paper and  
 could not find significant differences in the temperature range of interest.} 
}

We first compute the maximum temperature at the inner boundary of a HB 
(the position of the reverse shock) and  locate then the outer radius at which the temperature
 has decreased to $10^5$\,K. In young, relatively cool HBs, our procedure typically results 
 in about $30$ radial steps and an equal amount of sub-spectra. For evolved HBs with maximum
  temperatures of several MK, the steep temperature decrease towards the outer regions yields
  up to several hundred sub-spectra. These sub-spectra are then merged into an integrated,
  pseudo-observed spectrum but neglecting all the observational complications such as
  extinction and instrumental properties.   We restrict ourselves to the 
 spectral window between $5\,\AA$ and $40\,\AA$, according to the \textit{Chandra} observations
  of BD$+30$ taken by \citet{2009ApJ...690..440Y}, and used a spectral resolution of 
  $0.01\,\AA$ (10\,times better than \emph{Chandra} can provide) if not stated otherwise.

\begin{figure}
\includegraphics[width=0.99\linewidth]{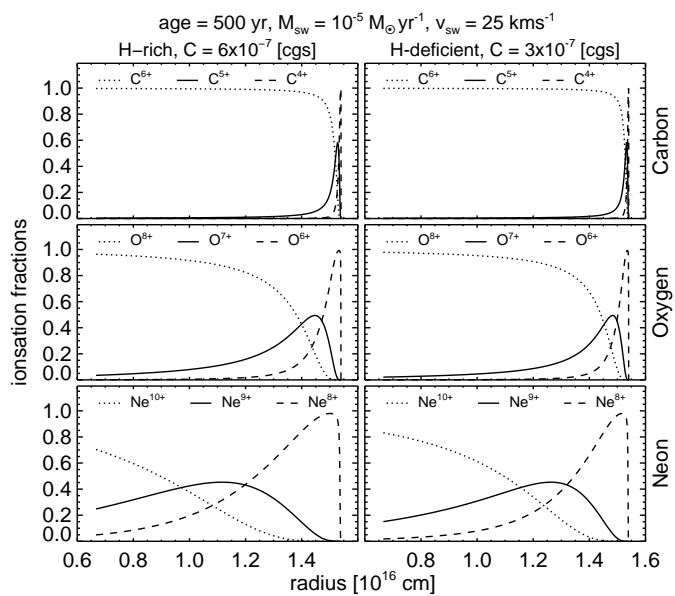}
\caption{\label{fig:ion_fracfig2}
         Radial distribution of ionic fractions of carbon (\emph{top}), oxygen
         (\emph{middle}), and neon (\emph{bottom}) inside the two bubbles shown 
         in Fig.~\ref{fig:bubble.example}. 
         \emph{Left column}: PN composition, \emph{right column}: WR composition.
         }
\end{figure}

   The results of our post-processing by means of the CHIANTI software are illustrated in
   Figs.~\ref{fig:bubble.example} and \ref{fig:ion_fracfig2}.  
   The latter figure illustrates how the ions are distributed within the two
   bubbles displayed in Fig.~\ref{fig:bubble.example} according to
   their respective temperature profiles:   In the bubble with WR composition, the 
   overall degree of ionisation is higher because of the generally higher bubble temperature 
   (cf. Fig.~\ref{fig:bubble.example}), although the characteristic temperature $\Tx$ is lower: 
   2.4 vs. 2.9~MK.\footnote
{For the definition of the characteristic X-ray temperature, $ \tx $, see Eq.~(\ref{eq:tx})
 in Sect.~\ref{subsec:measurements}. }

   In Fig.~\ref{fig:bubble.example}, the middle and right panels display the
   \changed{intrinsic} surface brightnesses and 
   flux distributions of the respective bubbles shown in the left
   panels.  One notices immediately the large difference of the strength of the X-ray
   emission:    The continuum emission of the hydrogen-deficient WR bubble is between 
   one to two orders of magnitudes higher than that of the hydrogen-rich PN  bubble 
   because the mean ion charge of the WR mixture  is much higher, 4.5 as compared to 
   1.4 for PN matter \citepalias[see][]{sandin16}.  
    
   The surface brightness is strongly limb brightened in the WR case because of the much
   steeper density decrease towards the conduction front, which in turn reflects the
   run of the electron temperature for lower conduction efficiency.  Furthermore,
   the line ``forest'' appears weaker for the WR composition because the abundance ratios
   of all elements to those of carbon, oxygen, and neon are considerably reduced, even if
   the former have solar abundances (see the discussion of Table~\ref{tab:abundances} in
   the previous Sect.~\ref{subsec:parameters}).

\changed{Figure~\ref{fig:bubble.example} provides obviously a simple explanation for the high 
   X-ray detection rate of planetary nebulae with [WR]-type nuclei:  Provided the 
   respective wind-blown bubbles consist of hydrogen-poor but helium-, carbon-, and 
   oxygen-rich gas, their intrinsic X-ray intensities are much higher than those of their
   hydrogen-rich counterparts.}

\begin{figure*} 
\centering 
\includegraphics[width=0.98\textwidth]{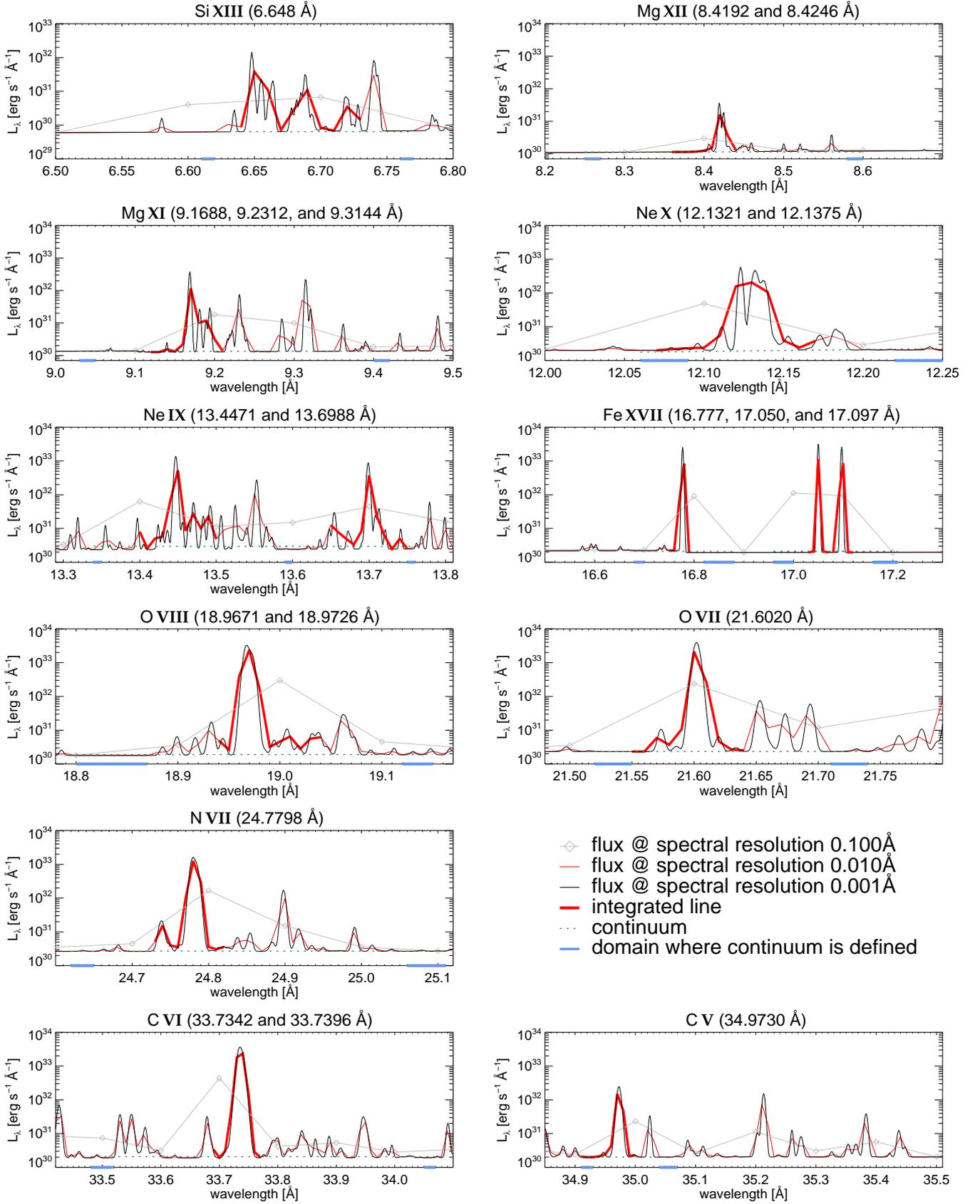} 
  \caption{\label{fig:HR_contamination}
  Zoom into spectral windows of the CHIANTI output which were considered for our analyses. 
   Note the logarithmic scale on the ordinate and the numerous \changed{contributing} lines. 
   Without loss of generality, the bubble used is that shown in the top panels of
   Fig.~\ref{fig:bubble.example}.  The hydrogen-rich PN case has been chosen 
   in order to highlighten the \ion{N}{vii} line which is virtually absent
   for the WR composition.  Shown are the spectra at different resolutions (see legend), the
   spectral range and resolution chosen for the line integration (red), 
   the continuum (short-dashed), and the
   domains where the continuum is defined for the integrations (blue, on the abscissae).   
           }  
\end{figure*}

\subsection{\changed{The characteristic {X-ray} temperature and line measurements} }
\label{subsec:measurements}

\changed{The characteristic temperature of the X-ray emitting plasma of a HB can be defined as}
\begin{equation}\label{eq:tx}
T_\mathrm{X} = \frac{4\pi}{L_\mathrm{X}} \int_{r_1}^{r_2}  r^2 \, T_\mathrm{e}(r) \, \eta_\mathrm{X}(r)\,\mathrm{d}r \,.
\end{equation}
\noindent
Here, $T_\mathrm{e}(r)$ is the electron temperature between \changed{inner bubble radius
 $r_1$ (position of the reverse wind shock) and outer bubble radius $r_2$ 
 (conduction front),}
\begin{equation}\label{eq:lx}
L_\mathrm{X} = 4\pi \int_{r_1}^{r_2}  r^2 \, \eta_\mathrm{X}(r)\,\mathrm{d}r 
\end{equation}
\noindent
is the X-ray luminosity, and 
\begin{equation}
\eta_\mathrm{X}(r) = \int_{E_1}^{E_2}   \eta(T_\mathrm{e}(r), n_\mathrm{e}(r), E)\,   \mathrm{d}E 
\end{equation}
   is the volume emissivity in the energy band $E_1 - E_2$ 
   \citep[equations~17--19 in][]{2008A&A...489..173S}.   
   The temperature $\tx$ as defined here is the typical plasma temperature where
   most of the X-ray emission is generated.  
   \changed{For our ZP96 bubbles, the radial temperature profile, $ T_{\rm e}(r) $, is 
   determined by heat conduction.}
    
   The definition of $ \tx $ in Eq.~(\ref{eq:tx}) implies that its value depends, via the
   emissivity, on the chemistry, even if the temperature profile is the same. 
   However, another chemistry also implies a different conduction efficiency and hence a
   different temperature profile inside the bubble. 
   This has been demonstrated in Fig.~\ref{fig:bubble.example} above:  
   the bubble with hydrogen-poor WR composition
   has a steeper temperature gradient because of its lower conduction efficiency, 
   hence more matter (and emissivity) is concentrated towards
   the conduction front where the electron temperature is lower.  
   Thus, according to Eq.~(\ref{eq:tx}), the bubble with WR composition has a lower $ \tx $ 
   value than the bubble with PN composition (cf. Fig.~\ref{fig:bubble.example}). 
 
    A similar effect occurs for  bubbles with an inhomogeneous chemical composition: 
   the value of $\Tx$ depends on the position of the chemical discontinuity in bubbles where 
   the inner region consists of original WR matter but the outer part of evaporated 
   hydrogen-rich PN matter.  The reason here are mainly the very different emissivities of 
   the WR and PN elemental mixtures.  Bubbles with inhomogeneous composition will be dicussed 
   in Sect.~\ref{sec:inhomogeneous.bubble} in detail.

\begin{table}
\centering
  \caption{Lines used for our line ratio analysis (see also Fig.~\ref{fig:HR_contamination}). 
          }
  \label{tab:lines}
  \begin{tabular}{l c}

    \hline\hline \noalign{\smallskip}

    Ion & Wavelength [$\AA$]\\

    \hline\noalign{\smallskip}

    \ion{Si}{xiii}   & 6.6480 \\
    \ion{Mg}{xii}    & 8.4192 \\
    \ion{Mg}{xii}    & 8.4246 \\
    \ion{Mg}{xi}     & 9.1688 \\
    \ion{Ne}{x}      & 12.1321~~ \\
    \ion{Ne}{x}      & 12.1375~~ \\
    \ion{Ne}{ix}     & 13.4471~~ \\
    \ion{Ne}{ix}     & 13.6988~~ \\
    \ion{Fe}{xvii}   & 16.7757~~ \\
    \ion{Fe}{xvii}   & 17.0510~~ \\
    \ion{Fe}{xvii}   & 17.0960~~ \\
    \ion{O}{viii}    & 18.9671~~ \\
    \ion{O}{viii}    & 18.9726~~ \\
    \ion{O}{vii}     & 21.6020~~ \\
    \ion{N}{vii}     & 24.7793~~ \\
    \ion{N}{vii}     & 24.7847~~ \\
    \ion{C}{vi}      & 33.7342~~ \\
    \ion{C}{vi}      & 33.7396~~ \\
    \ion{C}{v}       & 34.9730~~ \\ 
    \hline

  \end{tabular}
  \tablefoot{ 
         Lines of identical (or nearly identical) wavelengths are summed up;  for the
         two separated \ion{Ne}{ix} lines, we always take their mean value.
           }   
           \vspace{-2mm}
\end{table}

  We use \changed{appropriate line ratios} to analyse our synthesised HB model spectra 
  in \changed{terms of their $ \tx $ values.} 
  As we start with a fully parametrised model, we know the HB temperature
 distribution and the chemical composition, either homogeneously hydrogen-poor or -rich or
 inhomogeneously with a chemical discontinuity.

 We also know possible sources of line \changed{blending} from the line list. 
 As we want to test our model on BD$+30$, we focus our line ratio measurements on features 
 that have been observed or constrained by \citet[][table~2 therein]{2009ApJ...690..440Y}. 
 These lines are \ion{Mg}{xii}, \ion{Mg}{xi}, \ion{Ne}{x}, \ion{Ne}{ix},  
\ion{O}{viii}, \ion{O}{vii}, \ion{N}{vii}, \ion{C}{vi}, and \ion{C}{v} (Table~\ref{tab:lines}).
 In Fig.~\ref{fig:HR_contamination} we show the spectral windows of these lines. 
 The spectral resolution was set to $0.001\,\AA$ for clarity, but in our analyses we applied a
 resolution of $0.01\,\AA$, which is still about an order of magnitude finer than currently
 accessible by observations \citep{2009ApJ...690..440Y}.

\begin{figure*}
\sidecaption
\includegraphics[width=12.50cm]{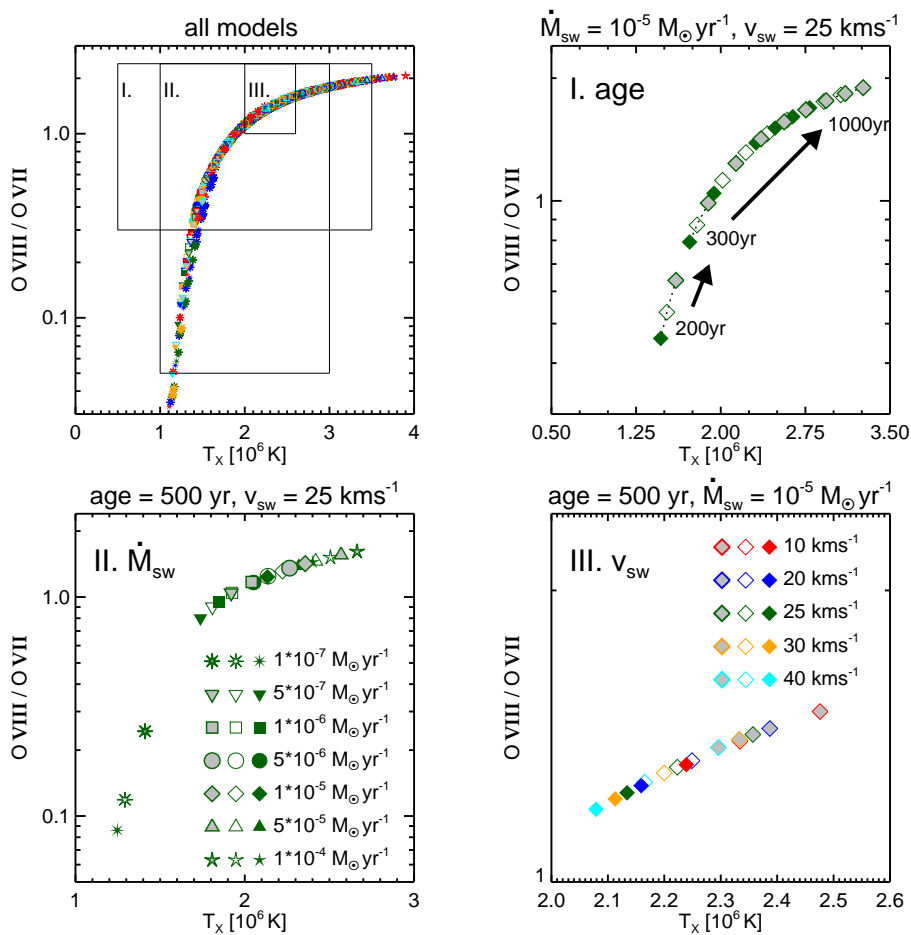}
\caption{\label{fig.models}            
         Dependence of the bubble properties in terms of the line ratio 
         \ion{O}{viii}/\ion{O}{vii} (18.97\,\AA/21.60\,\AA) vs. $\tx$ on age
         (\emph{panel~I}), and slow-wind properties (\emph{panels II and III}).
         The fixed model parameters are indicated at the top of the respective panels. 
         In each of these three panels we distinguish between three values of the
         conduction parameter $ C $:   
         $3.0{\times}10^{-7}$ (grey), $4.5{\times}10^{-7}$ (open),
         and $6.0{\times}10^{-7}$ erg\,cm$^{-1}$\,s$^{-1}$\,K$^{7/2}$  (filled),
         as indicated in the legends.  The \emph{top left panel} displays all computed
         bubble models, and the annotated areas (I, II, and III) indicate  
         the plot ranges shown in the respective panels I, II, and III.
         Without any loss of generality, we selected the WR model shown in 
         Fig.~\ref{fig:bubble.example}, which is close to the middle of the parameter space
         listed in Table~\ref{tab:grid}, as the reference model.       
         The bubbles' chemical composition corresponds to the WR case of
         Table~\ref{tab:abundances}. 
         }
\end{figure*}

  The line luminosity has been determined by integrating over a $0.1\,\AA$ window around 
  these lines to produce results similar to \textit{Chandra} observations
  \citep{2009ApJ...690..440Y}. 
\changed{The 0.01~\AA\ resolution ensures that the line in question is properly resolved,
  while the  choice of the 0.1\,\AA\ integration window represents the lower spectral
  resolution provided by the \textit{Chandra} LETG observations.}
   Hence, the two \ion{Mg}{xii} lines, the two \ion{Ne}{x} lines, the two \ion{O}{viii} lines, 
   the two \ion{N}{vii} lines, and the two \ion{C}{vi} lines were not resolved. 
  Throughout this paper, \ion{Fe}{xvii} measurements refer to the center \ion{Fe}{xvii} line 
  at $17.05\,\AA$, because this is the strongest emission line of this ionisation state. 
  For \ion{Ne}{ix}, we decided to work with the average of the two available lines, 
  as they are of similar strength. Line centers are taken from the CHIANTI database. 

 Figure~\ref{fig:HR_contamination} illustrates that \changed{contamination by overlapping 
 blend lines} is an issue \changed{for the 0.1~\AA\ \textit{Chandra} spectral resolution}. 
 As an example, consider the \changed{left} panel in the second row, where we target at the
 \ion{Ne}{x} lines at $12.1321\,\AA$ and $12.1375\,\AA$. In this high-resolution 
 representation, we identify \ion{Ne}{ix}\,d ($12.1113\,\AA$), \ion{Fe}{xvii} ($12.1230\,\AA$),
 and a blend of \ion{Fe}{xxi}$^*$ ($12.1551\,\AA$), \ion{Ni}{xx} ($12.1570\,\AA$), 
 \ion{Mn}{xxiii} ($12.1586\,\AA$), \ion{Fe}{xxi}$^*$ ($12.1589\,\AA$), and \ion{Fe}{xxiii}
 ($12.1612\,\AA$), all of which contribute to our line measurement of the two \ion{Ne}{x}
 lines.\footnote
{The CHIANTI manual says: ``Lines marked with a * do not have correspondent observed energy
levels and have approximate wavelengths.''} 
 Line \changed{blending} will unavoidably lead to imprecise line ratio measurements, because
 this melange of ions is distributed over a range of HB radii, leading to different ionisation
 equilibria (see Fig.~\ref{fig:ioneq}). When applied to observations with lower resolution
 \citep[as in][]{2009ApJ...690..440Y}, such values must be treated with caution if
 contamination is not discussed. In this particular case, we have verified that contamination
 under the \ion{Ne}{x} multiplet around $12.13\,\AA$ is weak in the cases we consider. 
 But note the logarithmic ordinate scale in Fig.~\ref{fig:HR_contamination}.

\subsection{Properties of the ZP96 bubble models}
\label{subsec:modelsprop}

  In this subsection, we will illustrate how the spectral appearence, and hence the 
  line ratios, depend on the bubble parameters.  
  All the models are assumed to have the WR chemical composition from
  Table~\ref{tab:abundances}.  We selected the oxygen line 
  ratio \ion{O}{viii}/\ion{O}{vii}  and plotted it in Fig.~\ref{fig.models} over the
  characteristic X-ray temperature $\tx$ computed according to Eq.~(\ref{eq:tx}).
  Using all models listed in Table~\ref{tab:grid} (top left panel),
   we see that they degenerate into virtually one single sequence over
  $\tx$, which allows us to determine a well-defined characteristic bubble temperature, 
  independently of the choice of the parameter set.  
  
  Starting from a model with mean properties (as used in  Fig.\,\ref{fig:bubble.example}), 
  the influence of changing age and slow-wind properties, both for three thermal conduction
  coefficients $C$, is illustrated in the other panels of this figure:  
\begin{itemize}
\item[--] evolution with age, the slow-wind parameters fixed ({top right});  
\item[--] evolution with $\dot{M}_{\rm sw}$, $\varv_{\rm sw}$ and age fixed 
          ({bottom left});
\item[--] evolution with $\varv_{\rm sw}$, $\dot{M}_{\rm sw}$ and age fixed 
         ({bottom right}).
\end{itemize}  
  First of all, we notice that a higher conduction efficiency leads to 
  higher characteristic X-ray tempereratures $\tx$.  For the temperature and
  conduction efficiency ranges considered here the corresponding $\tx$ variation is 
  limited to 0.4--0.5 MK.

   The dependences on the other parameters are quite different, and their ranges for the
   respective parameter spaces are indicated by the areas labelled ``I'', ``II'', and ``III''
   in the top left panel of Fig.~\ref{fig.models}.  The most important parameter is,
   of course, the age (top right panel): between 200 and 1000 years, and for the 
   annotated slow-wind parameters, $\tx$ increases from about 1.3 MK to up to 3.25~MK.
   Next comes the slow-wind mass-loss rate (bottom left panel): For a variation
   from $10^{-7}$ to $10^{-4}$ \ml\ at given age,   
   $\tx$ increases only by a factor of two, from about 1.3~MK to 2.7~MK!  
   The influence on slow-wind speed is depicted in the bottom right panel of
   Fig.~\ref{fig.models}:  For a reasonable velocity \emph{increase} from 10 to 40 \kms, 
   $\tx$ \emph{decreases} from 2.4~MK to 2.1~MK, only.
   All these results refer to the models with fixed parameters as indicated above each 
   individual panel.

   This general dependence of $\Tx$ is evident from the solution given in   
   \citet[][equation~12 therein]{1996A&A...309..648Z} where one sees that $\Tx$ is a
   monotonously increasing function of the density at the outer bubble boundary 
    via ``slow wind'' mass-loss rate and/or speed.
   
   For application to real objects, the parameters discussed above have to be specified,
   at least in principle.   The heat conduction parameter is fixed by the chemical
   composition,  but the slow-wind parameters must be estimated from observations.
   We note that they are not to be confused with the values of the former slow AGB-wind.
   Instead, we need the corresponding values ahead of the rim shock, i.e. at the
   rim/shell interface.  For the case of BD+30, \citet{1999MNRAS.309..731B} 
   measured 28 \kms\ from [\ion{N}{ii}] and 36 \kms\ 
   from [\ion{O}{iii}], respectively, which, however, are ``mean bulk'' velocities
   based on the line-peak separation.   We assume that $v_{\rm sw}$ is not far from
   either of these values.  To get the value of $\dot{M}_{\rm sw}$, a detailed
   density and velocity model of the nebula of BD+30 is necessary but not yet available.
   We will see in the next section that the range of values used in Fig.~\ref{fig.models} is
   representative.   
     
\begin{figure}
\vskip-6mm \hskip-7mm
\includegraphics*[width=1.12\linewidth]{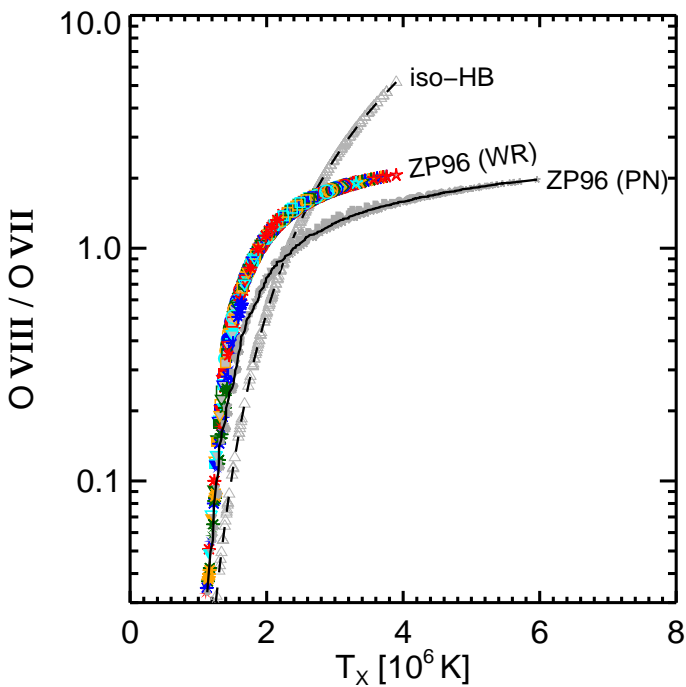} 
\vskip-2mm
\caption{\label{fig.wr.pn}   
        The same as in the top left panel of Fig.~\ref{fig.models} but now including 
        also bubble models with hydrogen-rich PN-matter (symbols with black 
        central line, $C = 6.0{\times}10^{-7}$~erg\,cm$^{-1}$\,s$^{-1}$\,K$^{7/2}$), 
        \changed{and single-temperature bubbles (iso-HB, small
        triangles).}       
        }
\vspace{-2mm}
\end{figure} 

\subsection{\changed{Single}-temperature plasma models}
\label{subsec:iso.bubbles}      

  For a comparison between our ZP96 models and the isothermal approach, we constructed
  bubbles with a constant electron temperature in the following way:  Given a ZP96
  bubble with the typical heat-conduction temperature profile characterised by $\tx$
  and a (constant) pressure $p$, a new bubble (``iso-HB'') with \emph{constant} temperature 
  at the value of $\tx$ and the same constant value of $p$ was then constructed and
  its X-ray spectrum computed in the same fashion as before.
  \changed{Note that these single-temperature bubbles have also a spatially constant density.}
    
\changed{The run of the \ion{O}{viii}/\ion{O}{vii} line ratio with $ \tx $ predicted by our
   iso-HBs is also displayed in Fig.~\ref{fig.wr.pn} and compared with the predictions of the 
   ZP96 bubbles with both the WR (same as in Fig.~\ref{fig.models}) and PN composition. 
   We note that the iso-HB predict line ratios independently of the 
   plasma's chemical composition.    The iso-HB \ion{O}{viii}/\ion{O}{vii} line ratio 
   depends differently on $ \tx $, and this behaviour reflects the different radial temperature
   profiles:  constant temperature (with hence constant ionisation fractions, too) vs. 
   heat-conduction temperature profile with stratified ionisation. 
   Below ${ \tx \approx 2.6 }$~MK (WR) or ${ \tx \approx 2.3 }$~MK (PN), the bubble's
   characteristic temperature is slightly overestimated, above severely underestimated if a 
   single-temperature plasma model is used for the interpretation of the 
   \ion{O}{viii}/\ion{O}{vii} ratio.}   
   
\changed{A qualitativly similar behaviour is found for the \nex/ \neix\ line ratio, 
    although the difference between the iso- and ZP96-HBs is higher, to the extent that the 
    use of iso-temperature models leads always to an overestimate of the characteristic plasma
    temperature below about 4.5~MK.}   
   
\changed{The difference between bubbles with WR and PN compositions has already been explained
    above (Sect.~\ref{subsec:parameters}) in conjunction with the definition of $ \tx $
    according to Eq.~(\ref{eq:tx}).  In general, at given \ion{O}{viii}/\ion{O}{vii} 
    line ratio, the PN bubbles have a higher $ \tx $ value, but the degree of this 
    difference depends on the value of the line ratio:  
    At the lowest line ratios, i.e. very young bubbles, the
    difference is virtually vanishing, but it increases steadily and becomes eventually
    more than 2~MK for the 1000-yr old models.}

\section{The applicability of the ZP96 bubbles} 
\label{sec:test}   

  One of the necessary assumptions to compute the structures of heat-conducting bubbles
  analytically is the neglect of radiative cooling.  The \changed{influence of 
  heat conduction is} twofold:
\begin{enumerate}
\item  a possible steepening of the temperature gradient towards the conduction front, and
\item  a delay of bubble formation.
\end{enumerate}

  The first item affects only the outermost, coolest layers close 
  to the conduction front where the density (and hence cooling) is highest. 
  However, once the bubble is formed, densities are low and the cooling time scale becomes
  long, whereas heat conduction has an ``instantaneous'' effect on the bubble structure
  \citepalias{1996A&A...309..648Z}.  We thus do not expect any significant effect on the
  bubble structure (radial temperature and density profile) by neglecting radiative cooling.
  
  Concerning the second item, radiative cooling is quite important:  While our ZP96
  bubbles form immediately at age zero (with very small wind speed), the formation of a 
  hydrodynamical bubble is postponed by radiative cooling until the cooling time exceeds 
  the ``crossing time'' of the free wind \citep[see, e.g., discussion in][]{ML.02}.   
  For normal PN-matter, this delay is modest and corresponds to a wind speed of about
  170~\kms\ according to \citet{KB.90}, which has been confirmed by the 
  radiation-hydrodynamics models of \citet{peretal.04}.  
  Heat conduction has only a minor impact on formation and evolution of a wind-blown
  bubble \citep[cf. Fig.~4 in][]{2008A&A...489..173S}.  
   
  For hydrogen-deficient and carbon-rich matter, radiative cooling is much more efficient, 
  and a wind-blown bubble does not form before the wind speed reaches about 500~\kms,
  heat conduction \emph{not\/} included \citep{ML.02}.
  Our detailed radiation-hydrodynamics simulation with heat-conduction reveal an
  even later development of a bubble formed out of a hydrogen-poor, carbon-rich 
  wind \citepalias{sandin16}:  Wind speeds of about 1000\,\kms\ 
  (corresponding to ${\teff \simeq 50\,000}$~K) 
  are required to overcome the very efficient radiative cooling caused 
  by the higher bubble densities (orders of magnitude) in the heat-conducting case. 

   Despite these differences between  analytical and hydrodynamical X-ray bubbles, the 
   application of the ZP-bubbles for the analysis the X-ray spectrum of real bubbles is 
   still possible provided the analytical bubbles are comparable in size and X-ray 
   luminosity to the bubble of BD\,+30.
   The nebula of BD\,+30 has a kinematical age of about 800 years
   \citep{2002AJ....123.2676L}, thus the nebula is rather young and quite small, 
   and so is also the X-ray emitting bubble.\footnote
{The X-ray bubble is younger than the nebula because of the
 aforementioned time delay of bubble formation due to line cooling.}
       With a distance to BD+30 of 1300 pc, the bubble radius is 0.013 pc, or 
       $4{\times} 10^{16}$ cm \citep[cf.][]{2008ApJ...672..957K}.\footnote
%
       {The value of $R_{\rm B} = 0.023$ pc = $7\times\! 10^{16}$~cm (distance = 1200 pc) 
       given in table~1 of \citet{2008ApJ...672..957K} is actually the diameter.} 
%
     Based on this distance, \citet{ruiz.13}
     estimated an X-ray luminosity of about 2.7$\times 10^{32}$~erg\,s$^{-1}$.       
     From their high-resolution observations, \citet{2009ApJ...690..440Y} determined 
     an X-ray luminosity (7.4\ldots 8.6)$\times 10^{32}$ erg\,s$^{-1}$ for a 
     distance of 1200 pc. 
 The lower luminosity given in \citet{ruiz.13} goes back to \citet{2000ApJ...545L..57K} and 
 is based on a much lower column density of intervening matter than found by
 \citet{2009ApJ...690..440Y}.
    From cuts along the minor axis using HST images of BD\,+30 one can estimate a 
    relative \changed{rim thickness of 
    ${\Delta\theta_{\rm rim}/\theta_{\rm rim} \simeq 0.15\ldots 025}$ (FWHM),} 
    \changed{just about the limit for justifying} 
    the thin-shell approximation for the expanding bubble/rim system of BD\,+30.   

\begin{figure}                
\vskip-2mm
\hskip-4mm
\includegraphics[width=1.05\linewidth]{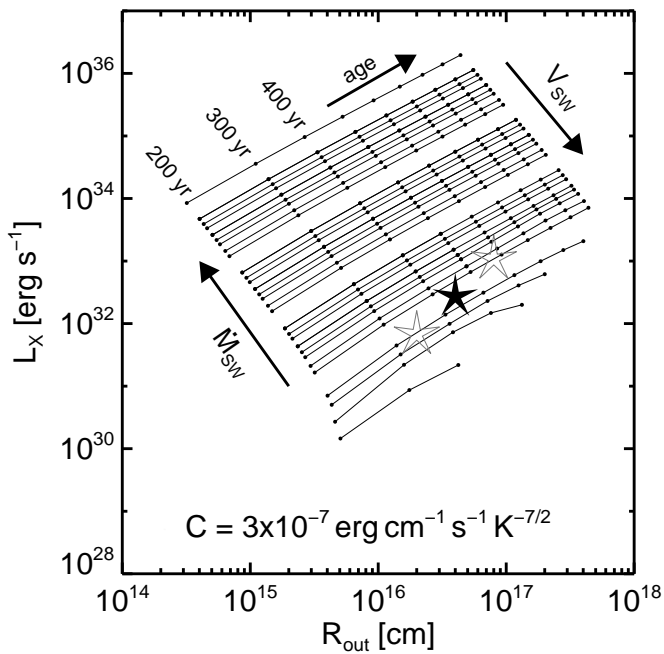}
\vskip -1mm
\caption{\label{fig:lx.rout}
         X-ray luminosity (5--40 \AA, 0.3--2.5 keV) vs. bubble size for the bubble subgrid 
         with WR composition and ${C= 3{\times} 10^{-7}}$ erg\,cm$^{-2}$\,s$^{-1}$\,K$^{-7/2}$,
         and for ages from 200 to 1000~yr.  
         Black dots indicate individual bubble models.  The slow wind, $\dot{M}_{\rm sw}$,
         ranges from 10$^{-7}$ to $10^{-4}$ \ml, the slow-wind velocity, $V_{\rm sw}$, 
         from 10 to 40 \kms\ (cf. Table~\ref{tab:grid}).  
         The adopted values for BD\,+30 (see text) are marked by the filled star. 
         The open stars indicate the positions if the assumed distance 
         to BD\,+30 is in- or decreased by a factor of two. 
         }
\vspace{-2mm}
\end{figure}  

  Figure~\ref{fig:lx.rout} illustrates the ranges in size and X-ray luminosity
  of our ZP96 bubbles with WR composition.  The ages run from
  200 up to 1000 years with monotonously increasing X-ray luminosities.  Furthermore, 
  $\Lx$ increases with    
  $\dot{M}_{\rm sw}$ and decreases with  the 
  flow velocity $v_{\rm sw}$, i.e. increases with upstream density,
  as indicated by the arrows in the figure.         
   Middle-aged (300--600 yr) bubbles with low $\dot{M}_{\rm sw}$ and relatively
   high $v_{\rm sw}$ cover well the observed bubble parameters for BD\,+30 with respect 
   to size and X-ray luminosity.
   At the lowest slow wind mass-loss rates and highest slow-wind velocities, the assumptions
   inherent to the analytical solutions 
   (see Sect.~\ref{subsec:structure}) break down for higher ages, and the models are 
   thus not shown in Fig.~\ref{fig:lx.rout}.

\section{Results for BD\,+30\degr3639}
\label{sec:results}

   In this section we re-analyse the existing 
   \changed{flux calibrated (in $ \rm  photons\,cm^{-2}\, s^{-1}$)}
   high-resolution X-ray line spectrum of BD+30
   published in \citet{2009ApJ...690..440Y} by means of the bubble models
   presented in the previous sections.  The line photon fluxes listed in their
   table~2 have to be converted into units of erg\,s$^{-1}$\AA$^{-1}$ by
\begin{equation}
E = \frac{x}{\lambda} 1.98648 \times 10^{-8}  
\end{equation}
\noindent
\citep{1976asqu.book.....A}, where $E$ (in erg\,s$^{-1}$) is the line luminosity
(that is, the area under the red segment in a line shown in Fig.~\ref{fig:HR_contamination}),
 $x$ is the number of photons per second \citep[e.g. from Table 2 in][]{2009ApJ...690..440Y}, 
and $\lambda$ is the wavelength in units of $\AA$.\footnote
{As an example, $40{\times}10^{-6}\,\mathrm{cm}^{-2}\,\mathrm{s}^{-1}$ photons detected at
  $18.966\,\AA$ from \ion{O}{viii} and $38.2{\times}10^{-6}\,\mathrm{cm}^{-2}\,\mathrm{s}^{-1}$
  photons observed at $21.607\,\AA$ for the \ion{O}{vii} line \citep{2009ApJ...690..440Y}
  translate into a line ratio of
\begin{equation}\label{eq:energy}
\frac{E_\mathrm{\ion{O}{viii}}}{E_\mathrm{\ion{O}{vii}}} = 
\frac{40/18.966}{38.2/21.602} = 1.194 
\end{equation}
in terms of energy units, \changed{instead of 1.047.}}
   
   We assume in a first approximation that all ZP96-bubbles have the homogeneous WR composition 
   listed in Table~\ref{tab:abundances}, which means that they are void of ``evaporated'' 
   PN matter.  We emphasise that the assumption of a homogeneous chemical composition of 
   the emitting plasma, whether normal or more ``exotic'', is the standard assumption 
   used in all analyses of the X-ray emission from wind-blown bubbles conducted so far.
   
   A general uncertainty of all plasma studies in the X-ray regime is the amount of 
   absorption by intervening matter, characterised by the neutral hydrogen column density 
   $ \nh $.   In the case of BD\,+30, the case is even more complicated:
   The absorption appears to be variable across the bubble's image \citep{kastner.02}.
   Given this situation, it is not astonishing that the value of $ \nh $ varies considerably
   from study to study, as listed in table~4 of 
   \citet[][and references therein]{2009ApJ...690..440Y}.  The value 
   of $ \nh $ varies between 0.10$\times 10^{22}$ and 0.24$\times 10^{22}\rm\ cm^{-2},$  
   depending on the method used.  The low values around 0.10$\times 10^{22}$ cm$^{-2}$
   are from optical data converted to $ \nh, $ whereas higher values of $\nh$ are 
   determined simultaneously with the other plasma parameters by matching 
   \changed{one-temperature} plasma models to the observations.
   
   The task that we have to solve by comparing our heat-conducting ZP96-bubble models 
   to the X-ray emission of BD\,+30's hot bubble is then the following:  
\begin{itemize}
\item   to determine the amount of intervening absorbing matter, i.e. $ \nh $,
\item   to fix the characteristic bubble temperature $\tx$,
\item   get abundance ratios for the main constituents of the bubble, and
\item   derive the amount of evaporated hydrogen-rich nebular matter, if any.  
\end{itemize}    

\subsection{The column density $ \nh $ and the characteristic temperature $ \tx $}
\label{subsec:NH}

\begin{figure}    
\includegraphics*[trim= 0.7cm 30cm 20cm 0.2cm, width=1.02\linewidth]
                 {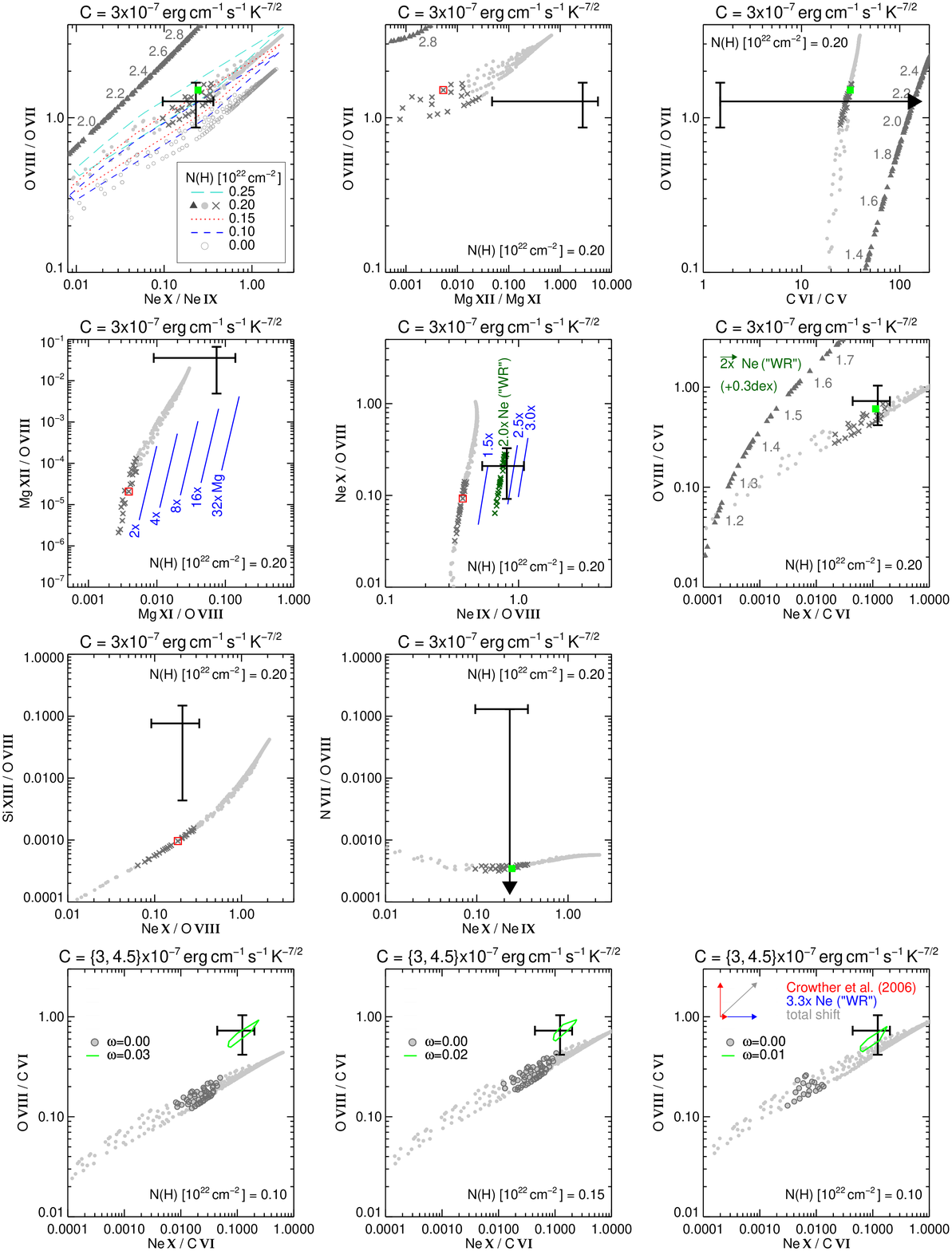}                            
\caption{\label{fig:w.extinction}
         Line ratio of \oviii/\ovii\ over line ratio \nex/\neix\ for BD\,+30 (error cross)
         after \citet[][table~2]{2009ApJ...690..440Y} and for ZP96 heat-conducting bubbles 
         with homogeneous hydrogen-poor WR composition
         subject to various absorptions as indicated in the legend.  Individual models 
         are only shown for ${ \nh = 0 }$ (circles) and 
         $ \nh = 0.20\times\! 10^{22}\rm\ cm^{-2} $, the value adopted (gray dots).  
         For other values of $ \nh $, only envelopes are given in order to avoid confusion.
         Evolution proceeds from low to high line ratios, i.e. from low to high $ \tx $.         
         The oldest (1000~yr) and hottest (4~MK) bubble models have always neon line ratios of
         about 2, virtually independent of $ \nh $.         
         The crosses are the models that are lying within BD\,+30's error box.  
         The filled square (green) marks our ``best-fit'' bubble model, 
         \changed{presented in Sect.~\ref{subsec:best-fit}.}
         The filled triangles are bubbles with constant temperatures (iso-HB, for 
         $ \nh = 0.20\times\! 10^{22}\rm\ cm^{-2} $ only, see text for details).
         \changed{The numbers along the iso-HB sequence indicate the bubble temperatures
         in MK.}    
         }
\end{figure}

   We follow the method of determining the column density of the intervening 
   matter from the X-ray spectrum itself.  However, since the extinction is 
   wavelength-dependent, it happens that abundance ratios, e.g. of C/O, derived from lines 
   with considerably wavelength separation do depend on the chosen value of $ \nh $.  
   In other words,  abundance ratios cannot be determined independently of $ \nh. $
  
   Therefore, we used abundance-independent line ratios for the $ \nh $ determination.      
   In order to be consistent with \citet{2009ApJ...690..440Y}, we used    
   for the wavelength-dependent absorption of our model spectra the standard, 
   solar abundance model of \citet{MMc.83}.  Our method of deriving $ \nh $ is illustrated
   in Fig.~\ref{fig:w.extinction}, where the absorbed temperature-sensitive line ratios
   \oviii/\ovii\ and \nex/\neix\ of our bubble models are compared with \BD's
   observed line ratios.  These line ratios are independent of the abundances, but 
   unfortunately not very sensitive to the value of $\nh$ because of the rather small
   wavelength separation of the applied lines.  Nevertheless, we believe that this is the 
   only acceptable method for high-resolution spectra because chemical abundances are, 
   in principle, not known a priori.     
      
    According to Fig.~\ref{fig:w.extinction}, the line ratios are changed by the amount of
    intervening gas such that all models are shifted upwards with increasing
    $ \nh $, i.e. only the \oviii/\ovii\ line ratio is really sensitive to $ \nh $.  
    The  models with $ \nh = 0.20\times 10^{22}\rm\ cm^{-2}$ (gray dots)
    match the observation best, and the cloud of 32 crosses mark the bubbles whose oxygen 
    and neon line ratios are within \BD's error box.   Lower and higher values of $\nh$ 
    appear to be acceptable as well, and we adopt  
    ${ \nh = 0.20^{+0.05}_{-0.10}\times 10^{22}\rm\ cm^{-2}}$, a result which is, within the
    errors, consistent with the determinations of \citet{2009ApJ...690..440Y} and
    \citet{2009ApJ...695..834N} of $ \nh = (0.24\pm 0.04)\times 10^{22}\rm\ cm^{-2}$.
    The extinction in the visible is reported to be $ A_V \simeq 1\rm\ mag $, 
    which then corresponds to $ \nh \simeq 0.22\times 10^{22}\rm\ cm^{-2}$ if the 
    conversion of \citet{G.75} is applied.
    
\begin{figure*}
\sidecaption
\includegraphics*[trim= 10.5cm 30cm 0.2cm 0.2cm, width=0.70\linewidth]
                  {iso-TX_bei_Linienverhaeltnissen.eps}       
\caption{\label{fig:w.Mg.C}
         Same as in Fig.~\ref{fig:w.extinction} but now for \oviii/\ovii\ over 
         \mgxii/\mgxi\ (\emph{left}) and over \cvi/\cv\ (\emph{right}), and only for 
         ${ \nh = 0.20{\times} 10^{22}}\rm\ cm^{-2}$.  The symbols have the same meaning as
         in Fig.~\ref{fig:w.extinction}.  
         The position of the ``best-fit'' models is either
         the filled (green, right) or crossed square (left).
         } 
\end{figure*} 

\begin{figure*}
\sidecaption
{\includegraphics*[trim= 0.7cm 20.2cm 10.5cm 10cm, width=0.7\linewidth]
                  {iso-TX_bei_Linienverhaeltnissen.eps} }
\caption{\label{fig:abund.Mg.Ne}
         Absorbed line ratios of \mgxii/\oviii\ over \mgxi/\oviii\ (\emph{left}) and
         \nex/\oviii\ over \neix/\oviii\ for $ \nh = 0.20{\times} 10^{22}\rm\ cm^{-2}$
         and the respective observed line ratios for \BD\ (error cross).  The symbols have 
         the same meaning as in Fig.~\ref{fig:w.extinction} (or Fig.~\ref{fig:w.Mg.C}).  
         The parallel lines indicate the shift of the crossed models (and the 
         ``best-fit'' model as well) if the original bubble abundances of Mg (\emph{left}) or 
         Ne (\emph{right}) are multiplied by the given factor.  The corresponding shifts of 
         the remaining bubble models are not shown in order to avoid confusion.         
         } 
\end{figure*}
    
    The 32 bubble models marked by crosses in Fig.~\ref{fig:w.extinction} have quite different
    ages, sizes, and X-ray luminosities and may not always fit the observed parameters of 
    \BD's bubble.  We thus select models which comply with the following criteria for
    luminosity and size:  ${ \Lx \simeq 10^{32}\ldots 10^{33}\rm\ erg\,s^{-1} }$ and 
    ${ R_{\rm out} \simeq 2\times 10^{16} \ldots 8\times 10^{16}\rm\ cm }$, which embrace 
    approximately the observed values for the bubble of \BD\  (cf. Fig.~\ref{fig:lx.rout}).   
    With these constraints, we are left with 10 bubble models
    in the range of 1.68 to 1.89~MK, and we adopt ${ \tx = (1.8\pm0.1~ }$MK for the bubble
    of \BD, a value nearly independent of $ \nh $.      
    This $ \tx $ value  corresponds well with that of \citet{2009ApJ...690..440Y}
    who found  $ 1.7\pm0.4~ $MK for the low-temperature plasma component.    
    
   The iso-HBs are plotted also in Fig.~\ref{fig:w.extinction} (filled triangles), 
   but only for the ${ \nh = 0.20\times\! 10^{22}\rm\ cm^{-2}}$ case \changed{because the
   dependence on $ N_{\rm H} $ is similar.}
   The  positions of these 
   bubbles with respect to the observations confirm the conclusion of 
   \citet{2009ApJ...690..440Y} that plasma models with a \changed{single} temperature
   are unable to describe the X-ray emission of \BD's bubble adequately.
   However, we need ``iso-HB's'' with rather high temperatures of $ 2.3\pm 0.2 $~MK and  
   $ 2.9\pm 0.3 $~MK for matching the observed \oviii/\ovii\ and \nex/\neix\ line ratio,
   respectively \changed{(Fig.~\ref{fig:w.extinction}).}   
   We consider it a \changed{real} success of our analytical ZP96 bubbles with heat
   conduction that they match both the oxygen \emph{and} neon line ratios with a \emph{single}
   value of $ \tx $.  
           
    For completeness, we consider also other temperature-sensitive line ratios, i.e.
    \cvi/\cv\ and \mgxii/\mgxi, although \citet{2009ApJ...690..440Y}'s measurements are very
    uncertain (Mg) or constrained by an upper limit (\cv).    The magnesium lines trace
    the hottest (\changed{$ \simeq\!10 $~MK, see Fig.~\ref{fig:ioneq})}, 
    innermost region of a bubble, and the failure of our ZP96 bubbles to
    match the observation despite their large errors (Fig.\ref{fig:w.Mg.C}, left panel) 
    can mean that
\begin{itemize}
\item  the innermost part of \BD's bubble is hotter than our models predict, 
       or that   
\item  the measurements are \changed{more uncertain than expected}, most probably 
      because of blending.
\end{itemize}    

\changed{The region immediately behind the wind shock can be hotter in reality than our
         ZP96 bubbles predict because the (standard) heat-conduction formulation may break
         down there due to heat-flux saturation \citep{cokee.77}. 
         This possibility has been discussed in some detail in  \citet{2008A&A...489..173S} 
         in conjunction with the numerical treatment of thermal conduction.}
  
   For the carbon line ratio, which is sensitive to the region behind the conduction front, 
   there appears no conflict between model predictions and the observation (right panel
   of Fig.~\ref{fig:w.Mg.C}).

\subsection{On the chemical composition of \BD's hot bubble}
\label{subsec:chem.comp}
 
\changed{
  In general, any abundance analysis of X-ray spectra
  suffers from the fact that the two usually most abundant elements, hydrogen and helium, are
  unobservable.  Instead, one has to make reasonable assumption about the abundance of these 
  two elements, otherwise only \emph{abundance ratios} can be derived.  
  In the case of a wind-blown bubble inside a planetary nebula with a 
  hydrogen-poor or \mbox{-free} [WR] central star it appears justified to assume that all or 
  at least most of the bubble contains also the hydrogen-poor material from the stellar
  surface.  This is especially important for the bubble of \BD\ because it is already clear
  from its low-resolution X-ray spectrum that the bubble plasma must be extremely enriched 
  in carbon and neon.  With this in mind, one has to accept that the absolute abundance values
  deduced for \BD's bubble by \citet{2009ApJ...690..440Y} and \citet{2009ApJ...695..834N} 
  are not meaningful. }
  
   The chemical composition of our ZP96-bubbles is given by the WR-mixture listed in
   Table~\ref{tab:abundances} which in turn is based on the work of
   \citet{2007ApJ...654.1068M}
   in which photospheric/wind lines of the central star have been analysed in detail. 
   As already mentioned in Sect.~\ref{subsec:parameters}, only He, C, and O are 
   provided by \citeauthor{2007ApJ...654.1068M}, and the abundances of the other elements,
   especially those of Ne and Mg, have been assumed by us.   
   Since hydrogen and helium lines are not observable in the
   X-ray range, only abundances relative to, e.g. oxygen, can be derived.      
   We use the \oviii\ (recombination) line 18.97~\AA\ as representative for oxygen because 
   $ \rm O^{8+} $ is the dominant ion inside the bubble, except closer to   
   the conduction front where temperatures are lower (cf. Fig.~\ref{fig:ion_fracfig2}).  
   This line is thus not very sensitive to possibly evaporated matter \changed{near the
   conduction front with a different oxygen content (for instance hydrogen-rich PN matter).}

\paragraph{Neon, magnesium, and silicon} 

    We \changed{illustrate} in Fig.~\ref{fig:abund.Mg.Ne} how the abundances of neon and
    magnesium relative to oxygen can be derived.   
    We begin with magnesium (left panel) because we have seen
    in Fig.~\ref{fig:w.Mg.C} above that our ZP96-models fail to reproduce the observed
    \mgxii/\mgxi\ line ratio.  First of all, only the hottest models with $ \tx \ga 3~ $MK
    can match the observed Mg/O line ratios. However, these models are far away from the
    observed line ratios in Figs.~\ref{fig:w.extinction} and \ref{fig:w.Mg.C} (left). 
    Increasing the magnesium abundance does not help either:  factors between 2 and 32
    provide consistence with the observed \mgxi/\oviii\ line ratio. \changed{A factor of 
    about 10 corresponds to the solar Mg/O reported by \citet{2009ApJ...690..440Y} and
    \citet{2009ApJ...695..834N}. Still, the observed \mgxii/\oviii\ line ratio remains
    unreachable.}  \changed{A reliable magnesium abundance cannot be determined, and we}
    thus do not consider Mg any further.    
         
    The situation is much better for neon, although our bubble models fail also to match 
    the neon/oxygen line ratios.  However, an increase of our initial neon
    abundance by factors between only 1.5\ldots 3 give good agreement with the observations. 
    We assume $\rm (Ne/O)_{\rm \BD} = (2.2\pm0.7)\times (Ne/O)_{\rm WR} = 0.66\pm0.20$.   
    \citet{2009ApJ...690..440Y} derived 
    $\rm (Ne/O)_{\rm \BD} = (2.4\pm0.5)\times (Ne/O)_{\odot} = 0.51\pm0.10$, in reasonable
    agreement.\footnote
    {Since \citet{2009ApJ...690..440Y} used the \citet{ag.89} solar oxygen abundance 
    $ \epsilon = 8.93$, their Ne/O value relative to the solar one has been corrected for
    the new solar oxygen abundance of $ \epsilon = 8.73$ \citep{lodders.10}. 
    }  
    Note that the neon abundance drawn from the diagram used in
    Fig.~\ref{fig:abund.Mg.Ne} is not very sensitive to the value of $ \nh $:  Assuming 
    even no absorption, the factor increases by about 50\,\% only.
    
\begin{figure}
{\includegraphics*[trim= 0.1cm 10.2cm 20.5cm 20cm, width=0.97\linewidth]
                  {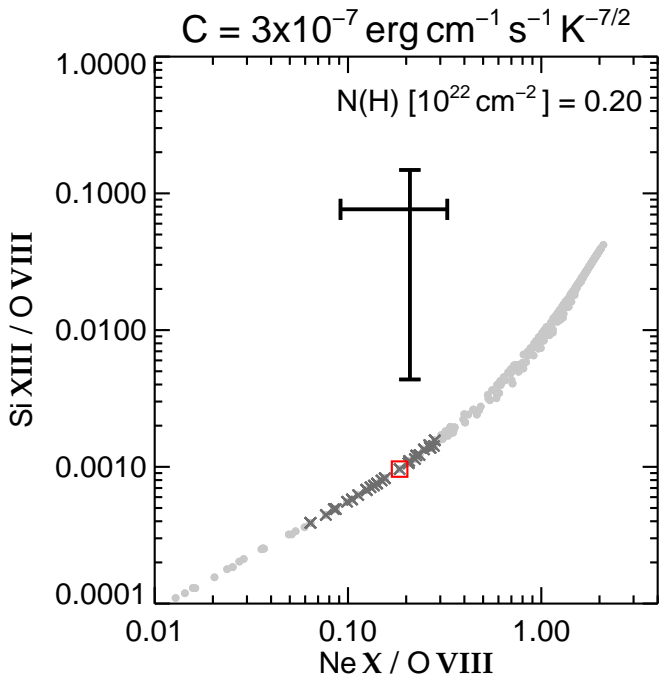} }
\caption{\label{fig:silicium}
         Same as in Fig.~\ref{fig:w.extinction} but for \ion{Si}{xiii}/\oviii\ over 
         \nex/\oviii\ and for $ \nh = 0.20\times\! 10^{22}\rm\ cm^{-2}$.  The gray dots and
         crosses mark the sequence of our ZP96-bubbles with the new enhanced neon abundance; 
         the crossed square indicates the ``best-fit'' model.
         } 
\vspace*{-2mm}         
\end{figure}

    The situation for silicon is similar to that for magnesium:
    The silicon abundances rests on one very weak line of \ion{Si}{xiii} only, and    
    \citet{2009ApJ...695..834N} found (also corrected for the solar oxygen abundance)
    $ \rm (Si/O)_{\BD} = (1.3\pm 0.7)\times (Si/O)_{\odot} \simeq 0.08 $, which is about
    1/8-th of their neon-to-oxygen ratio.   We cannot confirm this finding, as 
    Fig.~\ref{fig:silicium} shows.  We see that the silicon abundance used by us must be
    increased by about a factor of ten, at least.  But taking the observed 
    \ion{Si}{xiii}/\oviii\ line ratio at face value, we arrive at  
    ${\rm (Si/O)_{BD} \approx 100\times (Si/O)_{WR} \simeq 0.6 \simeq 9\times (Si/O)_\odot, }$
    i.e. silicon would be about equally abundant as neon, which is difficult to understand!

\begin{figure}
\vskip -1mm
\includegraphics*[trim= 20.6cm 20.2cm 0.cm 10cm, width=0.97\linewidth]
                  {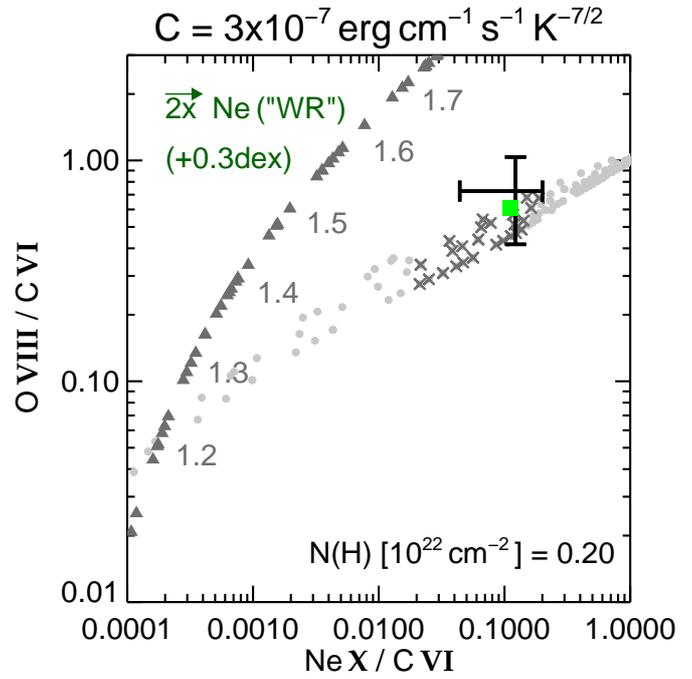}
\caption{\label{fig:abund.C} 
        Same as in Fig.~\ref{fig:w.extinction} but for \oviii/\cvi\ over \nex/\cvi\ and for
        $ \nh = 0.20\times 10^{22}\rm\ cm^{-2}$.  The neon abundance from 
        Table~\ref{tab:abundances} (WR) is increased by a factor of two. The horizontal
        shift of the bubbles from their original positions is indicated by the error 
        in the upper left corner of the plot.  The green square marks our ``best-fit'' bubble
        (see text).  Filled triangles represent again our iso-HB models, 
        \changed{shifted accordingly, too.}          
        }   
\vspace*{-2mm}          
\end{figure} 

\paragraph{Carbon}  
    
    Although the abundance of carbon (or C/O) inside the hot bubble of \BD\ is given by the 
    central star's photospherical composition, the study of the carbon lines accessible in 
    the X-ray regime (\cv\ and \cvi) is interesting in principle:
    The carbon lines are sensitive to extinction thanks to their rather high wavelengths, 
    and carbon gets fully ionised already closely behind the conduction front 
    (see Fig.~\ref{fig:ion_fracfig2}).   Because of the latter property,  the \cv\ line
    (not useful here) and also the \cvi\ lines are immediately influenced by even small
    amounts of evaporated matter with a different, i.e. normal, carbon content.

    Now the question to be anwered is, whether the oxygen/car\-bon and neon/carbon line ratios 
    are consistent with the respective photosperical abundance ratios for the derived value
    of $ \nh. $   If not, it means obviously that either the amount of absorption has been
    wrongly chosen or hydrogen-rich matter has evaporated into the bubble. 
  
   The case is illustrated in Fig.~\ref{fig:abund.C} where the necessary increase of the
   neon abundance (see Fig.~\ref{fig:abund.Mg.Ne}, right panel) has already been 
   taken into account.   We see that \changed{most} our ZP96-bubbles
   selected on the basis of Fig.~\ref{fig:w.extinction} are consistent with the
   observations, which can be interpreted such that the bubble of \BD\ has still the
   hydrogen-deficient WR composition of the stellar photosphere/wind.  Evaporated 
   hydrogen-rich nebular gas is either not present or its amount still unobservably small.
   We will come back to the latter point in the next section.

   \citet{2009ApJ...690..440Y} derived 
   a C/O ratio of $\rm (21\pm 10)\times (C/O)_{\odot}$, \citet{2009ApJ...695..834N} 
   from the same data $\rm (24\pm 5)\times (C/O)_{\odot}$, both values again corrected 
   for the lower solar oxygen abundance of  $ \epsilon = 8.73$.     
   The carbon/oxygen ratio of the WR mixture used by us is 12 (by number,
   Table~\ref{tab:abundances}).  Thus, we have 
   $ \rm (C/O)_{\BD} = 12 = 26\times (C/O)_{\odot} $, in good agreement with the 
   findings of \citeauthor{2009ApJ...690..440Y} and \citeauthor{2009ApJ...695..834N}.
   In this connection we remark that also in this case the iso-HB models are unable 
   to match both \changed{the observed \oviii/\cvi\ and \nex/\cvi\   
   line ratios} with a single temperature (Fig.~\ref{fig:abund.C}). 

   We mentioned already in Sect.~\ref{subsec:parameters} that the analysis of 
   \citet{CMS.06} provided a somewhat different chemistry at the stellar surface of \BD\
   than that given by \citet{2007ApJ...654.1068M} which we have used here:
   Helium and oxygen are higher at the expense of carbon which has now a mass fraction 
   of 0.38 only.  Specifically, the oxygen
   abundance is nearly doubled, from 0.06 (0.05 from \citealt{leuetal.96}) to 0.10 (mass
   fractions).   These changes of the carbon and oxygen abundances lead in 
   Fig.~\ref{fig:abund.C} to the following shift of the bubble positions: about 0.4 dex
   upwards and about 0.1 dex to the right.   Consequently most of the crossed models
   (including the ``best-fit'' model selected below) will leave the error box.   
   
   Even more disturbing is the fact that, because the Ne/O abundance ratio is fixed by
   the observation (cf. Fig.~\ref{fig:abund.Mg.Ne}, right panel), the neon mass fraction
   \changed{nearly doubles as well and becomes 0.09, an unreliably high value.}
   Mainly for the latter reason, we do not consider the \citet{CMS.06} abundances 
   any further.  We must consider, however, that a carbon abundance which is a bit lower
   by, say, about 10\,\% than the one used here would give an even better match to the
   observations.   Alternatively, a small increase of $ \nh\ $ to 
   0.22$ \times 10^{22}\rm\ cm^{-2} $ would do the same job without violating the
   constraints set by Fig.~\ref{fig:w.extinction}.

\paragraph{Iron} 

   We refrain from deriving the iron abundance because  \citet{2009ApJ...690..440Y} give
   only upper flux limits.   Nevertheless,  a Fe/O abundance ratio of about 0.1 solar
   was deduced and an iron deficiency claimed.   We comment that the low Fe/O abundance 
   ratio is more likely due to the highly increased oxygen abundance and not to an iron
   deficiency (cf. discussion in Sect.~\ref{subsec:parameters}).
     
\begin{figure}
\includegraphics*[trim= 10.1cm 10.2cm 10.4cm 20.1cm, width=0.99\linewidth]
                  {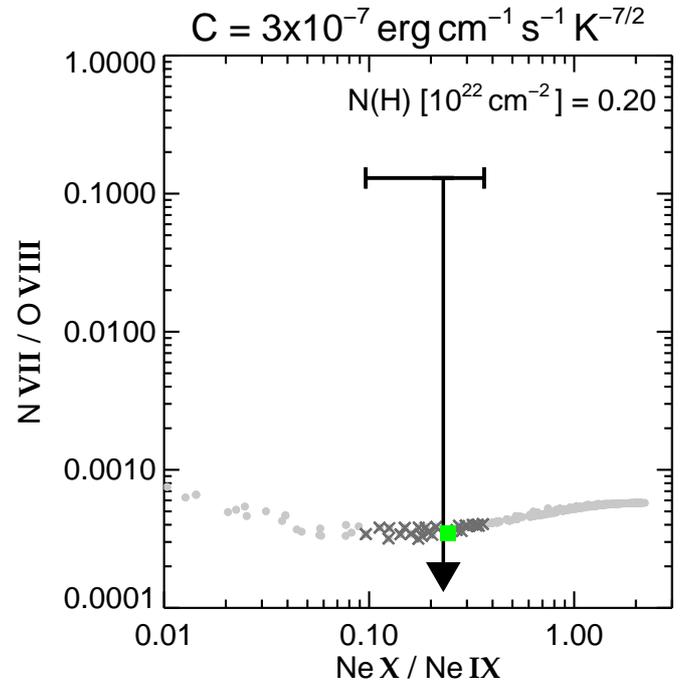} 
\caption{\label{fig:n}
         Upper limit of the observed line ratio \nvii/\oviii\ vs. \nex/\neix\  compared to 
         the model prediction.  Symbols have the same meaning as in the previous figures. 
         }
\end{figure}     

\paragraph{Nitrogen}  

   Knowing the abundance of nitrogen is crucial because it is a tracer of \BD's previous
   evolution.  We have mentioned in Sect.~\ref{subsec:parameters} that we assumed complete
   hydrogen and helium burning of nitrogen to neon after which the intershell matter 
   becomes exposed.   If such a scenario for the evolution of \BD\ is correct, the
   strong \nvii\ at 24.8~\AA\ line should be missing because our ZP96-models predict an
   \nvii/\oviii\ line ratio of well below $ 10^{-3} $ for our WR mixture
   with a N/O abundance ratio (by number) of ${ 2\times\! 10^{-4} }$ (see Fig.~\ref{fig:n}).   

   We note that \citet{2009ApJ...690..440Y} and \citet{2009ApJ...695..834N} derived 
   ${\rm (N/O)_{\BD} = (0.3\pm 0.4)\times (N/O)_{\odot} } $ and 
   $\rm (N/O)_{\BD} =\linebreak (0.4\pm 0.4)\times (N/O)_{\odot}  $, respectively,
   which is consistent with our assumption of virtually no nitrogen at all.
   Looking at the \emph{Chandra} LETG spectra of \BD's hot bubble publicly available\footnote
    {http://tgcat.mit.edu/tgData.php?q=95644\_3aec069391402323},
   we estimated a line flux ratio \nvii/\oviii\ of about 0.13, provided the weak feature seen 
   at 24.8~\AA\ is really the \nvii\ line. This value is used by us in Fig.~\ref{fig:n} as
   an upper limit.  Taken at face value, this upper line-ratio limit would roughly correspond 
   to a nitrogen mass fraction in the WR mixture of 0.003, close to the (about solar)
   nitrogen mass fraction in the PN mixture. 
   
   The upper limit for the bubble's nitrogen content is not helpful in clarifying the 
   evolutionary history of BD\,+30.   A better constraint of the nitrogen abundance is 
   urgently needed, such that a distinction between the different possible scenarios discussed
   in Sect.~\ref{subsec:parameters} can be made.

\subsection{Our ``best-fit'' model}
\label{subsec:best-fit}
  
    We have seen in Figs.~\ref{fig:w.extinction} and \ref{fig:abund.C} that quite a number 
    of model bubbles satisfy the temperature sensitive line ratios of oxygen and neon 
    in conjunction with the abundance sensitive carbon/oxygen and carbon/neon line ratios. 
    Without loss of generality, we selected a ZP96-bubble that is a compromise between
    Figs.~\ref{fig:w.extinction} and \ref{fig:abund.C} and is close to the error-cross
    centers of these figures: our ``best-fit'' bubble model (filled green square).
    The parameters of this model are listed in Table~\ref{tab:best-fit}.
     
\begin{table}
\caption{\label{tab:best-fit}  
         Parameters  of the ``best-fit'' bubble model and observed values of  
         BD\,+30\degr 3639's hot bubble. 
         }
\centering
\begin{tabular}{l @{\hspace{5mm}} l l}
\hline\hline\noalign{\smallskip}
    Parameter                                    & ``Best-fit''  & BD\,+30\degr 3639\\
\hline\noalign{\smallskip}
 $ C\rm\ (erg\,cm^{-1}\,s^{-1}\,K^{-7/2}) $      &  $ 3{\times}10^{-7} $   & --\\
 Age (yr)                                        &  500                    & --\\
 $ \dot{M}_{\rm sw}\rm\ (M_{\odot}\,yr^{-1}) $   &  $ 5{\times}10^{-7} $   & --\\
 $ v_{\rm sw} $ (\kms)                           &  40                     & --\\
 $ R_{\rm in}\rm\ (cm) $       &  $ 3.75{\times}10^{16} $                  & --\\
 $ R_{\rm out}\rm\ (cm) $      &  $ 4.89{\times}10^{16} $ & $ 4.0{\times}10^{16} $           \\
 $ \tx \rm\ (MK)$              &  1.82           & $ 2.3,\ \simeq\!1.7\ldots 2.7 $   \\
 $ \Lx \rm\ (erg\,s^{-1}) $    &  $ 8.23{\times}10^{32} $& $ (7.4\ldots 8.6)\times\!10^{32} $\\ 
 H                                               &  0.019                  & --   \\ 
 He                                              &  0.389                  & 0.43 \\ 
 C                                               &  0.472                  & 0.51 \\
 O                                               &  0.050                  & 0.06 \\
 Ne                                              &  0.046                  &  --  \\
\hline 
\end{tabular} 
\tablefoot{Only the mass fractions of the major elements H, He, C, O, and Ne are listed.
          Their values are essentially those of the WR mixture in Table~\ref{tab:abundances} 
          but renormalised because of the enhanced neon abundance.
          The hydrogen abundance must be understood as an upper limit only.  The abundances
          entries given for BD\,+30 refer to the stellar values of 
          \citet[][table~2 therein]{2007ApJ...654.1068M}.     
          The \changed{$ \tx $ and $ \Lx $} values for \BD\ are from
          \citet{2009ApJ...690..440Y}.  A distance of 1300~pc has been assumed .
          }  
\vspace*{-2mm}      
\end{table}
    
   We see from Table~\ref{tab:best-fit} that size and X-ray luminosity 
   of the ``best-fit'' model are lying well within the limits given by observations 
   (cf. Fig.~\ref{fig:lx.rout}).   
   However, we emphasise again that some of the parameters in Table~\ref{tab:best-fit} like
   those that describe the outer boundary conditions ($ \dot{M}_{\rm sw}$, $ v_{\rm sw} $)
   have no relation to the actual situation of the \BD's nebular 
   system.   Other combinations of $ \dot{M}_{\rm sw}$ and $ v_{\rm sw} $ are equally
   possible, but not listed here.   The only fitted parameters are $ \tx $ and the
   abundance ratio of Ne/O which is then used to derive the neon abundance mass fraction
   by applying the input abundances from the WR mixture used in the computations.  
      
\changed{We note that we cannot make any statement on the hydrogen content of \BD's bubble.
  The finite hydrogen content of the ``best-fit'' bubble listed in Table~\ref{tab:best-fit}
  is only meant as an upper limit based on analyses of \BD's stellar spectrum in the optical
  wavelength region \citep[e.g.,][]{leuetal.96}.  Most likely, the bubble of \BD\ is
  completely hydrogen-free.}

\section{Bubbles with inhomogeneous chemical composition}
\label{sec:inhomogeneous.bubble}  

   The mass budget of a HB with heat conduction is controlled by two contributions:  
   stellar wind matter from within passing through the reverse shock and gas from the
   environment ``evaporating'' through the conduction front \citep[cf.][]{borketal.90}.  
   In the framework of ZP96 it is implicitly assumed that the chemical composition
   is homogeneous throughout the HB, i.e. either hydrogen-rich (the normal case) or 
   hydrogen-deficient (as assumed here so far). 
   
   However, we know from observations that planetary nebulae
   with [WR]-type central stars do not share the stellar abundance pecularities
   \citep[see, e.g.,][]{giretal.07}.  Instead, the nebula chemistry is hydrogen-rich and
   indistinguishable from nebulae around normal, hydrogen-rich central stars.   
   Therefore, if we want to model realistic bubbles we have to consider that
   the evaporated gas is hydrogen-rich, i.e. with PN chemistry.  
   Depending on the age of the HB,  a certain fraction of the the bubble's outer mass shells
   behind the conduction front should consist of hydrogen-rich nebular gas heated and 
   evaporated across the conduction front.
 
\begin{figure}[h]
\vskip-3mm
\includegraphics[width=\linewidth]{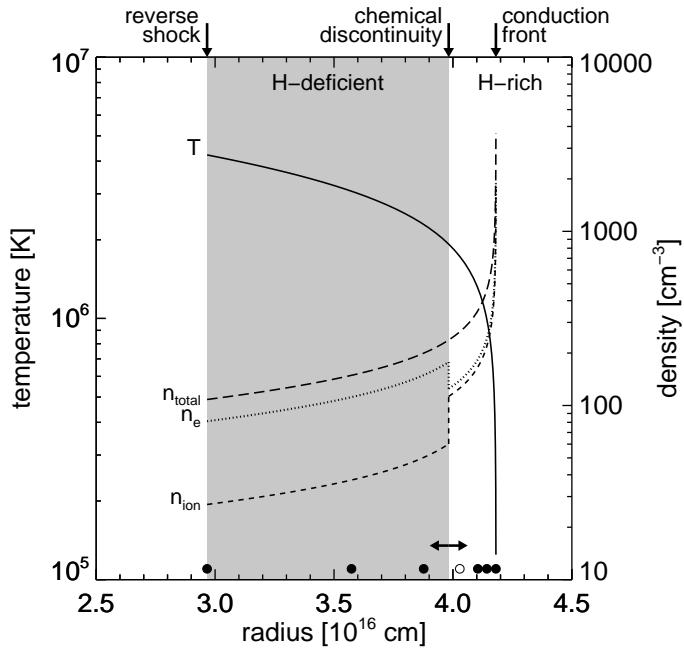}
\vskip-1mm
\caption{\label{fig.bubble}
         Physical structure of a ZP96 bubble of age = 500\ yr with a stratified 
         composition such that 
         ${\omega \equiv {M_{\rm PN} / (M_{\rm WR} + M_{\rm PN})} = 0.03}$, with 
         ${M_{\rm WR} = 5.59{\times} 10^{28}}$ g and ${M_{\rm PN} = 1.67{\times} 10^{27}}$ g,
         and with ${C= 3\times\! 10^{-7}}$ erg\,cm$^{-1}$\,s$^{-1}$\,K$^{7/2}$,
         ${\dot{M}_{\rm sw} = 5.0{\times}10^{-7}}$ \ml, and ${v_{\rm sw} = 25}$~\kms.         
         Shown are radial runs of electron,
         ion, total particle densities, and temperature.  The central star is 
         at the origin (radius = 0), the wind reverse shock (= inner bubble boundary) is at 
         2.97$\times 10^{16}$ cm, and the PN proper adjacent to the heat conduction front at
         $4.18{\times}10^{16}$ cm. 
         The compositions are those listed in
         Table~\ref{tab:abundances}, i.e. WR for hydrogen-deficient (grey) and PN
         for hydrogen-rich.    The chemical discontinuity is at  
         $3.98{\times}10^{16}$ cm, or $2.0{\times}10^{6}$~K.
         Note the logarithmic ordinate scales and the linear abscissa.  
         For the meaning of the dots along the
         abscissa, see Fig.~\ref{fig:intensity}.
         See also Sect.~\ref{subsec:mixing} for details. 
         }
\end{figure} 
 
\begin{figure}[t]
\vskip -1mm
\includegraphics[width=\linewidth]{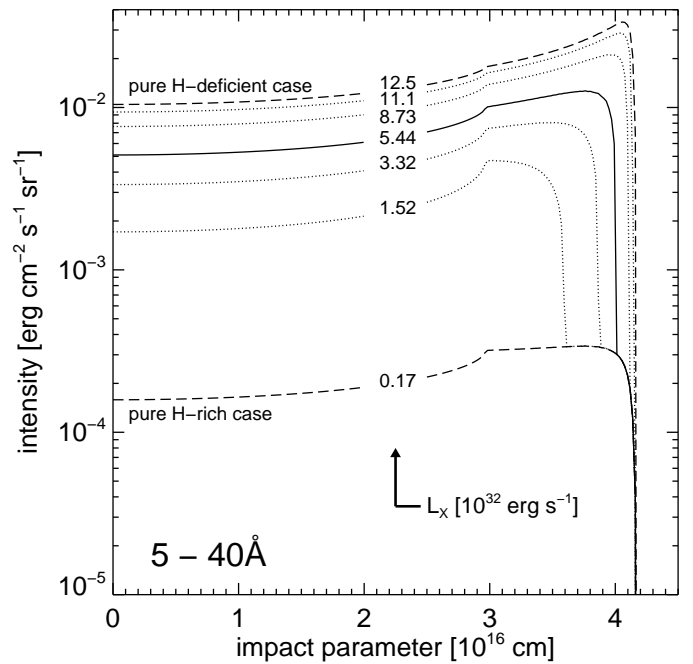}
\vskip-1mm
\caption{\label{fig:intensity}
         Intrinsic intensity distributions of X-ray emission between 5--40 \AA\
         for the bubble used in Fig.\,\ref{fig.bubble} but with 
         the chemical discontinuity located at different bubble radii, hence with
         different $\omega$.
         The individual intensity distributions correspond either to the specific bubble 
         model of Fig.\,\ref{fig.bubble} (solid) or to the positions of the chemical 
         discontinuities being at radii indicated by the filled circles in 
         Fig.\,\ref{fig.bubble} (dotted), whereas the two chemically homogeneous cases 
         PN and WR are rendered as dashed lines.
         The model belonging to the open circle in Fig.~\ref{fig.bubble} is not shown 
         because of its similarity to the ${\omega = 0.03}$ model.         
         The numbers at the curves indicate the X-ray luminosities (in $10^{32}$
         erg\,s$^{-1}$) computed for the given wavelength range.         
         }
\end{figure}
 
   In the following, we introduce and discuss heat-conducting bubbles which have a 
   inhomogeous (or stratified) chemical composition, i.e. the outer bubble region consist 
   of hydrogen-rich PN matter, while the remaining inner parts have still the original
   hydrogen-poor WR composition.

\subsection{Construction and properties of wind-blown bubbles with inhomogeneous chemical
            composition}
\label{subsec:mixing} 

   We constructed chemically inhomogeneous bubbles in the following way: 
   Beginning at the conduction front and keeping temperature distribution and
   pressure within the ZP96-bubble constant, we replaced shells of 
   hydrogen-deficient WR composition with shells of hydrogen-rich PN composition.   
   In order to speed up the analysis and since (i) changes are expected to be 
   most pronounced for shells with steep temperature gradients and (ii) BD+30 as a young 
   object is expected to have yet a limited amount of evaporated H-rich matter,
   we set the positions of the chemical discontinuity such as to form a geometric sequence 
   (illustrated by dots along the abscissa in Fig.\,\ref{fig.bubble}).
   
   Keeping the temperature distribution unchanged implies implicitly also no change
   of the conduction efficiency across the chemical discontinuity.  
   Since the fraction of evaporated hydrogen-rich matter is expected to be very
   small, if any at all, for the hot bubble of \BD, we generally kept the value of the heat 
   efficiency constant as is valid for the hydrogen-poor WR mixture, i.e. 
   ${C = 3.0\times\!10^{-7}}$ erg\,cm$^{-1}$\,s$^{-1}$\,K$^{7/2}$.    
   Additionally, we considered also bubbles with an intermediate value, i.e. 
    ${C = 4.5\times\!10^{-7}}$ erg\,cm$^{-1}$\,s$^{-1}$\,K$^{7/2}$.
   Anyway, as we have seen above (Fig.~\ref{fig.models}, upper right panel), the value 
   of $C$ has only a rather modest influence on the bubble temperature $ \tx $ and the
   correspondent temperature-sensitive line ratios.

\changed{How realistic is the assumption of a sharp discontinuity (or density jump) at the
   transition between both sets of compositions?  Our 1-D hydro models suggest a rather
   sharp transition region which is quite small compared to the bubble size          
 \citepalias[middle panel of figure~7 in][]{sandin16}.   Whether dynamical instabilities
  will break completely any chemical discontinuity set up by evaporation cannot 
  be answered at present.} 

   \changed{Our bubble modelling also ignores the time evolution} 
   during which the bubble mass increases by the evaporated gas.
   Instead, here we are just only replacing the WR by PN matter.  
   Nevertheless, as we will see below, chemically stratified bubbles constructed
   in this way can well be used as a diagnostic tool to investigate the 
   line emission of chemically stratified wind-blown bubbles and to address the
   question of whether heat conduction with concomitant PN-matter evaporation is 
   at work and how large the mass fraction of the evaporated PN matter really is.    
   
  As an example of our modelling, we show in Fig.~\ref{fig.bubble} the physical
  structure of a chemically stratified bubble of intermediate age with a 3\,\% mass fraction 
  of PN matter:  ${\omega \equiv {M_{\rm PN} / (M_{\rm WR} + M_{\rm PN})} = 0.03}$. 
  One sees clearly that the condition of constant pressure in conjunction with 
  the different element mixture leads to a jump 
  of electron and ion densities at the position of the chemical discontinuity. 

   Because we have seen  that hydrogen-rich and hydrogen-poor
   compositions have very different X-ray emission properties (Fig.~\ref{fig:bubble.example}), 
   a stratified chemical composition has a profound influence on the bubble's X-ray emission, 
   which is illustrated in the
   following two figures.  Figure~\ref{fig:intensity} shows how the X-ray
   surfaces brightness (or intensity) changes with the position of the chemical transition,
   with the other bubble parameters unchanged:  Since the 
   emissivity of the hydrogen-rich matter is much lower, the composition
   transition is marked by a huge intensity drop of up to two (!) orders of magnitude,
   depending on the the radial position of the WR/PN matter transition. 
   Correspondingly, the total X-ray luminosity decreases rapidly if the WR/PN discontinuity
   moves inwards.   
   \changed{The case of a purely hydrogen-rich bubble cannot be achieved in nature because 
   the stellar wind feeds continuously new WR matter into the bubble. }

\begin{figure*}
\sidecaption
\includegraphics[width=0.70\linewidth]{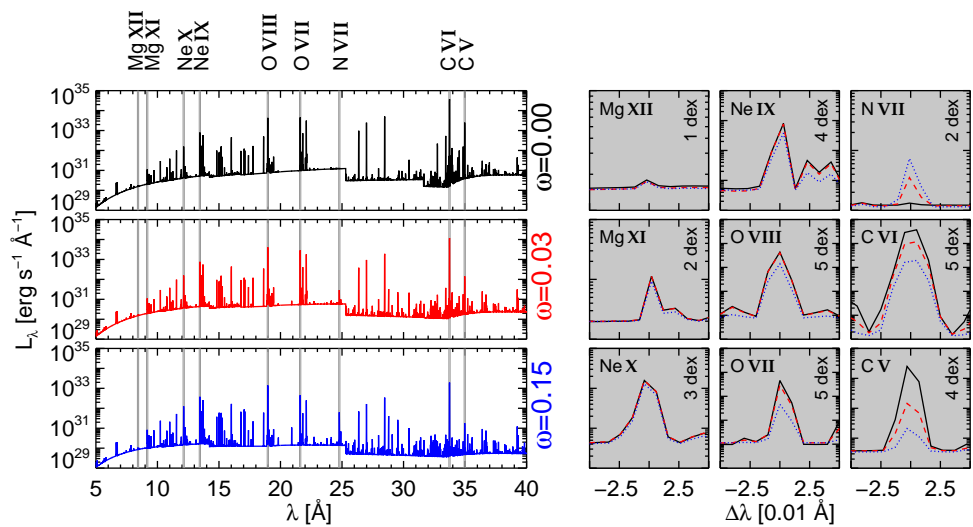}
\caption{\label{fig:spectra.inhomog}
       X-ray spectral luminosity (\emph{left}) and selected lines zoomed-in (\emph{right}) 
       of ZP96-bubbles with homogeneous WR composition (${\omega=0.00}$, solid/black) and
       inhomogeneous WR\,+\,PN composition (${\omega= 0.03}$, dashed/red, and 0.15,
       dotted/blue).  All three bubble models have the same dimension and temperature profile,
       and the one used in the \emph{middle-left} panel is the
       model used in Figs.~\ref{fig.bubble} and \ref{fig:intensity}.  
       The line-profile panels (grey) cover the wavelength range of 0.10 \AA\ each
       (i.e. the \emph{Chandra} resolution), but 
       the (logarithmic) ordinate range may be diffferent as to give a better match to the
       various line strengths.  Also, all lines are normalised to a single continuum level
       to ease comparison.  
         }
\vspace{-2mm}
\end{figure*}

   Figure~\ref{fig:spectra.inhomog} illustrates how the X-ray spectrum depends on the 
   position of the chemical discontinuity but
   with otherwise unchanged structural parameters.  The top-left panel displays the
   chemically homogeneous WR case (${\omega = 0.0}$), while the middle-left panel shows 
   the spectral luminosity distribution for the ${\omega = 0.03}$ case rendered
   in Figs.\,\ref{fig.bubble} and \ref{fig:intensity}.
    The spectrum where the chemical dicontinuity is placed even further inside 
    (${\omega = 0.15}$) is displayed in the bottom-left panel.
    Important emission lines are indicated above the panels, and the corresponding 
    high-resolution profiles are shown in the 3$\times$3 matrix of panels on the right. 
  
    The contamination by evaporated hydrogen-rich matter lowers the continuum emission
    of the bubble, preferably at longer wavelength where the emission of the cooler 
    hydrogen-rich PN-matter immediately behind the conduction front dominates. 
    For ${\omega = 0.15}$, this effect is also apparent at shorter wavelenghts. 
      
    The  influence of the position of the chemical discontinuity on the line emission 
    is seen in the 3$\times$3 panel matrix on the right part of Fig.~\ref{fig:spectra.inhomog}.
    The ionisation stratification imposed by the temperature gradient inside the bubble
    leads to line changes depending on the charge of the ion in question: The \ion{C}{v} 
    line nearly disappears for $\omega = 0.15$, while neon and magnesium lines remain 
    virtually unchanged.  The oxygen lines behave intermediately in the sense that the
    \oviii\ line is much less dependent on the amount of hydrogen-rich PN matter.
        
    Very interesting is the \ion{N}{vii} line emission at 24.78 \AA\  which is, of course, 
    absent in the ${\omega = 0.0}$ case.
   But even a small contribution of PN matter (here 3\,\%) makes this line visible
   (top-right panel in Fig.~\ref{fig:spectra.inhomog}).
   Obviously,  \ion{N}{vii} is a sensitive tracer of the contamination of originally
   hydrogen-deficient and nitrogen-free bubbles with hydrogen-rich matter by, e.g., 
   heat-conduction driven evaporation.  

\begin{figure}
\vskip -4mm
\includegraphics[width=1.03\linewidth]{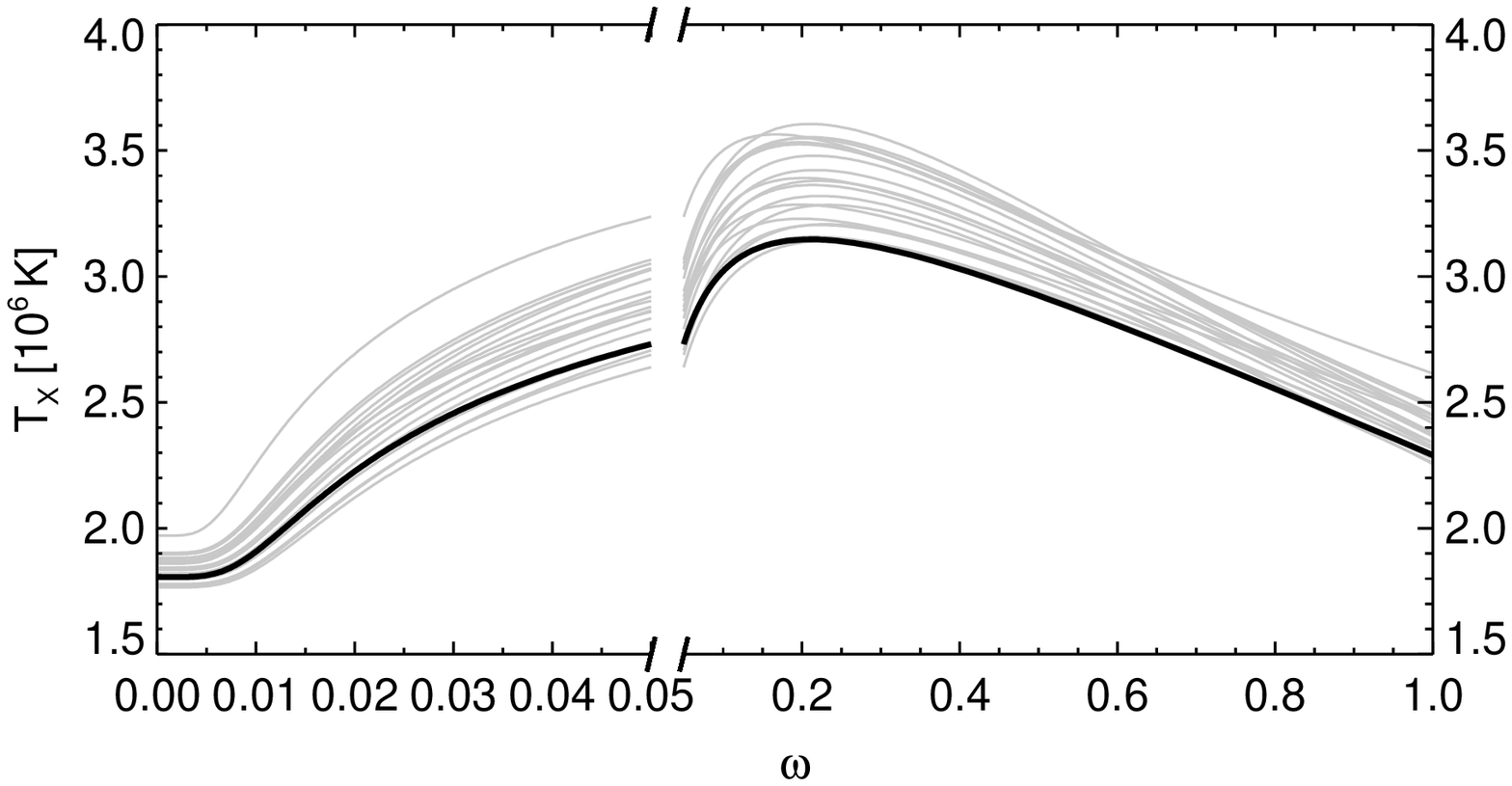}
\vskip-5mm
\includegraphics[width=1.03\linewidth]{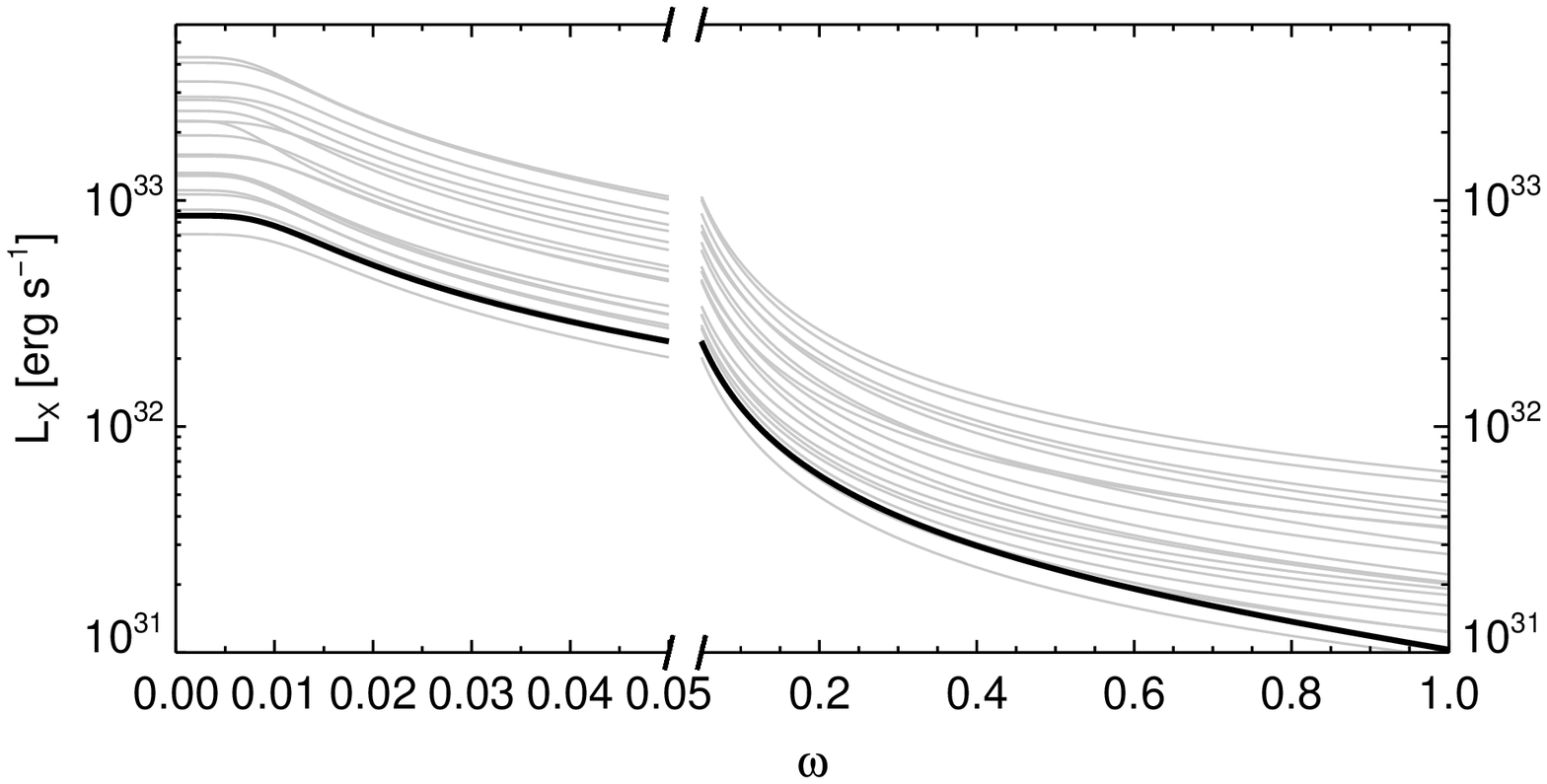}
\vskip-2mm
\caption{\label{fig:tx-variation}
         Characteristic X-ray temperature $\Tx$ (\emph{top}) and X-ray luminosity $L_{\rm X}$
         (\emph{bottom}) for a suite of chemically stratified heat-conduction bubbles
         constructed from pure WR ZP96-models that comply with both the observed oxygen 
         and neon line ratios of BD\,+30 (crosses in Fig.~\ref{fig:w.extinction}) vs. 
         ${\omega = M_{\rm PN} / (M_{\rm PN} + M_{\rm WR})}$.  
         The thick line corresponds to the model sequence originating from the 500-yr old 
         ``best-fit'' WR-model.  
         Note the break of the abscissa scale at $\omega = 0.05$.   \changed{The wavelength
         range considered is 5--40~\AA\ (0.3--2.5~keV.)}         
        }
\end{figure}

   From the intensity distributions shown in Fig.~\ref{fig:intensity}, it is evident that,
   although the bubble's temperature distribution remains unchanged, the characteristic
   X-ray temperature $\Tx$, as computed according to Eq.~(\ref{eq:tx}), does depend on the
   position of the chemical transition.  The reason are the different emissivities 
   of the WR and PN matter.    
   If the WR matter is replaced by hydrogen-rich PN matter with its lower emissivity, the
   weighting in Eq.~(\ref{eq:tx}) is more towards the inner, hotter regions, hence $\Tx$
   increases.  While the position of the chemical transition moves inwards, $\Tx$ will 
   go through a maximum since at some point the emission of the inner hydrogen-deficient
   bubble regions begins to decrease because of the low densities and falls below that 
   of the hydrogen-rich outer regions. Finally, $\Tx$ drops  to the value for a chemically
   homogeneous hydrogen-rich bubble.  
   
   This behaviour is illustrated in Fig.~\ref{fig:tx-variation}, where the run of the X-ray
   temperature vs. the mass fraction of hydrogen-rich matter, 
   ${\omega = M_{\rm PN} / (M_{\rm PN} + M_{\rm WR})}$,  is shown for a suite of originally 
   pure WR bubble models which satisfy the observed oxygen and neon line ratios (crosses in 
   Fig.~\ref{fig:w.extinction}).
   One sees that very small fractions of PN matter (${\omega \la 0.005}$) do not
   change the X-ray spectrum, i.e. $\Tx$, at all.  These mass shells are obviously still 
   too cool as to emit significantly in the soft X-ray range.
   But then $\Tx$ increases from below 2.0 MK
   to well above 3 MK (up to 3.5 MK) for ${\omega \simeq 0.2}$--0.3.    
   For larger fractions of PN matter, $\Tx$ decreases again somewhat, but even for the 
   pure PN case, $\Tx$ is higher than for the pure WR case (cf. Fig.~\ref{fig.wr.pn}
   and discussion concerning the behavour of $ \tx $ in Sect.~\ref{subsec:measurements}).
   In contrast, the X-ray luminosity decreases steadily with $\omega$, as seen in the
   bottom panel of Fig.~\ref{fig:tx-variation} (cf. also Fig.~\ref{fig:intensity}). 

   The strict correlations between temperature
   sensitive line ratios and $\Tx$ as in Figs.~\ref{fig.models}  and \ref{fig.wr.pn}
   do not exist anymore for chemically stratified bubbles.  
   Instead, these line ratios depend on the position of the chemical transition
   and the abundance differences on both sides of this transition, as we will see in the
   next section.

   \changed{We emphasise that the curves in both panels of Fig.~\ref{fig:tx-variation} are 
   not evolutionary sequences because the bubble masses increase steadily by evaporation and
   wind input. In reality, the dependance of $ \tx $ and $ \Lx $ on $ \omega $ is different 
   and can only be determined by hydrodynamical simulations.
   Cases with $ \omega \simeq 1 $ can certainly not occur in nature.}

\begin{figure*}
\sidecaption
{\includegraphics*[width=0.7\linewidth]
                {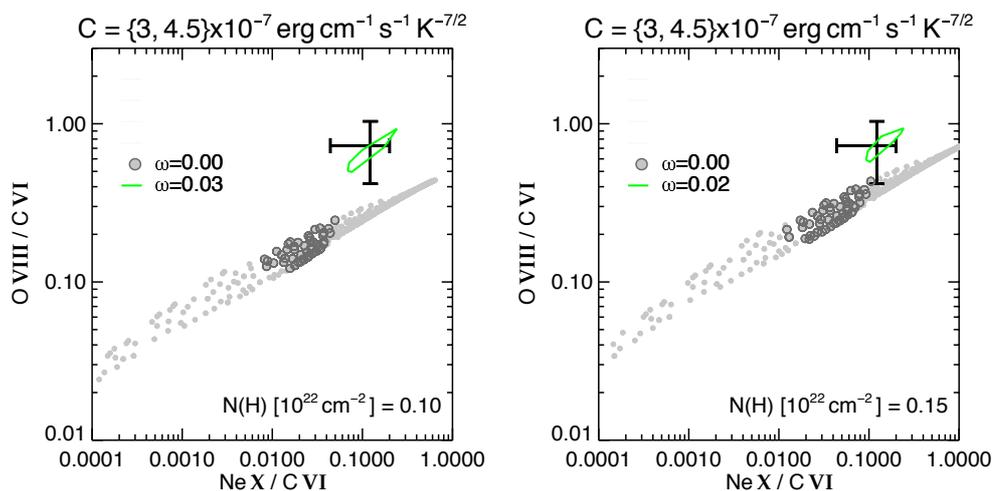} }
\caption{\label{fig:omega}
         Same as in Fig.~\ref{fig:abund.C} but now for bubble models with lower absorptions 
         as indicated in the respective panels.   Filled circles give the positions of the
         bubbles with homogeneous WR composition (${ \omega = 0 }$) which comply with the
         constraints from Fig.~\ref{fig:w.extinction} with the appropriate $ \nh $ value.   
         The green  lines enclose the corresponding positions if in the outer bubble parts
         the WR mixture is replaced by PN mixture (${ \omega = 0.03 }$, \emph{left}, and 
         $ \omega = 0.02 $, \emph{right}).      
         }
\vspace*{-2mm}
\end{figure*} 

\subsection{Constraints on the hydrogen-rich matter content of the hot bubble of BD\,+30}
\label{subsec:constraints.mixing}

  In this subsection, we apply our grid of chemically stratified HBs to
  estimate the degree of  chemical stratification within the young 
  X-ray emitting hot bubble of BD\,+30.  From our findings discussed above, three
  methods are, at least in principle, suited for this purpose: 
\begin{enumerate}
\item  
   Measuring a possible \changed{sharp} radial drop in the X-ray intensity distribution as 
   seen in Fig.~\ref{fig:intensity}.  However, the rather low numbers of X-ray photons 
   typical for \changed{existing observations of the objects in question}
   do not allow such kind of measurements.
  
\item    
   The \ion{N}{vii} line at 24.78\,\AA\ is obviously a very sensitive indicator:  
   It is virtually absent in the spectrum of a pure WR composition as  
   used here, but strong in the spectrum of a bubble with hydrogen-rich composition 
   (cf. Fig.~\ref{fig:spectra.inhomog}).  

   If BD\,+30 belongs to the class of  [WR]-type central stars that show photospheric nitrogen, 
   the \ion{N}{vii} line at 24.78\,\AA\ should 
   not be used at all as an indicator of element stratifications within wind-blown bubbles.

\item 
   Because of the ionic stratification caused by the heat conduction temperature profile
   (see Fig.~\ref{fig:ion_fracfig2}), 
   the reaction of the line strengths while the chemical discontinuity moves inwards
   depends on the degree of ionisation of the respective ions:  Lines coming from ions 
   with low degree of ionisation change \changed{early} (e.g. \cv\ and \cvi\ lines), those
   originating from ions with high degree of ionisation later (e.g. \oviii\ and \nex\ lines). 

\end{enumerate}   

   Since the methods under items one and two cannot be used for \BD\ because of the limited
   data quality, we are left with the facts described under item three and apply a diagram 
   as shown in Fig.~\ref{fig:abund.C} which has been used above to estimate the 
   carbon abundance of \BD's bubble.  Because of the relatively large wavelength separation
   of the used lines, such a diagram is sensitive to the selected hydrogen column density
   $ \nh $, and Fig.~\ref{fig:omega} gives two examples for smaller $ \nh $ than used in
   Fig.~\ref{fig:abund.C}. 
   
   This figure shows how sensitve the positions of the bubble models depend on the chosen
   value of $ \nh $.  For the low value of only $ 0.10\times 10^{22}\rm\ cm^{-2} $, our
   ZP96-bubbles with homogeneous WR composition fail completely to explain the observations.
   For the higher value of $ 0.15\times 10^{22}\rm\ cm^{-2} $, they just touch the error box.
   In both cases, chemically stratified models, constructed from the $ \omega = 0 $ models
   in the manner as described in the previous section, perform better: The \cvi\
   line emission gets weaker because of the lower line contribution from the PN matter 
   (cf. Fig.~\ref{fig:spectra.inhomog}, right grey panels), and the inhomogeneous models
   are shifted to the right and upwards in diagrams as shown in Fig.~\ref{fig:omega}, at 
   least as long the relative amount of PN matter is small and the oxygen and neon 
   lines remain virtually unchanged for small $ \omega $.  

   Figure~\ref{fig:omega} shows that agreement with the observation is achieved for 
   ${ \omega = 0.03 }$ in the case of small column density and $ \omega = 0.02 $ in the 
   case of a higher column density.   
   We note that for these relatively small amounts of evaporated matter
   the constraint posed in Fig.~\ref{fig:w.extinction} is still satisfied within the errors,
   but $ \tx $ and $ \Lx $ are a higher and lower, respectively, compared to the
   corresponding values of the ``best-fit'' bubble model (see Table~\ref{tab:best-fit}):  
   From the thick line in Fig.~\ref{fig:tx-variation}, we have
   ${ \tx = 2.2\rm\ MK }$ and $ \Lx = 5.0\times\!10^{32}\rm\ erg\,s^{-1} $ for 
   ${ \omega = 0.02 }$, and
   ${ \tx = 2.45\rm\ MK }$ and $ \Lx = 3.5\times\!10^{32}\rm\ erg\,s^{-1} $ for 
   ${ \omega = 0.03 }$.

\section{Discussion}       
\label{sec:discussion}     

   We have computed a grid of wind-blown ZP96 bubble models and used it as a tool to 
   investigate the X-ray emission-line spectrum of bubbles inside planetary nebulae. These
   bubble models provide a substantial improvement over the isothermal
   approach used to date for the line diagnostics.\footnote
   {\changed{The only exceptions are the works of \citet{2009ApJ...690..440Y} and 
   \citet{2009ApJ...695..834N} in which a two-temperature plasma model was applied for 
   getting an overall better spectral fit compared to a single-temperature plasma.}}
   Because the temperature profile of our bubble models is imposed by thermal conduction, 
   the temperature decreases in a typical way from the inner bubble boundary (the reverse 
   wind shock) towards the conduction front at the bubble/nebula interface.  Hence, 
   \changed{a corresponding density profile and ionisation stratification are established which
   both} are absent in \changed{single}-temperature plasma models.
   
   Although the physics of our model is somewhat simplified because radiative cooling 
   \changed{and dynamical effects are} neglected, we demontrated here that
   our ZP96 models can still be used to analyse the X-ray line spectra with respect
   to abundance ratios and possible chemical stratifications as long as no conclusions
   on the evolutionary stage of the object in question are made.
   In particular, the temperature profile imposed by heat conduction makes it possible to
   describe the observed values of temperature-sensitive line ratios from ions which
   reside in different parts of the bubble by a single bubble model. 

   As usual for studies of soft X-rays, also the results presented here are dependent
   on the absorption by intervening matter, characterised by the column density $ \nh $ of
   neutral hydrogen.   A certain range of $ \nh $ exist in the literature for \BD\ 
   which are based on \changed{one}-temperature plasma models which have been fitted to the
   observations by varying abundances {and} $ \nh $ \emph{simultaneously}.   
   We used a novel approach by keeping the abundances fixed and using line ratios for
   determining $ \nh $ which are only temperature-sensitive.   The advantage of our method
   is that $ \nh $ does not depend on abundances or vice versa, the disadvantage is that
   the line ratios used are only weakly dependent on absorption.

   In this context one should also mention that \citet{kastner.02} found evidence for
   nonuniform intranebular extinction in several planetary nebulae with diffuse X-ray emission, 
   also in the case of BD\,+30.  Accordingly, the X-ray emission from the bubble is not
   evenly distributed across the bubble's X-ray image.  Nevertheless, all existing analyses,
   including the present one, assume a constant, mean interstellar and intranebular extinction
   across BD\,+30's image. 
   
   Another new approach used in the present work is that we \emph{assumed} the chemical
   abundances inside \BD's bubble by using the results of existing photoshere/wind
   analyses.  These abundances predicted the correct observed line ratio of C/O, while
   the observed Ne/O line ratio demanded a considerably higher neon content   
   \changed{than originally assumed by us.}  Still, also our approach cannot test the 
   hydrogen and helium abundances, and other chemistries of \BD's bubble cannot be excluded 
   as long as C/O and Ne/O are maintained.  
   \changed{Because one cannot determine the reference abundances of hydrogen and/or helium
   in a wind-blown bubble from its X-ray spectrum, it is impossible to derive absolute
   abundances as table~3 of \citet{2009ApJ...690..440Y} suggests.}    
   
   \changed{We would like to} emphasise as well that, because the wind from
   the stellar surface feeds the bubble, any bubble chemistry other than \changed{that of the
   star} would be unrealistic.   Only the outermost bubble shells behind the
   conduction front may consist of hydrogen-rich matter with typical nebular (i.e. about
   solar) composition.  In the case of \BD, the evidence for evaporation is meager and 
   depends on the assumed value of $ \nh. $ \changed{In any case, the amount of 
   evaporated/mixed nebular matter, if there is any at all, must be very small.}
  
   We comment here that \citet{schoenetal.17} claimed the existence of evaporated 
   nebular matter in the bubble of \BD, viz. ${ \omega = 0.03 }$, using the same sort of
   diagrams as shown in Figs.~\ref{fig:abund.C} and \ref{fig:omega}.  
   This value, however, is based on no absorption and demonstrates clearly the importance 
   of a \changed{correctly} chosen value of $ \nh $.  

\changed{Our finding that the bubble of \BD\ contains virtually no hydrogen-rich nebular 
   matter also implies that there has been no significant mixing across the bubble-nebula
   interface.  This is somewhat surprising since the bubble's environment shows clearly
   signatures of dynamical instabilities.  It may indicate instead that mixing across the 
   bubble-nebula interface is less efficient than existing computations suggest.  The low
   characteristic X-ray temperature (1.8 vs. 17~MK as pedicted from observed wind speed and
   assuming an adiabatic wind shock) can thus only be the result of thermal conduction from
   the wind shock across the bubble range. We note in this context that heat conduction changes
   the bubble structure ``instantaneously'' 
   \citepalias[see equation~6 in][]{1996A&A...309..648Z} whereas mixing and evaporation 
   occur on much longer timescales.}\footnote
   {\changed{The \citet{TA.16} bubbles cannot be used for comparisons with \BD\
    because they are hydrogen-rich. As we show in our Figs.~\ref{fig:bubble.example} 
    and \ref{fig:intensity}, bubbles with WR composition are much more luminous than their
    hydrogen-rich counterparts (see also \citealt{steffen.12}).}}

\changed{The non-existence of nebular matter in the bubble is not necessarily an argument
  against the occurrence of thermal conduction for two reasons:  
  (i) It takes some time to accumulate an observable amount of nebular matter by evaporation,
      and 
  (ii) the high efficiency of radiation cooling inherent to hydrogen-poor and 
  carbon-/oxygen-rich matter may lead to a delay of evaporation as long as the stellar 
  wind power is still moderate (Sch\"onberner et al., in prep).} 

   Concerning the very high neon content of ${X_{\rm Ne} \simeq 0.05}$ found in 
   the bubble of BD\,+30, we note that high neon abundances at the surface of 
   [WR]-central stars are not unusual, although often not at this extreme level.
   \citet{LH.98} determined the neon content at the surfaces/in the wind of four cool 
   hydrogen-deficient central stars and found mass fractions of 0.02\ldots 0.04 (in one case
   even ${>\!0.04}$). 
   The recent study of five very hot hydrogen-deficient central stars by \citet{KBM.14} 
   indicates similar neon mass fractions: 0.01\ldots 0.04.  
   Burning of $^{14}$N, the ``ashes'' of
   CNO burning, into $^{22}$Ne during a thermal pulse is responsible for generating high
   neon abundances between the two burning shells of an AGB star.

   Stellar models predict a near one-to-one correspondence between the $^{22}$Ne 
   produced during a thermal pulse and the $^{14}$N hydrogen-burning ``ashes''.   
   The original abundances of the CNO matter in stellar 
   envelopes are, however, only able to generate a $^{22}$Ne mass fraction on the 1--2\,\%
   level (cf. WR mixture in Table~\ref{tab:abundances}).  A higher neon production is only
   possible if $^{12}$C from the intershell region is efficiently dredged-up into the 
   envelope (3rd dredge-up) in the aftermath of a thermal pulse and is later burned into
   \element[][14]N during the next pulse. 
     
   Indeed, the models of \citet{Kara.03} predict a maximum intershell mass fraction for
   \element[][22]Ne of 0.035 for initial masses of around 3~\Msun, rather independent of
   metallicity.  However, any prediction of the intershell neon abundance depends on the
   numerical treatment of convection:  Models which include convective overshoot at all
   convective/radiative boundaries have more efficient 3rd dredge-up and hence
   predict neon mass fractions of up to 0.05  (Herwig, priv. comm.).
             
\changed{Our analysis of BD\,+30's X-ray spectrum shows also that the neon content of
   the bubble (and thus also of the stellar surface)  is much higher than follows from the
   nebular abundance alone:}  0.05 vs. 0.02 (mass fractions, cf. Table~\ref{tab:abundances}).   
   The one-to-one correspondence predicted by AGB nucleosynthesis is obviously broken by a
   process (which may well be the last thermal pulse) that leads to a heavy 3rd dredge-up 
   (more carbon that can be burned to nitrogen \changed{and later to neon)} and
   eventually to the complete loss of the hydrogen-rich envelope 
   and the exposure of the hydrogen-depleted intershell region.   
   The youth of the nebula together with a possible very small amount of evaporated  
   hydrogen-rich nebular matter suggests that the ``separation'' between the hydrogen-rich
   stellar envelope and the virtually hydrogen-free intershell regions must have 
   occurred very recently when the object was still at or close to the tip of the AGB.

\section{Summary and conclusions}       
\label{sec:conclusions}

   We have presented and discussed in this paper a novel approach to analyse
   the X-ray line emission of wind-blown bubbles
   which is based on the analytical models developed by \citet{1996A&A...309..648Z}.    
   These models include thermal conduction and allow a time-dependent stellar wind with 
   respect to mass loss and velocity as inner boundary condition.  The property of the
   ambient medium, given by (AGB) mass-loss rate and (AGB) outflow speed, is assumed 
   to remain unchanged with time.       
   Once the boundary conditions have been specified, temperature and density structures 
   of a ZP96 bubble can easily be determined and, 
   for a given elemental composition, X-ray spectra be
   computed by means of the CHIANTI software package.  Time consuming
   hydrodynamical simulations can thus be avoided.   
   Another advantage is that any possible element stratification
   that is expected for hydrogen-deficient bubbles 
   around Wolf-Rayet central stars can \changed{be studied in a simplified approach.}       

   We applied a grid of bubble models with two compositions, one hydrogen-poor and 
   carbon/oxygen-rich (WR) and the other hydrogen-rich (PN), for a set of outer boundary
   conditions in order to be able to cover the X-ray spectrum 
   emitted from the hot bubble of \BD.   The dependence
   of the thermal conduction efficiency on the chemical composition (hydrogen-rich or
   hydrogen-poor), although small, is properly taken into account.      
   We also constructed a number of bubble models
   with a stratified chemical structure, where for given bubble size and radial temperature
   profile a certain outer mass fraction consists of hydrogen-rich PN matter instead of 
   WR matter.  The neglect of radiation cooling is unimportant for all aspects of the 
   line diagnostics. 
    
  Using a set of chemically homogeneous ZP96 bubble models with our hydrogen-deficient
  WR mixture and properties suited for applications to the high-resolution X-ray spectrum 
  of the bubble of BD\,+30, the following conclusions can be made:   
\begin{itemize}
\item   
         We determined the absorbing column density  $ \nh $ towards the bubble of \BD\
         by using only abundance-independent
         line ratios and found $ \nh = 0.20^{+0.05}_{-0.10}\times 10^{22}\rm\ cm^{-2}, $ 
         in good agreement with other values found in the literature, e.g. of
         \citet{2009ApJ...695..834N} and \citet{2009ApJ...690..440Y}. 

\item  
     Heat conduction imposes a temperature profile inside wind-blown bubbles which is 
     able to provide consistence between temperature sensitive line ratios of elements
     which reside preferentially in different parts of the bubble.     
     Thus a \emph{single} \changed{bubble model is able to reproduce simultaneously the     
     line ratios \oviii/\ovii\ and \nex/\neix.}
     \changed{The characteristic X-ray temperature ${ \tx = 1.8 }$~MK} is lower than the
     2.3~MK based on a \changed{one}-temperature plasma model of \citet{2009ApJ...690..440Y},  
     but \changed{within the errors equal to} the low-temperature value of their 
     two-component model, 1.7~MK. 
       
\item  The use of our WR abundance set derived from the \citet{2007ApJ...654.1068M} 
       photosphere/wind analyses for \BD\  (nearly identical with the work of 
       \citealt{leuetal.96}) yields line ratios for C/O and Ne/O consistent with the
       observations provided the neon abundance, not available from the photospher/wind, 
       is increased by a factor of about two. It is thus reasonable to conclude that the 
       bubble of \BD\ \changed{has the same chemical composition as the star, i.e.} 
       with fractions of He, C, O, and Ne as found by \citet{2007ApJ...654.1068M} (He, C, O)
       and this work (Ne),  \changed{and most likely no hydrogen at all 
       (see Table~\ref{tab:best-fit}).}
       
\item  Unfortunately, the abundance (relative to oxygen) of the important element nitrogen
      remains uncertain because of the weakness of the feature at 24.8~\AA.  
      We estimated an upper limit for nitrogen mass fraction which is considerably higher 
      then assumed in our WR composition.  The estimates of \citet{2009ApJ...695..834N} and
      \citet{2009ApJ...690..440Y} provide a similar value for N/O. However, their 
      error margins are consistent with the assumption of no nitrogen at all.   
       
\item  Also, the abundances (relative to oxygen) for Mg, Si, and Fe  
      could not be determined because of the weakness and uncertainty of the corresponding
      lines.  In the case of Mg, we showed explicitly that the measurements must be
      systematically wrong, most likely due to severe blending.    
       
\item  Finding a subsolar number ratio of an element with respect to oxygen, 
       does not necessarily mean that the abundance of this particular element is subsolar.
       Rather, the low element-to-oxygen ratio can be the 
      consequence of an increase of the oxygen abundance by helium burning 
      while the element in question remains unchanged by nuclear processing.  
 
\item  \changed{Although our analysis of \BD's X-ray spectrum is fully consistence with the
       assumption of a purely hydrogen-free bubble with WR composition (with increased Ne
       content), a small contamination by hydrogen-rich matter cannot fully be excluded:}
       At possibly lower hydrogen column densities as preferred here, the models predict 
       \cvi\ lines stronger than observed. 
       \changed{However, only} a very small amount of hydrogen-rich PN matter, not exceeding a
       3\,\% fraction of the bubble's mass \changed{for the (unrealistic) ${ N_{\rm H} = 0.10}$
       case,} suffices to reach agreement with the observed \cvi/\oviii\ line ratio.
 
\item
  \changed{Even with such a maximally possible} amount of hydrogen-rich matter
  contained in the bubble of BD+30, its X-ray emission is \changed{sill dominated by} the 
  hydrogen-poor and carbon- and oxygen-rich matter \changed{provided by the shocked 
  stellar wind.  \changed{No substantial mixing between bubble and nebular matter and/or
  evaporation of nebular matter has obviously occurred so far. } 
  Despite this, the observed characteristic temperature of the emitting plasma is a factor
  of ten lower than predicted for an adiabatic shock, i.e. 1.8 vs. 17~MK, and the only
  mechanism left for reducing the plasma temperature is thermal conduction
  across the bubble. This, of course, would then exclude the existence of a magnetic field. }
 
\item  The youth of the object (${\la\! 1000}$~yr) \changed{together with the practical 
       absence of evaporated or/and mixed nebular matter contamination of the bubble}
       suggests a scenario in which the 
       separation between the hydrogen-rich nebula and the nuclearly processed hydrogen-poor
       interior regions must have occurred quite recently, namely when the object was still 
       at or close to the tip of the AGB.
\end{itemize}  
              
   Our \changed{models of hot bubbles} 
   would be very useful for (re)analyses of existing X-ray spectra of low spectral resolution, 
   especially also for the chemically homogeneous bubbles around normal central stars.
   It is straightforward to degrade the spectral resolution numerically and to apply 
   extinction appropriate to the relevant hydrogen column density.  This kind of
   models would provide a more physically consistent diagnostic tool than the 
   \changed{single-temperature} plasma models used to date.

\begin{acknowledgements}
 The counsel of S.\,A. Zhekov helped us to reconstruct the mathematical framework
 of this study.
 R. Heller received funding from the Deutsche Forschungsgemeinschaft (SCHO 394/29-1) and
 from the German Space Agency (Deutsches Zentrum f\"ur Luft- und Raumfahrt) under PLATO
 Data Center grant  50OO1501. 
 This work has made use of NASA's Astrophysics Data System Bibliographic Services. 
CHIANTI is a colloborative project involving George Mason University, the University of
Michigan (USA) and the University of Cambridge (UK). 
\changed{We appreciate the anonymous referee's very careful reading of the 
  manuscript and her or his comments which helped considerably to improve the presentation of 
  this work.}
\end{acknowledgements}

\vspace{-.7cm}



\newpage

\appendix

\section{Ionisation fractions}                            
\label{app.sec:ioni.frac}

\begin{figure*}[t]
\centering
\scalebox{0.26}{\includegraphics*[trim=0cm 2.5cm 0cm 0cm,angle=90]{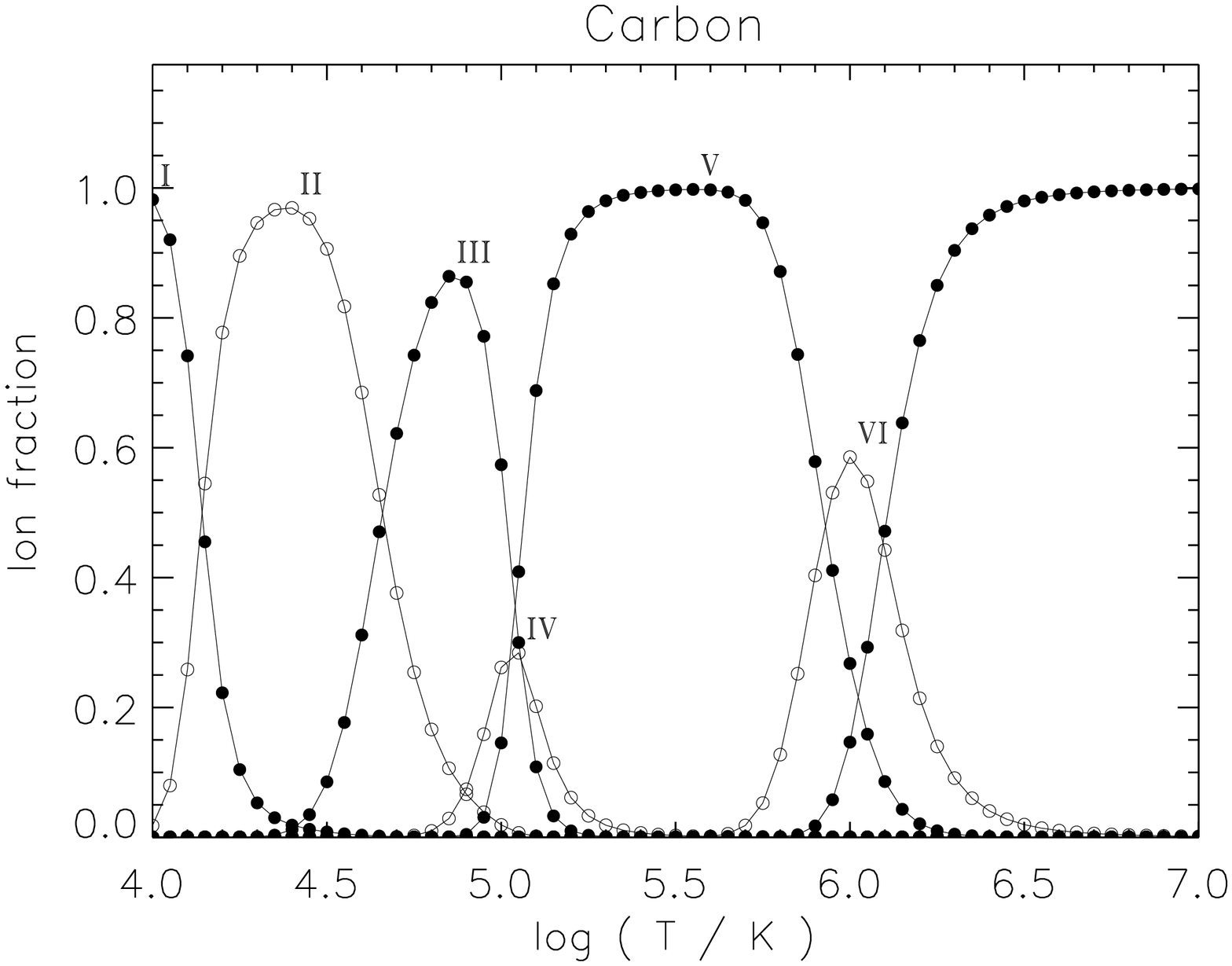}}
\hspace{0.23cm}
\scalebox{0.26}{\includegraphics*[trim=0cm 2.5cm 0cm 0cm,angle=90]{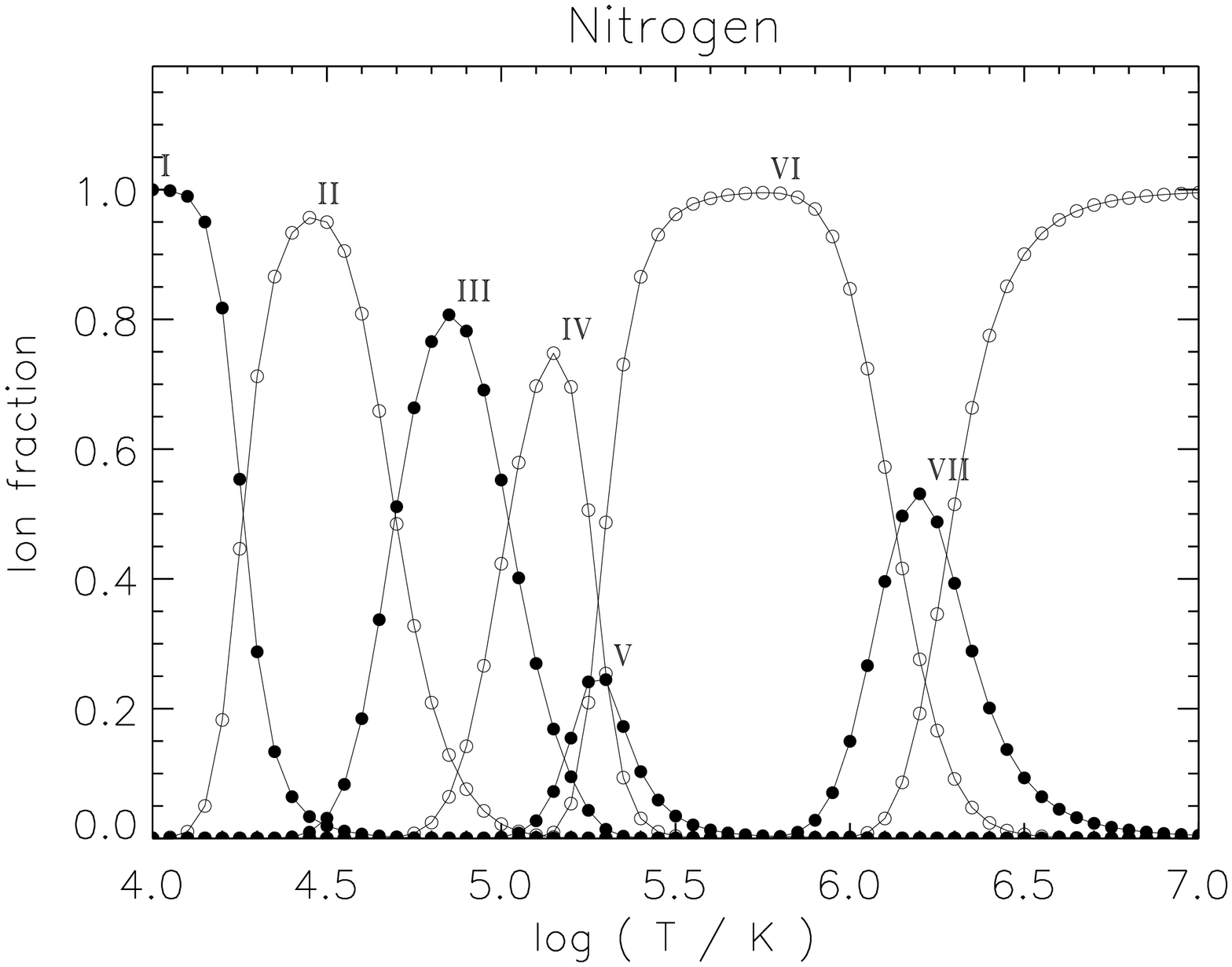}}
\hspace{0.23cm}
\scalebox{0.26}{\includegraphics*[trim=0cm 2.5cm 0cm 0cm,angle=90]{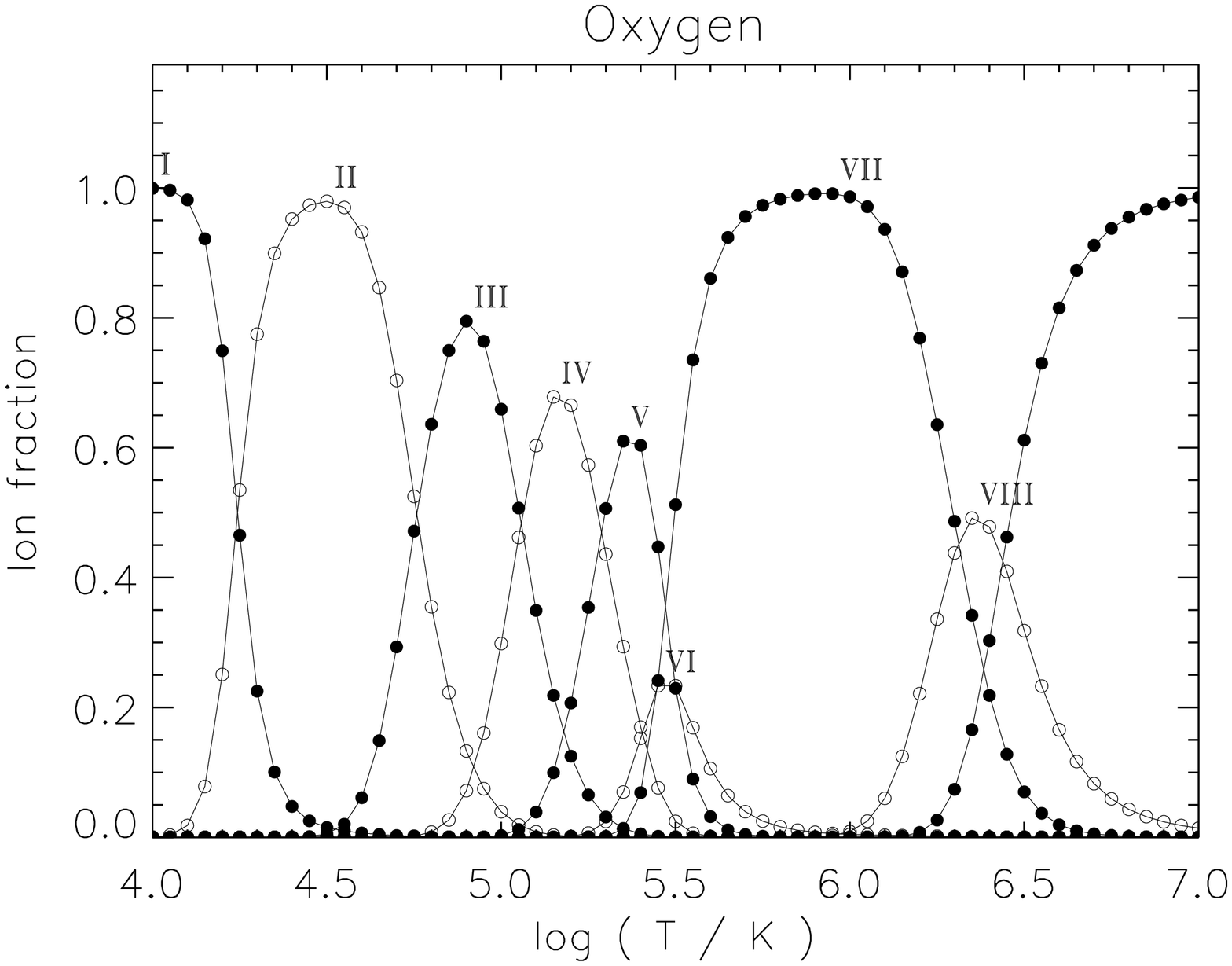}}\\
\vspace{0.0cm}
\scalebox{0.26}{\includegraphics*[trim=0cm 2.5cm 0cm 0cm,angle=90]{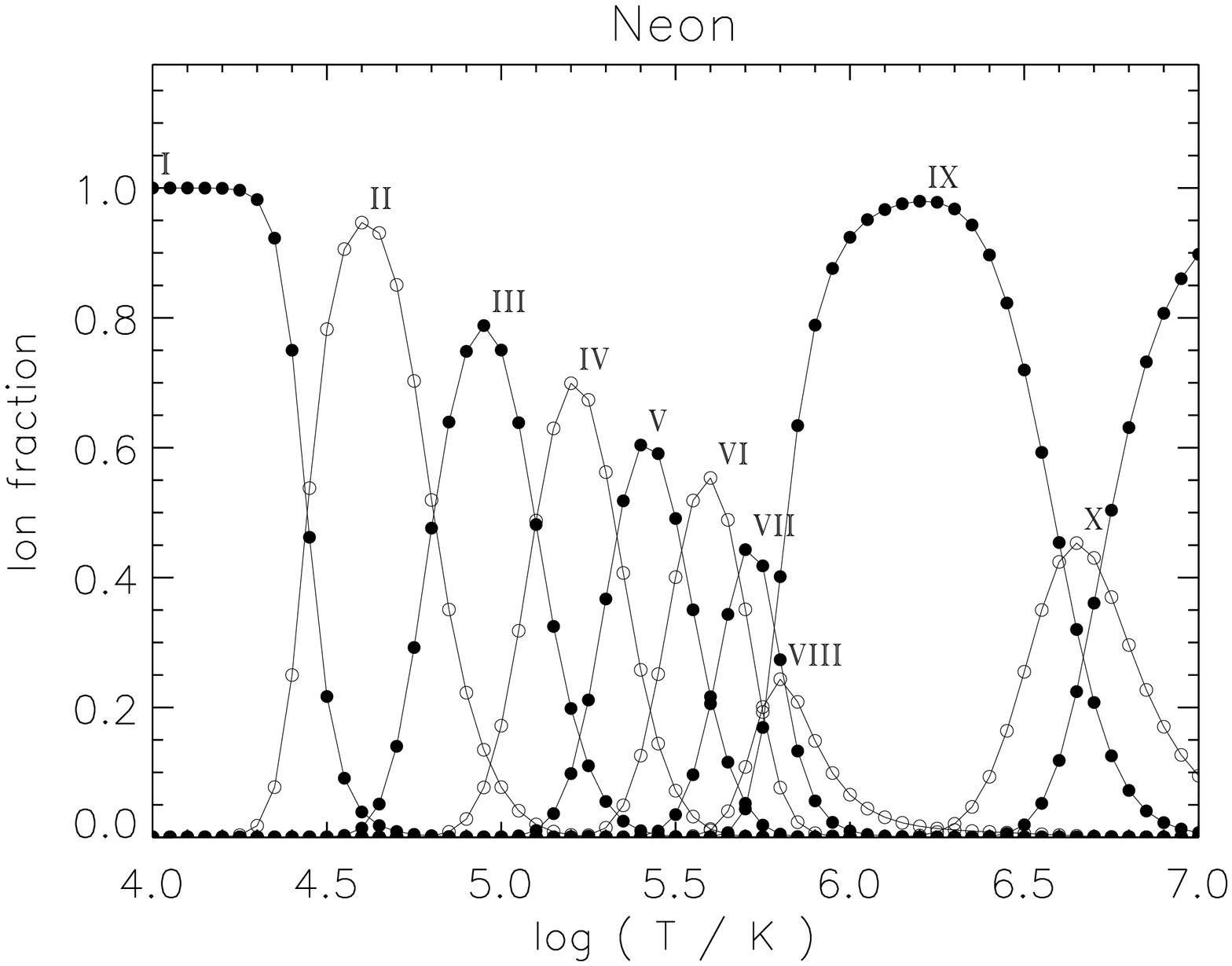}}
\hspace{0.23cm}
\scalebox{0.26}{\includegraphics*[trim=0cm 2.5cm 0cm 0cm,angle=90]{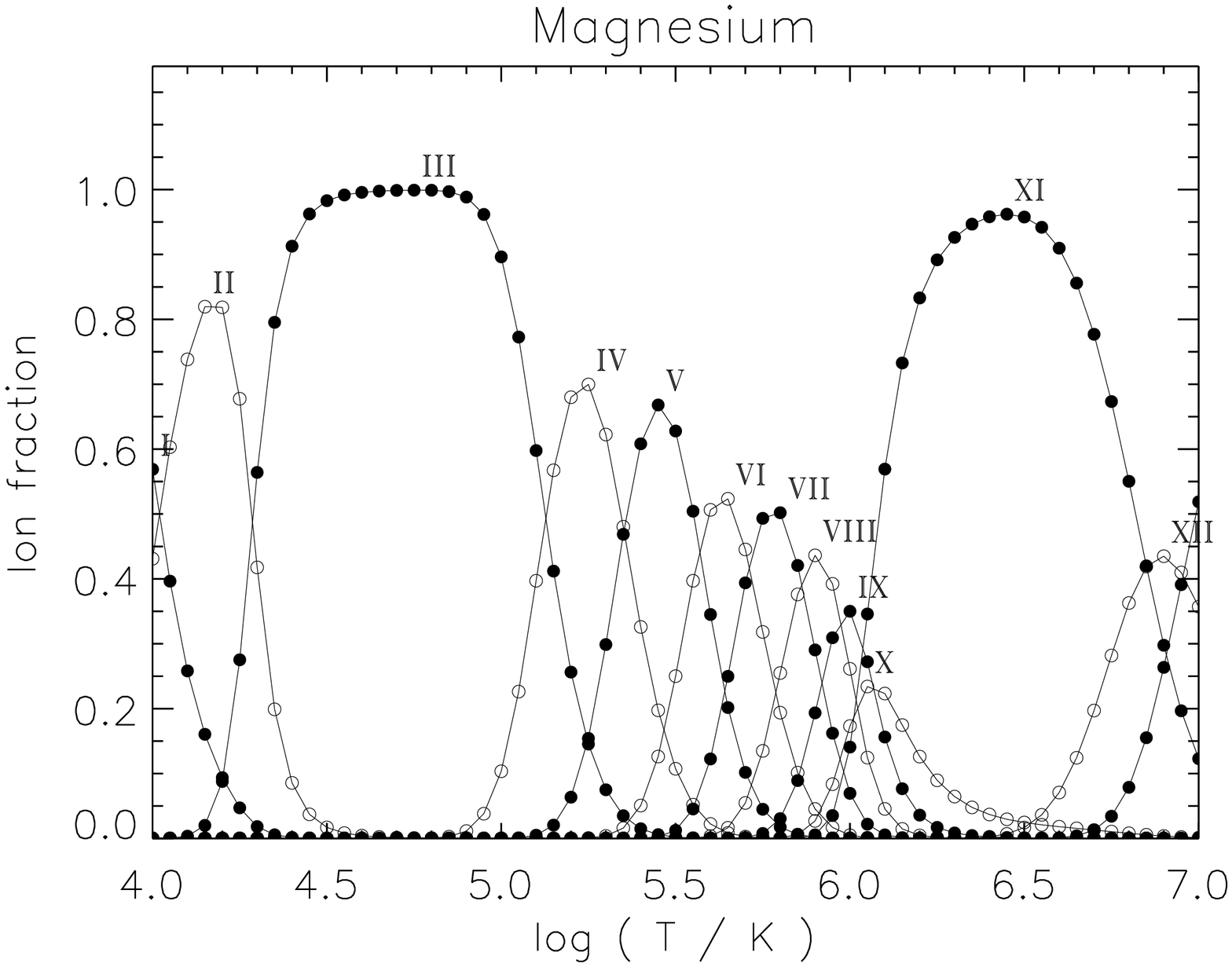}}
\hspace{0.23cm}
\scalebox{0.26}{\includegraphics*[trim=0cm 2.5cm 0cm 0cm,angle=90]{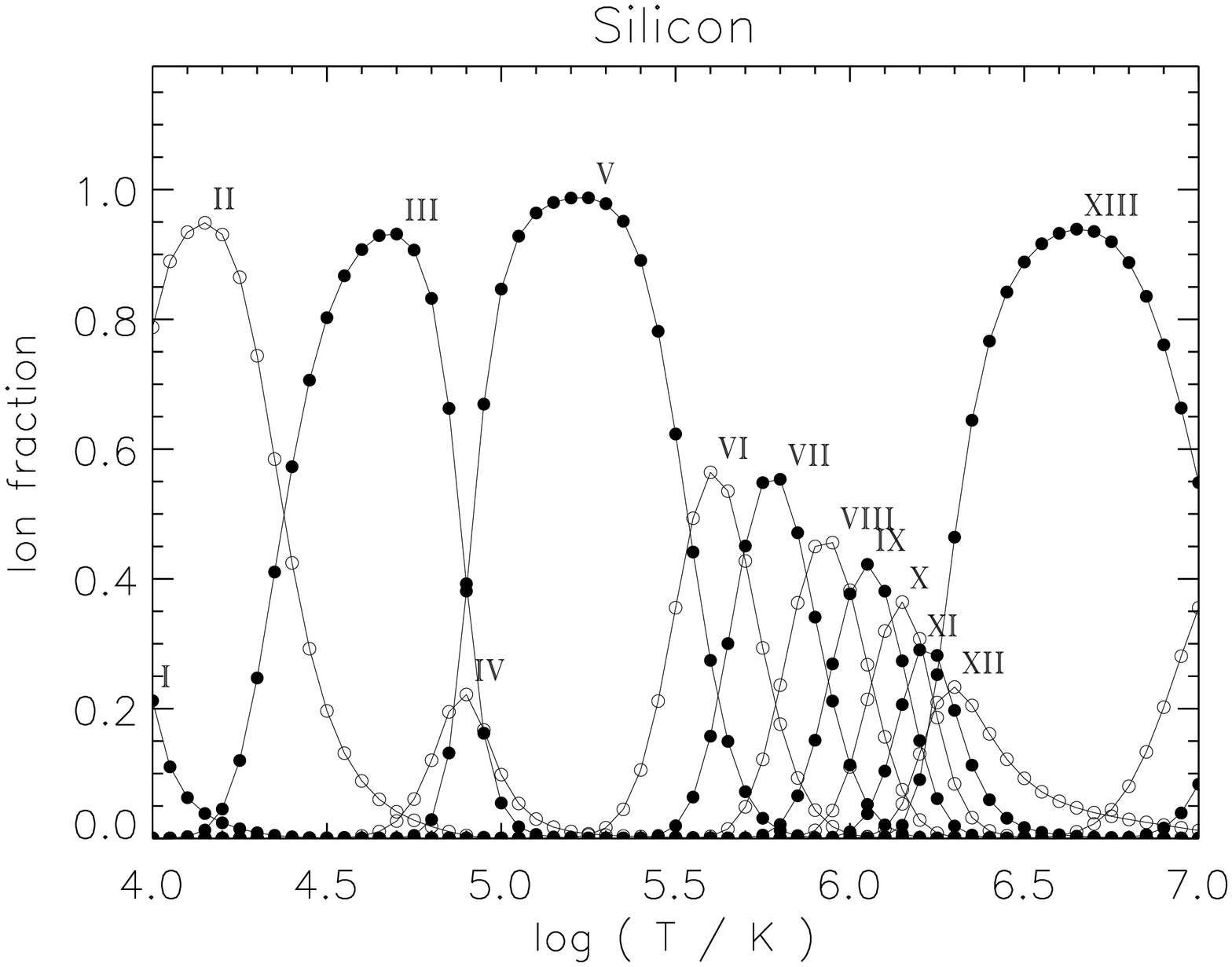}}
\vskip -1mm
\caption{Ionisation fractions of carbon (\emph{upper left}), nitrogen (\emph{upper middle}),
         oxygen (\emph{upper right}), neon (\emph{lower left}), magnesium 
         (\emph{lower middle}),  and silicon (\emph{lower right})
         as a function of temperature (from CHIANTI's default collisional ionisation 
         equilibria tables  ``chianti.ioneq''). For each element, the sum of all
         ionisation fractions at a given temperature equals one.    
         }
\label{fig:ioneq}
\vspace{-2mm}
\end{figure*}

\changed{Figure~\ref{fig:ioneq} displays the ionisation fractions of selected elements as a
         function of temperature as predicted by the Chianti software used by us.          
         At temperatures above roughly $10^6$\,K which typically occur in hot bubbles, 
         \ion{C}{v}, \ion{C}{vi}, \ion{N}{vi}, \ion{N}{vii},
         \ion{O}{vii}, \ion{O}{viii}, \ion{Ne}{ix}, \ion{Ne}{x},
         \ion{Mg}{xi}, and \ion{Mg}{xii} emission lines are important for the diagnostics.
        }

\begin{thebibliography}{}

\bibitem[{{Allen}(1976)}]{1976asqu.book.....A}
{Allen}, C.~W. 1976, {Astrophysical Quantities}

\bibitem[Anders \& Grevesse(1989)]{ag.89}
   Anders, E., \& Grevesse, N. 1989, Geochimica et Cosmochemica Acta, 53, 197

\bibitem[Arnaud et al.(1996)]{arnaudetal.96}
   Arnaud, K., Borkowski, K. J., \& Harrington, P. 1996, \apj, 462, L75

\bibitem[Bl\"ocker(1995)]{B.95}
   Bl\"ocker, T. 1995, \aap, 299, 755
   
\bibitem[Borkowski et al.(1990)]{borketal.90}
   Borkowski, K. J., Balbus, S. A., \& Fristrom, C. C. 1990, \apj, 355, 501   

\bibitem[{{Bryce} \& {Mellema}(1999)}]{1999MNRAS.309..731B}
  {Bryce}, M. \& {Mellema}, G. 1999, \mnras, 309, 731

\bibitem[{{Campbell}(1893)}]{1893PASP....5..204C}
  {Campbell}, W.~W. 1893, \pasp, 5, 204

\bibitem[Castor et al.(1975)]{castor.75}
   Castor, J., Mc Cray, R., \& Weaver, R. 1975, \apjl, L107

\bibitem[{{Chu} {et~al.}(2001){Chu}, {Guerrero}, {Gruendl}, {Williams}, \&
  {Kaler}}]{2001ApJ...553L..69C}
  {Chu}, Y.-H., {Guerrero}, M.~A., {Gruendl}, R.~A., {Williams}, R.~M., \&
  {Kaler}, J.~B. 2001, \apjl, 553, L69 
  
\bibitem[Cowie \& McKee(1977)]{cokee.77}
  Cowie, L. L., \& McKee, C. F. 1977, \apj, 211, 135  

\bibitem[Crowther et al.(2006)]{CMS.06}
   Crowther, P. A., Morris, P. W., \& Smith, J. D. 2006, \apj, 636, 1033

\bibitem[{{Dere} {et~al.}(1997){Dere}, {Landi}, {Mason}, {Monsignori Fossi}, \&
  {Young}}]{1997A&AS..125..149D}
{Dere}, K.~P., {Landi}, E., {Mason}, H.~E., {Monsignori Fossi}, B.~C., \&
  {Young}, P.~R. 1997, \aaps, 125, 149

\bibitem[{{Dere} {et~al.}(2009){Dere}, {Landi}, {Young}, {Del Zanna},
  {Landini}, \& {Mason}}]{2009A&A...498..915D}
{Dere}, K.~P., {Landi}, E., {Young}, P.~R., {et~al.} 2009, \aap, 498, 915

\bibitem[Freeman \& Kastner(2016)]{FK.16}
    Freeman, M. J., \& Kastner, J. H. 2016, \apjs, 226, 15
    
\bibitem[Freeman et al.(2014)]{2014ApJ...794...99F}
    Freeman, M. J., Montez Jr., R., Kastner, J. H., et al. 2014, \apj, 794, 99  

\bibitem[Girard et al.(2007)]{giretal.07}
    Girard, P., K\"oppen, J., \& Acker, A. 2007, \aap, 463, 265 

\bibitem[Gorenstein(1975)]{G.75}
   Gorenstein, P. 1975, \apj, 198, 95

\bibitem[{{Guerrero} {et~al.}(2001){Guerrero}, {Chu}, {Gruendl}, {Williams}, \&
  {Kaler}}]{2001ApJ...553L..55G}
{Guerrero}, M.~A., {Chu}, Y.-H., {Gruendl}, R.~A., {Williams}, R.~M., \&
  {Kaler}, J.~B. 2001, \apjl, 553, L55

\bibitem[{{Guerrero} {et~al.}(2002){Guerrero}, {Gruendl}, \&
  {Chu}}]{2002A&A...387L...1G}
{Guerrero}, M.~A., {Gruendl}, R.~A., \& {Chu}, Y.-H. 2002, \aap, 387, L1  

\bibitem[{{Guerrero} {et al.}(2012){Guerrero} {et al.}}]{guerrero.12}
   {Guerrero}, M. A., {Ruiz}, N., {Hamann}, W.-R., {et al.} 2012, \apj, 755, 129 
   
\bibitem[Hanami \& Sakashita(1987)]{HS.87}
   Hanami, H., \& Sakashita, S. 1987, \aap,181, 343   
  
\bibitem[Kahn \& Breitschwerdt(1990)]{KB.90}
   Kahn, F. D., \&  Breitschwerdt, D. 1990, \mnras, 242, 505 
   
\bibitem[Karakas(2003)]{Kara.03}
   Karakas, A. I. 2003, PhD thesis, Monash University, Australia   
   
\bibitem[Kastner et al.(2002)]{kastner.02}
   Kastner, J. H., Li, J., Vrtilek, S. D., et al. 2002, \apj, 581, 1225   
   
\bibitem[{{Kastner} {et~al.}(2008){Kastner}, {Montez}, {Balick}, \& {De
  Marco}}]{2008ApJ...672..957K}
{Kastner}, J.~H., {Montez}, Jr., R., {Balick}, B., \& {De Marco}, O. 2008,
  \apj, 672, 957
  
\bibitem[{{Kastner} {et~al.}(2012){Kastner}, {Montez}, {Balick}, {Frew},
  {Miszalski}, {Sahai}, {Blackman}, {Chu}, {De Marco}, {Frank}, {Guerrero},
  {Lopez}, {Rapson}, {Zijlstra}, {Behar}, {Bujarrabal}, {Corradi}, {Nordhaus},
  {Parker}, {Sandin}, {Sch{\"o}nberner}, {Soker}, {Sokoloski}, {Steffen},
  {Ueta}, \& {Villaver}}]{2012AJ....144...58K}
{Kastner}, J.~H., {Montez}, Jr., R., {Balick}, B., {et~al.} 2012, \aj, 144, 58

\bibitem[{{Kastner} {et~al.}(2000){Kastner}, {Soker}, {Vrtilek}, \&
  {Dgani}}]{2000ApJ...545L..57K}
{Kastner}, J.~H., {Soker}, N., {Vrtilek}, S.~D., \& {Dgani}, R. 2000, \apjl,
  545, L57
  
\bibitem[Keller et al.(2014)]{KBM.14}  
  Keller, G. R., Bianchi, L., \& Maciel, W. J. 2014, \mnras, 442, 1379  
  
\bibitem[Kingsburgh \& Barlow(1994)]{KB.94}
  Kingsburgh, R. L., \& Barlow, M. J. 1994, \mnras, 271, 257  
  
\bibitem[Kreysing et al.(1992)]{krey.92}
  Kreysing, H. C., Diesch, C., Zweigle, J., et al. 1992, \aap, 264, 623  
  
\bibitem[Leuenhagen \& Hamann(1998)]{LH.98}  
   Leuenhagen, U., \& Hamann, W.-R. 1998, \aap, 330, 265    
  
\bibitem[Leuenhagen et al.(1996)]{leuetal.96}
   Leuenhagen, U., Hamann, W.-R., \& Jefferey, C. S. 1996, \aap, 312, 167   

\bibitem[{{Li} {et~al.}(2002){Li}, {Harrington}, \&
  {Borkowski}}]{2002AJ....123.2676L}
{Li}, J., {Harrington}, J.~P., \& {Borkowski}, K.~J. 2002, \aj, 123, 2676

\bibitem[Lodders(2010)]{lodders.10}
  Lodders, K. 2010, in Principles and Perspectives in Cosmochemistry, 
          Astrophys. \& Space Science Proc., 16, 379

\bibitem[{{Marcolino} {et~al.}(2007){Marcolino}, {Hillier}, {de Araujo}, \&
  {Pereira}}]{2007ApJ...654.1068M}
{Marcolino}, W.~L.~F., {Hillier}, D.~J., {de Araujo}, F.~X., \& {Pereira},
  C.~B. 2007, \apj, 654, 1068

\bibitem[Mellema \& Lundqvist(2002)]{ML.02}
  Mellema, G., \& Lundqvist, P. 2002, \aap, 394, 901

\bibitem[{{Montez} \& {Kastner}(2013)}]{2013ApJ...766...26M}
{Montez}, Jr., R. \& {Kastner}, J.~H. 2013, \apj, 766, 26

\bibitem[{{Montez} {et~al.}(2010){Montez}, {De Marco}, {Kastner}, \&
  {Chu}}]{2010ApJ...721.1820M}
{Montez}, Jr., R., {De Marco}, O., {Kastner}, J.~H., \& {Chu}, Y.-H. 2010,
  \apj, 721, 1820

\bibitem[Montez et al.(2015)]{2015ApJ...800....8M}
   Montez, R., Jr., Kastner, J. H., Balick, B., et al. 2015, \apj, 800, 8

\bibitem[Morrison \& McCammon(1983)]{MMc.83}
   Morrison, R., \& McCammon, D. 1983, \apj, 270, 119

\bibitem[Murashima et al.(2006)]{muaretal.06}
  Murashima, M., Kokubun, K., Makishima, K, et al. 2006, \apj, 647, L131

\bibitem[{{Nordon} {et~al.}(2009){Nordon}, {Behar}, {Soker}, {Kastner}, \&
  {Yu}}]{2009ApJ...695..834N}
{Nordon}, R., {Behar}, E., {Soker}, N., {Kastner}, J.~H., \& {Yu}, Y.~S. 2009,
  \apj, 695, 834
  
\bibitem[Pauldrach et al.(1988)]{pauldrach.88}
   Pauldrach, A., Puls, J., Kudritzki, R.-P., M\'endez, R. H., \& Heap, S. R. 1988, 
      \aap, 207, 123
         
\bibitem[Perinotto et al.(2004)]{peretal.04}
   Perinotto, M., Sch\"onberner, D., Steffen, M., \& Calonaci, C. 2004, \aap, 414, 993
   
\bibitem[Pittard et al.(2001a)]{pit.01a}
   Pittard, J. M., Dyson, J. E., \&  Hartquist, T. W. 2001, \aap, 367, 1000  
   
\bibitem[Pittard et al.(2001b)]{pit.01b} 
   Pittard, J. M., Hartquist, T. W., \& Dyson, J. E. 2001, \aap, 373, 1043 

\bibitem[Pottasch \& Bernard-Salas(2006)]{pott.06}
   Pottasch, S. R., \& Bernard-Salas, J. 2006, \aap, 457, 189

\bibitem[{{Ruiz} {et~al.}(2011){Ruiz}, {Guerrero}, {Chu}, \&
  {Gruendl}}]{2011AJ....142...91R}
{Ruiz}, N., {Guerrero}, M.~A., {Chu}, Y.-H., \& {Gruendl}, R.~A. 2011, \aj,
  142, 91
  
\bibitem[Ruiz et al.(2013)]{ruiz.13}
   Ruiz, N., Chu, Y.-H., Gruendl, R. A., et al. 2013, \apj, 767, 35  

\bibitem[{{Sandin} {et~al.}(2012){Sandin}, {Steffen}, {Jacob},
  {Sch{\"o}nberner}, {R{\"u}hling}, {Hamann}, \& {Todt}}]{2012IAUS..283..494S}
{Sandin}, C., {Steffen}, M., {Jacob}, R., {et~al.} 2012, in Planetary Nebulae:
        An Eye to the Future, eds. A. Manchado, L. Stanghellini, \& D. Sch\"onberner, 
        IAU Symposium No. 283, 494
        
 \bibitem[{{Sandin} et~al.(2016)}]{sandin16}
{Sandin}, C., Steffen, M., Sch\"onberner, D., \& R\"uhling, U. 2016, \aap, 586, A57,
            (Paper I)       
        
\bibitem[Sch\"onberner et al.(2017)]{schoenetal.17}  
   Sch\"onberner, D., Jacob, R., Heller, R., \& Steffen, M. 2017, in Planetary Nebulae:
        Multi-wavelength Probes of Stellar and Galactic Evolution, eds. X. Liu, 
        L. Stanghellini, \& A. Karakas, IAU Symposium No. 323, 109       
        
\bibitem[Sch\"onberner et al.(2014)]{schoenetal.14} 
   Sch\"onberner, D., Jacob, R., Lehmann, H., et al. 2014, AN, 335, 378        
        
\bibitem[Sch\"onberner et al.(2005)]{schoenetal.05}
   Sch\"onberner, D., Jacob, R., Steffen, M., et al. 2005, \aap, 431, 963        

\bibitem[Spitzer(1962)]{spitzer}
   Spitzer, L. 1962, Physics of Fully Ionized Gases, 2nd edn. (J. Wiley \& Sons)
   
\bibitem[Steffen et al.(2012)]{steffen.12}
   Steffen, M., Sandin, C., Jacob, R,, \& Sch\"onberner, D. 2012, in Planetary Nebulae:
        An Eye to the Future, eds. A. Manchado, L. Stanghellini, \& D. Sch\"onberner, 
        IAU Symposium No. 283, 215  

\bibitem[{{Steffen} {et~al.}(2008){Steffen}, {Sch{\"o}nberner}, \&
  {Warmuth}}]{2008A&A...489..173S}
{Steffen}, M., {Sch{\"o}nberner}, D., \& {Warmuth}, A. 2008, \aap, 489, 173

\bibitem[Stute \& Sahai(2006)]{SS.06}
   Stute, M., \& Sahai, R. 2006, \apj, 651, 882

\bibitem[Toal\'a \& Arthur(2014)]{TA.14}
   Toal\'a, J. A., \& Arthur, S. J. 2014, \mnras, 443, 3486
   
\bibitem[Toal\'a \& Arthur(2016a)]{TA.16}  
   Toal\'a, J. A., \& Arthur, S. J. 2016a, \mnras, 463, 4438 
   
\bibitem[Toal\'a \& Arthur(2016b)]{TA.16b}  
   Toal\'a, J. A., \& Arthur, S. J. 2016b, \mnras, 464, 178   
   
\bibitem[{Todt} \& {Hamann}(2015)]{TH.15} 
   {Todt}, H., \& {Hamann}, W.-R. 2015, in Wolf-Rayet Stars, 
   Proc. of an Intern. Workshop,
   eds. W.-R. Hamann, A. Sander \& H. Todt,  Universit{\"a}tsverlag Potsdam, p. 253 
   
\bibitem[Villaver et al.(2002)]{villa.02}
   Villaver, E., Manchado, A., \& Garc\'ia-Segura, G. 2020, \apj, 581, 1204    

\bibitem[{{Weaver} {et~al.}(1977){Weaver}, {McCray}, {Castor}, {Shapiro}, \&
  {Moore}}]{1977ApJ...218..377W}
{Weaver}, R., {McCray}, R., {Castor}, J., {Shapiro}, P., \& {Moore}, R. 1977,
  \apj, 218, 377

\bibitem[{{Yu} {et~al.}(2009){Yu}, {Nordon}, {Kastner}, {Houck}, {Behar}, \&
  {Soker}}]{2009ApJ...690..440Y}
{Yu}, Y.~S., {Nordon}, R., {Kastner}, J.~H., {et~al.} 2009, \apj, 690, 440

\bibitem[Zhekov \& Perinotto(1996)]{1996A&A...309..648Z}
{Zhekov}, S.~A., \& {Perinotto}, M. 1996, \aap, 309, 648, (ZP96)

\bibitem[Zhekov \& Perinotto(1998)]{ZP.98}
  Zhekov, S. A., \& Perinotto, M. 1998, \aap, 334, 239 


\end{thebibliography}
\end{document}